\documentclass[final,5p,times,twocolumn]{elsarticle}
\usepackage{lineno,hyperref}
\usepackage{graphicx}
\usepackage{bm}
\usepackage{float}
\usepackage{epstopdf}
\usepackage{caption}
\usepackage{multirow}
\usepackage{amsmath}
\usepackage{amssymb}
\usepackage{xcolor}
\usepackage{booktabs}
\usepackage{algorithm}
\usepackage{algpseudocode}
\usepackage{amssymb,amsfonts}
\usepackage{amsmath}
\usepackage{xcolor}
\usepackage{longtable,pdflscape}
\usepackage{booktabs}
\usepackage{supertabular}
\usepackage{array}
\usepackage{graphicx}
\usepackage{subfigure}
\usepackage{dcolumn}
\usepackage{bm}
\usepackage[T1]{fontenc}
\usepackage{mathptmx}
\usepackage{multirow}
\usepackage{array}
\usepackage{float}
\usepackage{lscape}

\biboptions{sort&compress}
\modulolinenumbers[5]

\journal{}

\begin{document}

\begin{frontmatter}

\title{Hydrodynamic and thermodynamic non-equilibrium characteristics of shock waves: Insights from the discrete Boltzmann method}

 \author[label0,label1]{Dejia Zhang}
 \address[label0]{School of Physical Science and Technology, Guangxi University, Nanning 530004, China}
 \address[label1]{National Key Laboratory of Computational Physics, Institute of Applied Physics and Computational Mathematics, Beijing 100088, China}

 \author[label2]{Yanbiao Gan\corref{mycorrespondingauthor}}
\address[label2]{Hebei Key Laboratory of Trans-Media Aerial Underwater Vehicle, North China Institute of Aerospace Engineering, Langfang 065000, China}
\cortext[mycorrespondingauthor]{Corresponding author at: Hebei Key Laboratory of Trans-Media Aerial Underwater Vehicle, North China Institute of Aerospace Engineering, Langfang 065000, China}
\ead{Gan@nciae.edu.cn}

 \author[label3]{Bin Yang}
\address[label3]{School of Energy and Safety Engineering, Tianjin Chengjian University, Tianjin 300384, China}

 \author[labels]{Yiming Shan }
\address[labels]{College of Forensic Science, Criminal Investigation Police University of China, Shenyang 110854, China}

 \author[label1,labelx2,labelx3,labelx4]{Aiguo Xu \corref{mycorrespondingauthor}}

\address[labelx2]{State Key Laboratory of Explosion Science and Safety Protection, Beijing Institute of Technology, Beijing 100081, China}
\address[labelx3]{National Key Laboratory of Shock Wave and Detonation Physics, Mianyang 621999, China}
\address[labelx4]{HEDPS, Center for Applied Physics and Technology, and College of Engineering, Peking University, Beijing 100871, China}
\cortext[mycorrespondingauthor2]{Corresponding author at: National Key Laboratory of Computational Physics, Institute of Applied Physics and Computational Mathematics,
P. O. Box 8009-26, Beijing 100088, China}
\ead{Xu\_Aiguo@iapcm.ac.cn}

\begin{abstract}
Shock waves are typical non-equilibrium phenomena in nature and engineering, driven by hydrodynamic non-equilibrium (HNE) and thermodynamic non-equilibrium (TNE) effects. However, the mechanisms underlying these non-equilibrium effects are not fully understood. In this study, we develop the discrete Boltzmann method (DBM) by directly discretizing velocity space, allowing for the adequate capture of higher-order HNE and TNE effects. To reveal these mechanisms, we derive analytical solutions for distribution functions and TNE quantities at various orders using Chapman-Enskog analysis, 
although DBM simulations do not rely on these theoretical derivations.
Using argon shock structures as a case study, DBM simulations of interface profiles and thickness at the macroscopic level agree well with experimental data and direct simulation Monte Carlo results. At the mesoscopic level, DBM-derived distribution functions and TNE measures closely match their corresponding analytical solutions.
The effect of Mach number on HNE is analyzed by examining the shape and thickness of density, temperature, and velocity interfaces. Key findings include: (i) Mach number induces a two-stage effect on macroscopic quantities, influencing both interface smoothness and thickness, and (ii) as Mach number increases, the region of strong compressibility shifts from the outflow region to the inflow region.
As for TNE characteristics, increasing Mach number significantly amplifies TNE intensity and expands the non-equilibrium region. 
Distribution functions at different shock locations exhibit variations in amplitude, shape, and deviation from equilibrium, are analyzed theoretically. These findings highlight the close connection between macroscopic and mesoscopic non-equilibrium behaviors and emphasize that non-equilibrium manifestations depend on the analytical perspective.
This research provides kinetic insights into the multiscale nature and effects of non-equilibrium characteristics in shock waves, offering theoretical references for constructing kinetic models that describe different types and orders of non-equilibrium effects.

\end{abstract}

\begin{keyword}
 shock wave \sep thermodynamic non-equilibrium \sep hydrodynamic non-equilibrium \sep discrete Boltzmann method
\end{keyword}

\end{frontmatter}

\section{\label{sec:level1} Introduction}

Shock waves, a hallmark of supersonic flow, are prevalent in both natural
phenomena and engineering applications. Examples include collisionless
shocks during supernova remnant evolution \cite{Hwang2005}, laser-induced
shocks in inertial confinement fusion (ICF) \cite{Liu2023POP}, shocks
encountered in supersonic and hypersonic vehicles \cite{Gaitonde2013}, and
those generated by medical devices for kidney stone treatment \cite%
{Lingemen2009NRU}. With the rapid advancements in aerospace and
energy-related fields, shock wave research has gained increasing attention.
For example, shock wave/boundary layer interactions in
scramjet engines are extensively studied due to their critical role in
enhancing intake and combustion efficiency \cite%
{Gaitonde2023ARFM,JiangH2023AST,Bao2022POF}. Shock wave propagation and
evolution significantly affect fluid system performance. In supersonic
combustion systems, shock waves deform fuel droplets, thereby affecting fuel
mixing, combustion efficiency, and overall performance metrics \cite%
{Ranjan2011ARFM,Zhang2019CNF}. In shock tubes, interactions between shock
waves and walls or mechanical interfaces generate complex wave structures,
which in turn shape intricate flow dynamics \cite%
{Guo2024JFM,Ding2024JFM,Zhai2024SCPMA}. Moreover, the growing importance of
small-scale structures and rapid dynamics in engineering highlights the need
to investigate the internal structure of shock waves.. Cai \emph{et al.}
demonstrated that in indirectly driven laser ICF, increased ion-ion mean
free paths during gold-wall and target plasma interactions lead to
collisionless shock waves, significantly affecting implosion neutron yield
\cite{Cai2020HPPB}.

Shock wave flow, as a typical form of non-equilibrium flow, driven by
small-scale structures and rapid dynamic modes, has been extensively studied
in recent years due to its importance in various scientific and engineering
contexts \cite%
{Qiu2021PRE,Yang2025AST,ZhangJC2024AST,ZhangWQ2022AST,ZhangJC2025AST,Jin2025AST, GuoSQ2024AST,UY2023108520,GaoYG2022AST,GuoJH2019AST}%
. The shock wave thickness is several times larger than the mean molecular
free path, leading to a highly discrete internal structure that deviates
significantly from thermodynamic equilibrium. Mott-Smith first identified
that the strong non-equilibrium nature of shock waves results in bimodal
molecular velocity distributions within the shock region \cite%
{Mott-Smith1951}. Under these conditions, he derived an analytical solution
for the spatial structure of strong shock waves by solving the transport
equation. Bird later developed the direct simulation Monte Carlo (DSMC)
method to solve the Boltzmann equation using a Monte Carlo algorithm \cite%
{Bird1976book}. The DSMC method has been extensively validated for various
non-equilibrium flows, including shock wave structures. Alsmeyer measured
the density profiles of argon and nitrogen shock waves with Mach numbers
ranging from 1.5 to 10.0 using electron beam experiments \cite%
{Alsmeyer1976JFM}. His study compared the results with Bird's Monte Carlo
simulations, Mott-Smith's theory, the Navier-Stokes (NS) equations, and the
Burnett equations. Pham-Van-Diep \emph{et al.} measured the velocity
distribution function in a Mach 25 helium shock wave \cite{Pham1989Science}.
Their experimental results were the first to confirm the bimodal nature of
the velocity distribution predicted by Mott-Smith's theory.

Initially, the NS equations, based on continuity assumptions and
near-equilibrium approximations, were employed to model shock structures
\cite{Gilbarg1953JRMA}. However, the NS equations neglect second-order and
higher-order nonequilibrium effects, limiting their applicability to Mach
numbers below approximately 1.3 \cite{Alsmeyer1976JFM}. To extend the
applicability of continuum models, Foch applied the Burnett equations to
model shock structures at Mach numbers up to 1.9 \cite{Foch1973}. The
Burnett equations, derived from Chapman-Enskog (CE) theory, incorporate
second-order Knudsen (Kn) number effects, offering improved accuracy over
the NS equations \cite{Foch1973,Burnett1936}. Beyond continuum-based
approaches, kinetic methods based on the Boltzmann equation have emerged as
powerful tools for studying shock structures. These methods include the
direct simulation Monte Carlo (DSMC) method, improved Burnett and
super-Burnett equations from kinetic theory \cite{Zhao2014AST}, Grad's
moment method \cite{Grad1949}, regularized 13-moment equations \cite%
{Struchtrup2003POF,Torrilhon2004JFM}, and the lattice Boltzmann method \cite%
{2001-Succi-Boltzmann,Hejranfar2020AST,Feng2020PRE,Zhao2020POF,Huang2024POF,Fei2023JFM,
ChenZ2021PRE,Busuioc2024PRF}. Recent advances in
kinetic methods include the unified gas-kinetic scheme (UGKS) \cite%
{Xu2010JCP,Zhang2024CNSNS,ZhuYJ2019JCP}, discrete unified gas-kinetic scheme
(DUGKS) \cite{Guo2015PRE}, unified gas-kinetic wave-particle methods \cite%
{Liu2020JCP,Wei2024JCP,Liu2022AMS}, and kinetic models for monatomic gas
mixtures \cite{Li2024JFM}. Additionally, nonlinear coupled constitutive
relations \cite{Chen2021AST,Yang2025AST}, gas-kinetic unified algorithms
\cite{Li2008IJCFD,Li2015PAS}, and particle-on-demand-based kinetic schemes
\cite{Kallikounis2022PRE,JiY2024JFM} have further advanced shock structure
modeling. Other effective methods include the discrete velocity method \cite%
{Yang2022POF} and the discrete Boltzmann method \cite{Xu2024FOP}, both of
which provide valuable insights into the intricate dynamics of shock
structures under various flow regimes.

Shock waves are widely used as benchmarks to validate algorithm accuracy in
current studies. However, the non-equilibrium characteristics and underlying
mechanisms of shock waves, which critically influence fluid system
performance, remain poorly understood. Within the framework of DBM,
non-equilibrium effects can be classified into two categories: hydrodynamic
non-equilibrium (HNE) and thermodynamic non-equilibrium (TNE) \cite%
{Xu2024FOP}. HNE describes non-equilibrium through spatial variations in
macroscopic quantities (e.g., density, temperature, velocity, and pressure),
while TNE reflects mesoscopic kinetic characteristics, such as distribution
functions and TNE-specific quantities. TNE provides the physical foundation
for understanding HNE.

The discrete Boltzmann model (DBM) is a kinetic approach for modeling
discrete/non-equilibrium flows and analyzing complex physical fields \cite%
{Xu2024FOP,Xu2022CMK}. From a physical modeling perspective, DBM is a direct
kinetic approach that eliminates the need to derive and solve complex
hydrodynamic equations, such as those in the Burnett and Grad moment
methods. The physical function of DBM is equivalent to the evolution
equations, incorporating not only conserved moments but also some closely
related non-conserved moments. The DBM simulation does not rely on the CE
multiscale analysis. In DBM, CE analysis serves two purposes: (i)
identifying non-conserved kinetic moments relevant to system states and
features, and (ii) offering an intuitive understanding of discrete/non-equilibrium
mechanisms. 
For complex physical field analysis, DBM provides a set of
analytical schemes capable of checking, describing, manifesting and analyzing the discrete/nonequilibrium states and effects,
which are difficult to capture using NS model and other kinetic methods. DBM
has been extensively applied to non-equilibrium phenomena, including
multiphase flows \cite{2015-Gan-SM-PS,Gan2023JFM}, hydrodynamic
instabilities \cite%
{Chen2022PRE,Shan2023CTP,Lai2016PRE,Li2024POF,ChenJ2025FOP}, microscale
flows \cite{ZhangYD2023POF}, combustion and detonation dynamics \cite%
{Lin2018CNF}, and plasma systems \cite{Song2024POF}. Numerous studies have
highlighted the critical role of thermodynamic non-equilibrium (TNE)
behavior in determining system performance \cite%
{ZhangYD2023POF,Chen2022PRE,Shan2023CTP,Lai2016PRE}. Notable examples
include: Chen \emph{et al.} examined how the shock Mach number influences
the interplay between Rayleigh-Taylor and Richtmyer-Meshkov (RM)
instabilities, analyzing its correlation with TNE and HNE behaviors \cite%
{Chen2018POF}. Song \emph{et al.} proposed that non-equilibrium
characteristic quantities serve as physical criteria for assessing whether a
magnetic field can prevent interface inversion in plasma RM instability
systems \cite{Song2024POF}. Zhang \emph{et al.} investigated how shock waves
influence entropy production during shock-bubble interactions \cite%
{Zhang2023POF}. Gan \emph{et al.} investigated HNE-TNE interactions in phase
separation and demonstrated that TNE intensity provides a robust criterion
for distinguishing spinodal decomposition from domain growth \cite%
{2015-Gan-SM-PS,Gan2023JFM}.

To further extract, illustrate and study the HNE and TNE effects overlooked
by other simulation methods, we develop a DBM model that incorporates
sufficiently higher-order Knudsen number effect, and investigate the HNE and
TNE behaviors inside argon normal shock structures with Mach numbers ranging
from 1.2 to 12.0. The remainder of this paper is organized as follows. The
remainder of this paper is organized as follows. Section \ref{MMCPFA}
outlines the DBM modeling framework and the analysis scheme for TNE
characteristics. Section \ref{CE-analysis} presents the derivation of TNE
quantities. Section \ref{Numerical-results} details the simulation setup and
presents the numerical results of the internal shock structure. Finally,
Section \ref{sec:conclusion} summarizes the key findings of this study.

\section{DBM modeling and complex physical field analysis method}

\label{MMCPFA}

\subsection{DBM equation for normal shock wave}



For a normal shock wave propagating along the $x$-axis, the simplified
Boltzmann equation with the Shakhov collision model is employed:
\begin{equation}
\frac{\partial f}{\partial t}+ v_{x} \cdot\frac{\partial f}{\partial x}=-%
\frac{1}{\tau}(f-f^{s}),  \label{Eq.fs}
\end{equation}
with the Shakhov distribution function
\begin{equation}
f^s = f^{eq} + {f^{eq}} \cdot \left[ (1 - \Pr ) \cdot c_x q_x \cdot \frac{
\frac{c^2 + \eta^2}{RT} - (n + 3) }{(n + 3) pRT} \right].
\end{equation}
and the equilibrium distribution function
\begin{equation}
f^{eq}=\rho {\left(\frac{1}{{2\pi RT}}\right)^{(1 + n)/2}} \exp \left( -
\frac{{{c_x^2} + {\eta ^2}}}{{2RT}}\right).  \label{Eq.feq}
\end{equation}
Here, $f$ represents the distribution function, and $r_{\alpha}$ and $%
v_{\alpha}$ denote the fluid's spatial position and particle velocity in the
$\alpha$-direction, respectively. The parameter $\tau$ represents the
relaxation time of molecular collisions, defined as the reciprocal of the
collision frequency. The variables $\rho$, $u_x$, $T$, and $p$ denote the
fluid's mass density, velocity, temperature, and pressure, respectively. The
ideal gas equation of state, $p=\rho R T$, is used to simulate the argon
shock wave, where $R$ is the gas constant. $q_x$ represents the heat flux.
The parameter $\eta$ accounts for additional degrees of freedom beyond
translational motion, described by $n$. For instance, when $n=2$, $%
\eta^2=\eta_1^2+\eta_2^2$.
When the Prandtl number is $\Pr = 1$, the collision operator in Eq. \eqref{Eq.fs} is simplified to the BGK operator.
In fact, the BGK operator used in the research field of non-equilibrium flow is modified by the mean field theory \cite{Xu2024FOP,Gan2023JFM}.

Despite spatial simplification, the distribution function $f$ remains
high-dimensional, expressed as $f = f(x, v_x, \eta)$. To address this, we
introduce two reduced distribution functions. Specifically, the evolution of
Eq. (\ref{Eq.fs}) can be transformed into the evolution of two reduced
distribution functions, as:
\begin{equation}
\frac{\partial }{{\partial t}}\left \{ {%
\begin{array}{ccccccccccccccccccc}
g   \\
h 
\end{array}%
} \right \} + {v_x} \cdot \frac{\partial }{{\partial x}}\left \{ {%
\begin{array}{c}
g \\
h%
\end{array}%
} \right \} = - \frac{1}{\tau }\left \{ {%
\begin{array}{c}
{g - {g^s}} \\
{h - {h^s}}%
\end{array}%
} \right \}.  \label{Eq.gh}
\end{equation}

The two reduced distribution functions are defined as:
\begin{equation}
g = \int {f d\eta},
\end{equation}
and
\begin{equation}
h = \int {f \frac{{\eta^2}}{2} d\eta}.
\end{equation}
Physically, the reduced distribution function $g$ describes the evolution of
density and velocity, while $h$ captures the influence of additional degrees
of freedom on temperature.
When $f = f^{eq}$, we get
\begin{equation}
g^{eq} = \int {f^{eq} d\eta} = \rho \left( \frac{1}{{2\pi RT}} \right)^{1/2}
\exp\left( - \frac{{c_x^2}}{{2RT}} \right),  \label{Eq.geq}
\end{equation}
and
\begin{equation}
h^{eq} = \int {f^{eq} \frac{{\eta^2}}{2} d\eta} = \frac{{nRT}}{2} g^{eq}.
\end{equation}
The expressions of $g^{s}$ and $h^{s}$ in Eq. (\ref{Eq.gh}) are
\begin{equation}
g^{s}={g^{eq}} + {g^{eq}}\left[ {(1 - \Pr )\cdot{c_x }{q_x } \cdot \frac{{%
\frac{{{c_x^2} }}{{RT}} - 3 }}{{( n + 3)pRT}}} \right],
\end{equation}
and
\begin{equation}
h^{s}={h^{eq}} + {h^{eq}}\left[ {(1 - \Pr )\cdot{c_x }{q_x } \cdot \frac{{%
\frac{{{c_x^2} }}{{RT}} - 1}}{{( n + 3)pRT}}} \right].
\end{equation}

Macroscopic quantities are obtained by evaluating three conserved kinetic
moments of the distribution function,
\begin{equation}
\rho = \int {gd{v_x}},
\end{equation}
\begin{equation}
\rho {u_x } = \int {g{v_x}d{v_x}},
\end{equation}
and
\begin{equation}
\frac{{1 + n}}{2}\rho RT = \int {(g\frac{{c_x^2}}{2} + h)d{v_x}}.
\end{equation}
The viscous stress and heat flux are given by
\begin{equation}
{\Pi } = \int {(g - {g^{eq}})} {c_x}{c_x}d{v_x},
\end{equation}
and
\begin{equation}
{q} = \int {\left[ {(g - {g^{eq}})\frac{{{c_x}^2}}{2} + (h - {h^{eq}})} %
\right]} {c_x}d{v_x}.
\end{equation}


\subsection{DBM with higher-order TNE effects}

\label{higer-order-TNE}

In this section, we aim to maximize the DBM model's capability in capturing
non-equilibrium effects. To achieve this, two strategies exist: one based on
Chapman-Enskog (CE) analysis and the other independent of it \cite{Xu2024FOP}%
.

The first strategy enhances DBM's capability to capture non-equilibrium
effects, starting from the equilibrium state. Kinetic theory characterizes a
system through the distribution function $f$ and its kinetic moments.
Consequently, DBM preserves the kinetic moments governing physical behavior
before and after velocity space discretization:
\begin{equation}
\int{f}\bm{\Psi }\left( \bm{v}, \eta \right)d\bm{v}d\eta = \sum{{{f}_{i}}}%
\bm{\Psi }\left( {{\bm{v}}_{i}, \eta_i} \right).
\end{equation}
CE analysis provides an efficient approach for identifying the kinetic
moments essential for capturing non-equilibrium effects. Retaining
additional kinetic moments improves DBM's ability to capture non-equilibrium
effects and extends its applicability to higher Knudsen numbers \cite%
{Gan2018PRE,Gan2023JFM}.

Specifically, for continuum flows ($\mathrm{Kn} < 0.001$), seven kinetic
moments ($M_{0}$, $M_{1}$, $M_{2,0}$, $M_{2}$, $M_{3,1}$, $M_{3}$, and $%
M_{4,2}$) must be retained before and after discretization. In the
transition regime ($0.1 < \mathrm{Kn} < 10$), second- and higher-order TNE
effects become significant and must be considered. To capture second-order
TNE effects, two additional kinetic moments, $M_{4}$ and $M_{5,3}$, need to
be retained. For third-order TNE effects, two additional kinetic moments, $%
M_{5}$ and $M_{6,4}$, should also be included. Further details on DBM
modeling using CE analysis are provided in Section \ref{CE-analysis}. As
discretization levels and non-equilibrium effects increase, DBM's complexity
grows more slowly than that of kinetic macroscopic modeling (e.g., deriving
and solving extended hydrodynamic equations), as it requires only a limited
number of additional kinetic moments. Therefore, this method is both
straightforward and computationally efficient. Notably, this method retains
a limited set of kinetic moments. While these moments have clear physical
meanings, $f_i$ itself lacks direct physical interpretation.


An alternative approach to describing non-equilibrium flows in DBM involves
directly discretizing the particle velocity space with a sufficiently large
number of grid points, rather than relying on a fixed discrete velocity
stencil. This method maximizes the retention of non-conserved kinetic
moments, enhancing the accuracy of non-equilibrium effect characterization.
Building on this foundation, Zhang \emph{et al.} developed a steady-state
DBM tailored for non-equilibrium flows at the micro-nanoscale \cite%
{Zhang2023POF}. The model effectively captures gas flow behaviors across a
broad range of rarefaction parameters, spanning from slip flow to free
molecular flow. Since this approach retains a potentially infinite number of
kinetic moments, the discrete particle velocity $\bm{v}_i$ closely
approximates the true particle velocity $\bm{v}$, and the discrete
distribution function $f_i$ accurately represents the continuous
distribution function $f$. 
\emph{ Consequently, not only do the kinetic moments of $%
f_i$ retain clear physical meanings, but $f_i$ itself also carries a direct
physical interpretation.}

In the steady-state DBM, the time derivative of $f$ is set to zero, which
allows for a more in-depth exploration of non-equilibrium effects at the
expense of the model's applicability over extended time spans. In contrast,
the time-dependent DBM reduces the system's descriptive capability from
potentially infinite kinetic moments of $f$ to a finite set of moments.
Thus, the time-dependent and steady-state DBMs are complementary. Based on
these approaches, this paper extends the steady-state DBM to an
unsteady-state version that captures non-equilibrium flows across the entire
time domain. The extended model achieves high accuracy in describing TNE
effects and enables the study of the system's kinetic characteristics.

\begin{table*}[t]
\centering
\begin{tabular}{m{5cm}<{\centering}m{9cm}<{\centering}}
\hline
Perspectives & Physical meanings \\ \hline
$k$th-order Kn number effects & Retention of terms up to the $k$th-order of
the Knudsen number in CE analysis. \\
$g-g^{eq}$ & Deviation of the DF $g$ from the equilibrium DF $g^{eq}$. \\
$g^{(k)}$ & $k$th-order deviation of the DF $g$ from the equilibrium DF $%
g^{eq}$. \\
$\Delta_{m}^{*}$, $\Delta_{m,n}^{*}$ & Non-conserved central kinetic moments
of $(g-g^{eq})$, representing the total summation of all orders of TNE
effects. \\
$\Delta_{2}^{*}$ & Non-organized momentum flux (NOMF), or the internal
energy in the $x$ degree of freedom. \\
$\Delta_{3,1}^{*}$ & Non-organized energy flux (NOEF), or heat conduction in
the $x$ direction. \\
$\Delta_{m}^{*(k)}$, $\Delta_{m,n}^{*(k)}$ & Non-conserved central kinetic
moments of $(g-g^{eq})$, representing the $k$-order TNE effects. \\
$\Delta_{m}$, $\Delta_{m,n}$ & Non-conserved kinetic moments of $g-g^{eq}$,
representing the total summation of all orders of HNE+TNE effects. \\
$\Delta_{m}^{(k)}$, $\Delta_{m,n}^{(k)}$ & Non-conserved kinetic moments of $%
(g-g^{eq})$, representing the $k$-order HNE+TNE effects. \\
$\bm{S}_{\text{TNE}} = \{\mathrm{Kn}, g-g^{eq},g^{(k)},\Delta_{2}^{*},
\Delta_{3,1}^{*}, \Delta_{3}^{*}, \Delta_{4,2}^{*}, \Delta_{4}^{*},
\Delta_{5,3}^{*}, \Delta_{5}^{*}, \Delta_{6,4}^{*}\}$ & Multi-perspective,
cross-dimensional description of non-equilibrium states and behaviors. \\
\hline
\end{tabular}%
\caption{ Perspectives used to characterize the non-equilibrium states and
behaviors in this paper, where DF denotes the distribution function. }
\label{table0}
\end{table*}

\subsection{Scheme for extracting and analyzing TNE characteristics}

The Kn number is commonly used to characterize the degree of non-equilibrium
in complex flows. However, the Kn number, whether local or global, is a
coarse-grained quantity that cannot fully capture all non-equilibrium
characteristics of a flow system. Recent studies show that relying solely on
the Kn number gives an incomplete and potentially misleading view of
non-equilibrium phenomena \cite{Xu2024FOP,Gan2023JFM,Zhang2022POF,Gan2018PRE}%
. Therefore, in addition to traditional physical quantities used in fluid
modeling, such as the Kn number, macroscopic gradients, relaxation time $\tau
$, and the distribution function, introducing additional TNE characteristic
quantities is crucial. These quantities provide a more detailed and
comprehensive understanding of the complex features of non-equilibrium
systems.

In the framework of DBM, the non-conserved kinetic moments of ($f-f^{eq}$)
can be utilized to describe both the state of a system deviating from
continuum/equilibrium and the effects resulting from this deviation \cite%
{xu2012lattice}. To illustrate this, we take the TNE quantities in the $x$
direction as an example:
\begin{equation}
\Delta_{m}^{*} = \int {(g - {g^{eq}})} {c^m_x}d{v_x},  \label{Eq.deltam*}
\end{equation}
and
\begin{equation}
\Delta_{m,n}^{*} = \int {\left[ {(g - {g^{eq}})\frac{{{c}_x^{m-n}}}{2} + (h
- {h^{eq}})} \right]} {c^n_x}d{v_x},  \label{Eq.deltamn*}
\end{equation}
where $m$ indicates the total power of $c_x$ and $n$ denotes the contraction
of the $m$-th tensor into an $n$-th order tensor. When $m=2$, the TNE
quantity $\Delta_{2}^{*}$, representing the flux of momentum (known as the
non-organized momentum flux, NOMF), is given by
\begin{equation}
\Delta_{2}^{*} = \int {(g - {g^{eq}})} {c^2_x}d{v_x}.  \label{Eq.delta2*}
\end{equation}
In addition to characterizing momentum flux, $\Delta_{2}^{*}$ also
represents the internal energy associated with the $x$-direction degree of
freedom. For $m=3$ and $n=1$, the TNE quantity $\Delta_{3,1}^{*}$,
describing the flux of total energy (known as non-organized energy flux,
NOEF) is expressed as
\begin{equation}
\Delta_{3,1}^{*} = \int {\left[ {(g - {g^{eq}})\frac{{{c}_x^2}}{2} + (h - {%
h^{eq}})} \right]} {c_x}d{v_x}.  \label{Eq.delta31*}
\end{equation}

The TNE quantity $\Delta_{2}^{*}$ corresponds to the generalized viscous
stress $\Pi$ in hydrodynamic descriptions, and $\Delta_{3,1}^{*}$
corresponds to the generalized heat flux $q$. TNE analysis provides insights
into the mechanisms underlying HNE descriptions. To explore TNE behaviors
further, higher-order TNE quantities, including $\Delta_{3}^{*}$, $%
\Delta_{4}^{*}$, $\Delta_{5}^{*}$, $\Delta_{4,2}^{*}$, $\Delta_{5,3}^{*}$,
and $\Delta_{6,4}^{*}$, are defined. Each TNE quantity characterizes the
system's non-equilibrium properties from a unique perspective. 
A natural step is to define a generalized TNE vector, $\bm{S}_{TNE} = \{\Delta_{2}^{*},
\Delta_{3,1}^{*}, \Delta_{3}^{*}, \Delta_{4,2}^{*}, \Delta_{4}^{*},
\Delta_{5,3}^{*}, \cdots\}$, providing a multi-perspective, cross-dimensional
description of non-equilibrium states and behaviors.

The definitions in Eqs. (\ref{Eq.deltam*}) and (\ref{Eq.deltamn*}) represent
the total summation of all orders of TNE, as $g-g^{eq} =
g^{(1)}+g^{(2)}+g^{(3)}+\cdots$. When the fluid is in equilibrium, $g
\approx g^{eq}$, and the TNE in Eqs. (\ref{Eq.deltam*}) and (\ref%
{Eq.deltamn*}) are zero. For the first-order TNE (the first-order Kn
number), $g-g^{eq} \approx g^{(1)}$. The first-order TNE quantities, $%
\Delta_{m}^{*(1)}$ and $\Delta_{m,n}^{*(1)}$, are obtained. In this case, $%
\Delta_{m}^{*}$ and $\Delta_{m,n}^{*}$ are TNE quantities up to the
first-order TNE. When $g-g^{eq} \approx g^{(1)}+g^{(2)}$, we have $%
\Delta_{m}^{*}=\Delta_{m}^{*(1)}+\Delta_{m}^{*(2)}$ and $\Delta_{m,n}^{*}=%
\Delta_{m,n}^{*(1)}+\Delta_{m,n}^{*(2)}$, where $\Delta_{m}^{*(2)}$ and $%
\Delta_{m,n}^{*(2)}$ are second-order TNE quantities, and $\Delta_{m}^{*}$
and $\Delta_{m,n}^{*}$ are TNE quantities up to the second-order TNE.
Similarly, $\Delta_{m}^{*(3)}$ and $\Delta_{m,n}^{*(3)}$ are third-order TNE
quantities. For clarity, we list in Table \ref{table0}.

\section{ CE multi-scale analysis and derivation of TNE measures}

\label{CE-analysis}

This section performs CE multi-scale analysis to reveal the driving
mechanisms of non-equilibrium flows and derives analytical expressions for
TNE quantities. In fact, DBM modeling does not rely on CE analysis, and
numerical simulations can be conducted independently.

\subsection{Hydrodynamic equations}

We begin with the Shakhov-Boltzamann equation \eqref{Eq.fs}. In CE analysis,
the distribution functions are expanded around the equilibrium distribution
function
\begin{equation}
f = f^{eq} + {\text{Kn }}f^{(1)} + {\text{Kn}}^2 f^{(2)} + \cdot \cdot \cdot,
\end{equation}
and
\begin{equation}
f^{s}={f^{eq}} + \mathrm{Kn }{f^{s(1)}} + \mathrm{K}{\mathrm{n}^2}{f^{s(2)}}
+ \cdot \cdot \cdot.
\end{equation}
The temporal and spatial derivatives are also expanded as
\begin{equation}
\frac{\partial } {{\partial t}} = {\text{Kn}}\frac{\partial } {{\partial t_1
}} + {\text{Kn}}^2 \frac{\partial } {{\partial t_2 }} + \cdot \cdot \cdot,
\end{equation}
and
\begin{equation}
\frac{\partial } {{\partial x }} = {\text{Kn}}\frac{\partial } {{\partial
x_1 }} + \cdot \cdot \cdot.
\end{equation}

\begin{table*}[p]
\centering
\begin{tabular}{m{2.2cm}<{\centering}m{2.7cm}<{\centering}m{11.5cm}<{\centering}}
\hline
TNE quantities & Physical meanings & Analytical expressions \\ \hline
\multirow{4}{*}{\vspace{0.2cm}\centering $\Delta_{2}^{*}$ } & %
\multirow{4}{*}{\vspace{0.2cm}\centering NOMF } & $%
\begin{array}{c}
\Delta_{2}^{*(1)} = -\frac{4  }{3}\tau \rho T \frac{\partial u_x}{\partial x}
\\
\end{array}
$ \\
&  & $%
\begin{array}{c}
\Delta_{2}^{*(2)} = - \frac{4 }{3 \Pr}\tau^{2} \bigg\{
(\Pr - 1) \rho T \frac{\partial^{2} T}{\partial x^{2}} + \Pr T^2 \frac{%
\partial^{2} \rho}{\partial x^{2}} - \Pr \frac{T^2}{\rho} \left( \frac{%
\partial \rho}{\partial x} \right)^2 \\
+ (\Pr - 1) T \frac{\partial T}{\partial x} \frac{\partial \rho}{\partial x}
+ \rho \left[ \frac{\Pr T}{3} \left( \frac{\partial u_x}{\partial x}
\right)^2 - \left( \frac{\partial T}{\partial x} \right)^2 \right] \bigg\}%
\end{array}
$ \\ \hline
\multirow{5}{*}{\vspace{0.5cm}\centering $\Delta_{3,1}^{*}$ } & %
\multirow{5}{*}{\vspace{0.5cm}\centering NOEF } & $%
\begin{array}{c}
\Delta_{3,1}^{*(1)} = -\frac{5   }{2 \Pr}\tau\rho T \frac{\partial T}{%
\partial x} \\
\Delta_{3,1}^{*(2)} = \frac{14 \tau^{2}}{3 \Pr^2} \left[ \frac{2}{7}%
\left(\Pr-\frac{5}{4}\right) T\frac{\partial^{2}u_x}{\partial x^{2}}%
+\left(\Pr-\frac{1}{14}\right)\frac{\partial u_x}{\partial x} \frac{\partial
T}{\partial x} \right] \\
\Delta _{3,1}^{\ast(3) }=-\frac{77 }{3 \Pr^3}\tau^{3} \frac{1}{\rho^2} %
\bigg\{ -\frac{4 }{77}\left(\Pr^{2}-7 \Pr+\frac{25}{8}\right) \rho^{3}T^{2}
\frac{\partial^{3}T}{\partial x^{3}} + \frac{4 }{77} \left(\Pr-\frac{5}{4}%
\right) \Pr \rho^{3} T^{2} \mathit{u_x} \frac{\partial^{3}\mathit{u_x}}{%
\partial x^{3}} \\
-\frac{18 }{77}\rho^{2}T \frac{\partial^{2}T}{\partial x^{2}} \left[ \rho
\left(\Pr^{2}-\frac{100}{9} \Pr+\frac{25}{9}\right) \frac{\partial T}{%
\partial x}+\frac{4 }{9}T \left(\Pr^{2}-4 \Pr+\frac{25}{16}\right) \left(%
\frac{\partial \rho}{\partial x}\right)-\frac{5 }{9}\Pr  \left(\Pr+\frac{3}{10}\right) \rho\mathit{u_x}  \frac{\partial
\mathit{u_x}}{\partial x}%
\right] \\
+ \frac{2 }{7}\Pr \rho^{2}T \left[ \rho \mathit{u_x} \left(\Pr-\frac{1}{2}%
\right) \frac{\partial T}{\partial x} + \frac{50 }{33}[ \frac{3 }{25}%
\left(\Pr-\frac{5}{4}\right) \mathit{u_x} \frac{\partial \rho}{\partial x}%
+\rho \frac{\partial \mathit{u_x}}{\partial x} \left(\Pr-\frac{22}{25}%
\right) ] T \right] \frac{\partial^{2}\mathit{u_x}}{\partial x^{2}} \\
-\frac{26 }{77}\rho T^{2} \left[ \rho \left(\Pr^{2}-\frac{2}{13} \Pr+\frac{25%
}{52}\right) \frac{\partial T}{\partial x}+\frac{2 }{13}\Pr^{2} ( \rho
\mathit{u_x} \frac{\partial \mathit{u_x}}{\partial x}-2\frac{\partial \rho}{%
\partial x} T) \right] \frac{d^{2}\rho}{d x^{2}} +\frac{4 }{77}%
\Pr^{2}\rho^{2} T^{3} \frac{\partial^{3}\rho}{\partial x^{3}} \\
+ \rho^{3} \left(\Pr -\frac{25}{154}\right) \left(\frac{\partial T}{\partial
x}\right)^{3} + \frac{2 }{11}\left[-\frac{9 }{7}T \left(\Pr^{2}-\frac{85}{18}
\Pr+\frac{25}{18}\right) \frac{\partial \rho}{\partial x}+\Pr\left(\Pr+\frac{%
3}{14}\right) \rho \mathit{u_x} \frac{\partial \mathit{u_x}}{\partial x} %
\right] \rho^{2} \left(\frac{\partial T}{\partial x}\right)^{2} \\
+ \frac{62 }{77}\left[  \left(\frac{11 \Pr^{2}}{31}+\frac{25}{124}\right)T
\left(\frac{\partial \rho}{\partial x}\right)^{2}+\frac{5 \Pr }{31}\frac{%
\partial \mathit{u_x}}{\partial x} \left(\Pr+\frac{3}{10}\right)\rho \mathit{u_x}
 \frac{\partial \rho}{\partial x}+\Pr \rho^{2}  \left(\Pr-\frac{14}{93}\right)\left(\frac{\partial \mathit{u_x}}{%
\partial x}\right)^{2}\right] 
\rho T \left(\frac{\partial T}{\partial x}\right) \\
-\frac{4 \Pr^{2}}{231}T \left[\rho^{3} \mathit{u_x} \left(\frac{\partial
\mathit{u_x}}{\partial x}\right)^{3} -3 \rho^{2} T \frac{\partial \rho}{%
\partial x}\left(\frac{\partial \mathit{u_x}}{\partial x}\right)^{2}-3 \rho
T \mathit{u_x} \frac{\partial \mathit{u_x}}{\partial x} \left(\frac{\partial
\rho}{\partial x}\right)^{2}+3 T^{2} \left(\frac{\partial \rho}{\partial x}%
\right)^{3}\right] \bigg\}%
\end{array}
$ \\ \hline
\multirow{3}{*}{\vspace{0.5cm}\centering $\Delta_{3}^{*}$ } & %
\multirow{3}{*}{\vspace{0.5cm}\centering flux of $\Delta_{2}^{*}$ } & $%
\begin{array}{c}
\Delta_{3}^{*(1)} = -3 \tau\rho T  \frac{\partial T}{\partial x}-\frac{6}{5}%
\left(\Pr-1\right)\Delta_{3,1}^{*(1)} \\
\end{array}
$ \\
&  & $%
\begin{array}{c}
\Delta_{3}^{*(2)} = \frac{8  }{\Pr}\tau^{2} \rho T \left[ \frac{1}{2}
\left(\Pr-\frac{1}{2}\right) T \frac{\partial^{2}\mathit{u_x}}{\partial x^{2}%
}+ \left(\Pr+\frac{1}{4}\right)\frac{\partial T}{\partial x}\frac{\partial
\mathit{u_x}}{\partial x}\right] - \frac{6}{5}(\Pr-1)\Delta_{3,1}^{*(2)}%
\end{array}
$ \\ \hline
\multirow{4}{*}{\vspace{0.5cm}\centering $\Delta_{4}^{*}$ } & %
\multirow{4}{*}{\vspace{0.5cm}\centering flux of $\Delta_{3}^{*}$ } & $%
\begin{array}{c}
\Delta_{4}^{*(1)} = -8 \tau \rho  T^{2} \frac{\partial \mathit{u_x}}{\partial
x} \\
\end{array}
$ \\
&  & $%
\begin{array}{c}
\Delta_{4}^{*(2)} = \frac{38  }{\Pr }\tau^{2} \bigg\{  -\frac{4}{19}%
\left(\Pr-\frac{5}{2}\right) \rho T^{2} \frac{\partial ^{2}T}{\partial x^{2}}%
-\frac{4}{19}\Pr T^{3} \frac{\partial^{2}\rho}{\partial x^{2}}  +\frac{4}{19}\Pr%
\left(\frac{\partial \rho}{\partial x}\right)^{2} \frac{T^{3}}{\rho}  \\
-\frac{4}{19}\left(\Pr-1\right) T^{2} \frac{\partial T}{\partial x}  \frac{%
\partial \rho}{\partial x} +\rho T \left[ \frac{4}{19}\Pr T \left( \frac{%
\partial \mathit{u_x}}{\partial x}\right)^{2} +\left(\frac{\partial T}{%
\partial x}\right)^{2}\right] \bigg\}%
\end{array}
$ \\ \hline
\multirow{3}{*}{\vspace{0.5cm}\centering $\Delta_{5}^{*}$ } & %
\multirow{3}{*}{\vspace{0.5cm}\centering flux of $\Delta_{4}^{*}$ } & $%
\begin{array}{c}
\Delta_{5}^{*(1)} = 40 \tau \rho T \mathit{u_x}  \left(\frac{1}{3}\mathit{u_x}%
^{2}+T\right) \frac{\partial \mathit{u_x}}{\partial x}-30 \left[ \frac{%
\partial T}{\partial x} \tau \rho  T+\frac{2}{5}\left(\Pr-1\right)\Delta_{3}^{(1)*}
\right] \left(\mathit{u_x}^{2}+T \right) \\
\end{array}
$ \\
&  & $%
\begin{array}{c}
\Delta_{5}^{*(2)} = \frac{120  }{\Pr}\tau^{2} \left[ \frac{1 }{3} \left(\Pr-\frac{1}{2}\right) \rho T^{3}
 \frac{\partial^{2}\mathit{u_x}}{\partial x^{2}}%
+\left(\Pr+\frac{1}{2}\right)\rho T^{2} \frac{\partial \mathit{u_x}}{\partial x} \frac{\partial T}{%
\partial x}  \right] -12(\Pr-1)T%
\Delta_{3,1}^{*(2)}%
\end{array}
$ \\ \hline
\multirow{4}{*}{\vspace{0.5cm}\centering $\Delta_{4,2}^{*}$ } & %
\multirow{4}{*}{\vspace{0.5cm}\centering flux of $\Delta_{3,1}^{*}$ } & $%
\begin{array}{c}
\Delta_{4,2}^{*(1)} = -\frac{14 }{3}\tau \rho T^{2} \frac{\partial \mathit{%
u_x}}{\partial x} \\
\end{array}
$ \\
&  & $%
\begin{array}{c}
\Delta_{4,2}^{*(2)} = \frac{77  }{3 \Pr }\tau^{2} \bigg\{  -\frac{2 }{11}\left(\Pr-\frac{19}{7}\right)
\rho T  \frac{\partial^{2}T}{\partial x^{2}} -%
\frac{2 \Pr }{11}T^{2}\frac{\partial^{2}\rho}{\partial x^{2}} +\frac{2 \Pr}{%
11 }\frac{T^{2}}{\rho^2}\left(\frac{\partial \rho}{\partial x}\right)^{2} \\
-\frac{2 }{11}\left(\Pr-1\right)T\frac{\partial T}{\partial x}  \frac{%
\partial \rho}{\partial x} + \rho T \left[ \frac{6T}{77}(\frac{\partial u_{x}%
}{\partial x})^{2}\Pr + (\frac{\partial T}{\partial x})^{2} \right] \bigg\}%
\end{array}
$ \\ \hline
\multirow{3}{*}{\vspace{0.5cm}\centering $\Delta_{5,3}^{*}$ } & %
\multirow{3}{*}{\vspace{0.5cm}\centering flux of $\Delta_{4,2}^{*}$ } & $%
\begin{array}{c}
\Delta_{5,3}^{*(1)} = - 21 \tau \rho T^{2} \frac{\partial T}{\partial x} -%
\frac{42}{5} \left(\Pr-1\right) T \Delta_{3,1}^{*(1)} \\
\end{array}
$ \\
&  & $%
\begin{array}{c}
\Delta_{5,3}^{*(2)} = \frac{72  }{\Pr} \tau^{2} \left[\frac{11 }{36}\left(\Pr-%
\frac{7}{11}\right) \rho T^{3} \frac{\partial^{2}\mathit{u_x}}{\partial x^{2}%
}+ \rho T^{2}\left(\Pr+\frac{5}{12}\right) \frac{\partial \mathit{u_x}}{\partial x} \frac{\partial T}{%
\partial x} \right] - \frac{42}{5}%
(\Pr-1)T\Delta_{3,1}^{*(2)}%
\end{array}
$ \\ \hline
\multirow{4}{*}{\vspace{0.5cm}\centering $\Delta_{6,4}^{*}$ } & %
\multirow{4}{*}{\vspace{0.5cm}\centering flux of $\Delta_{5,3}^{*}$ } & $%
\begin{array}{c}
\Delta_{6,4}^{*(1)} = -36\tau \rho T^{3}  \frac{\partial \mathit{u_x}}{%
\partial x} \\
\end{array}
$ \\
&  & $%
\begin{array}{c}
\Delta_{6,4}^{(2)*} = \frac{441  }{\Pr } \tau^{2} \bigg\{  -\frac{4 }{49}%
\left(\Pr-\frac{25}{6}\right) \rho T^{3} \frac{\partial^{2}T}{\partial x^{2}}%
-\frac{4 \Pr }{49} T^{4} \frac{\partial^{2}\rho}{\partial x^{2}} +\frac{4\Pr%
}{49}\frac{T^{4}}{\rho} (\frac{\partial \rho}{\partial x})^{2}  \\
-\frac{4 }{49}\left(\Pr-1\right)T^{3}\frac{\partial T}{\partial x} \frac{%
\partial \rho}{\partial x}+\rho T^{2} \left[\frac{68}{441}\Pr T \left(\frac{%
\partial \mathit{u_x}}{\partial x}\right)^{2} +\left(\frac{\partial T}{%
\partial x}\right)^{2}\right] \bigg\}%
\end{array}
$ \\ \hline
\end{tabular}%
\caption{Analytical expressions for various orders of TNE quantities.}
\label{table1}
\end{table*}

Substituting the expanded distribution functions and derivatives into Eq. (%
\ref{Eq.fs}), we obtain:
\begin{equation}
\begin{aligned} ({\rm{Kn}}\frac{\partial }{{\partial {t_1}}} &+
{\rm{K}}{{\rm{n}}^2}\frac{\partial }{{\partial {t_2}}} + \cdots )({f^{eq}} +
{\rm{Kn }}{f^{(1)}} + {\rm{K}}{{\rm{n}}^2}{f^{(2)}}{\rm{ + }} \cdots ) \\
&{v_x} {\rm{Kn}}\frac{\partial }{{\partial {x_1}}}({f^{eq}} + {\rm{Kn
}}{f^{(1)}} + {\rm{K}}{{\rm{n}}^2}{f^{(2)}}{\rm{ + }} \cdots ) \\ &= -
\frac{{\rm{1}}}{\tau }[({f^{eq}} + {\rm{Kn }}{f^{(1)}} +
{\rm{K}}{{\rm{n}}^2}{f^{(2)}}{\rm{ + }} \cdots ) \\ &- ({f^{eq}} + {\rm{Kn
}}{f^{s(1)}} + {\rm{K}}{{\rm{n}}^2}{f^{s(2)}} + \cdot \cdot \cdot )]
\end{aligned}.  \label{Eq.quan}
\end{equation}

Extracting the first-order terms of the Kn number yields
\begin{equation}
\frac{\partial f^{eq}}{\partial t_1 }+ v_{x} \frac{\partial f^{eq}} {%
\partial x }=-\frac{1}{\tau}( f^{(1)}-f^{s(1)}).  \label{Eq.first-order}
\end{equation}
In the following and here $x_1$ is replaced by $x$. By applying the first
three orders of conserved kinetic moment operators, namely, $%
\int dv_x d\eta $, $\int v_x dv_x d\eta $, and $\int \frac{1}{2} (v_x^{2} +
\eta^2) dv_x d\eta $ into Eq. (\ref{Eq.first-order}), we can
derive the Euler equations:
\begin{equation}
\frac{\partial \rho}{\partial t_1} + \frac{\partial (\rho u_{x})}{\partial x}%
=0,  \label{Eq.Euler-midu-1}
\end{equation}
\begin{equation}
\frac{\partial \rho u_{x}}{\partial t_1} + \frac{\partial (\rho u_{x}u_{x} +
\rho R T )}{\partial x}=0,  \label{Eq.Euler-dongliang-1}
\end{equation}
\begin{equation}
\frac{\partial e}{\partial t_1} + \frac{\partial ( e u_{x} + \rho R T u_{x} )%
}{\partial x}=0,  \label{Eq.Euler-nengliang-1}
\end{equation}
where $e=\rho(\frac{1+n}{2}R T+\frac{1}{2}u_x^{2})$ represents the total
energy density of the fluid.

In the above derivation, the first five kinetic moments, $M_0(f^{eq})$, $%
M_1(f^{eq})$, $M_2(f^{eq})$, $M_{2,0}(f^{eq})$ and $M_{3,1}(f^{eq})$, are
needed . Their expressions are as follows:
\begin{equation}
{\emph{M}}_{0}(f^{eq})=\int f^{eq} dv_x d\eta=\rho \text{,}  \label{eq.36}
\end{equation}%
\begin{equation}
{\emph{M}}_{1}(f^{eq})=\int f^{eq}v_x dv_x d\eta=\rho u_x \text{,}
\label{M1}
\end{equation}%
\begin{equation}
{\emph{M}}_{2,0}(f^{eq})=\int \frac{1}{2} f^{eq} (v_{x}^{2}+\eta^2) dv_x
d\eta=\frac{1}{2}\rho [ (1+n)R T+u_{x}^{2}]\text{,}  \label{M20}
\end{equation}%
\begin{equation}
{\emph{M}}_{2}(f^{eq})=\int f^{eq} v_x v_x dv_x d\eta =\rho (R T+u_{x}^{2})%
\text{,}  \label{M2}
\end{equation}%
\begin{equation}
{\emph{M}}_{3,1}(f^{eq})=\int \frac{1}{2} f^{eq} v_{x} (v_{x}^{2}+\eta^2)
dv_x d\eta=\frac{1}{2}\rho u_x[(n+3)R T+u_x^{2}].  \label{M31}
\end{equation}

From CE analysis, incorporating all orders of TNE effects into Eqs. %
\eqref{Eq.Euler-midu-1} to \eqref{Eq.Euler-nengliang-1}, yields the
completed hydrodynamic equations
\begin{equation}
\frac{\partial \rho}{\partial t_1} + \frac{\partial (\rho u_{x})}{\partial x}%
=0,  \label{Eq.quan-midu-1}
\end{equation}
\begin{equation}
\frac{\partial \rho u_{x}}{\partial t_1} + \frac{\partial (\rho u_{x}u_{x} +
\rho R T )}{\partial x}+\frac{\partial \Delta_{2}}{\partial x}=0,
\label{Eq.quan-dongliang-1}
\end{equation}
\begin{equation}
\frac{\partial e}{\partial t_1} + \frac{\partial ( e u_{x} + \rho R T u_{x} )%
}{\partial x}+\frac{\partial \Delta_{3,1}}{\partial x}=0.
\label{Eq.quan-nengliang}
\end{equation}
where $\Delta_{2}$ and $\Delta_{3,1}$ are the thermo-hydrodynamic
non-equilibrium (THNE) quantities. Their expressions are
\begin{equation}
\Delta_{2} = \int {({f^{(1)}+f^{(2)}+\cdots + f^{(k)}})} {v^2_x}d{v_x}d\eta,
\end{equation}
and
\begin{equation}
\Delta_{3,1} = \int {\left[ ({{f^{(1)}+f^{(2)}+\cdots + f^{(k)}) }\frac{{{v}%
_x^{2}+\eta^{2}}}{2} } \right]} {v_x}d{v_x}d\eta.
\end{equation}
Unlike $\Delta_{m}^{*}$ and $\Delta_{m,n}^{*}$ which solely measure the TNE
effects, $\Delta_{m}$ and $\Delta_{m,n}$ represent the combined effects of
HNE and TNE.

When considering only the first-order Kn number effects, the generalized
hydrodynamic equations, Eqs. (\ref{Eq.quan-midu-1}) to (\ref%
{Eq.quan-nengliang}), reduce to the NS equations, with linear constitutive
relations, $\Delta_{2}=\Delta_{2}^{(1)}=\Delta_{2}^{*(1)}$ and $%
\Delta_{3,1}=\Delta_{3,1}^{(1)}=\Delta_{3,1}^{*(1)}+u_x\Delta_{2}^{*(1)}$.
When the second-order Kn number effects are preserved, the nonlinear
constitutive relations in the Burnett equations are: $\Delta_{2}=%
\Delta_{2}^{(1)}+\Delta_{2}^{(2)}=\Delta_{2}^{*(1)}+\Delta_{2}^{*(2)}$ and $%
\Delta_{3,1}=\Delta_{3,1}^{(1)}+\Delta_{3,1}^{(2)}
=\Delta_{3,1}^{*(1)}+\Delta_{3,1}^{*(2)}+u_x\Delta_{2}^{*(1)}+u_x%
\Delta_{2}^{*(2)}$. For the third-order case, the nonlinear constitutive
relations in the super-Burnett equations are: $\Delta_{2}=\Delta_{2}^{(1)}+%
\Delta_{2}^{(2)}+\Delta_{2}^{(3)}$ and $\Delta_{3,1}=\Delta_{3,1}^{(1)}+%
\Delta_{3,1}^{(2)}+\Delta_{3,1}^{(3)}$.
The derivation of TNE quantities is provided in Section \ref{analysis-TNE}.

\subsection{Distribution functions}

From Eqs. (\ref{Eq.Euler-midu-1}) to \ref{Eq.Euler-nengliang-1}, the
temporal derivatives of the macroscopic quantities can be expressed in terms
of spatial derivatives.
\begin{equation}
\frac{\partial \rho}{\partial t_1}=-u_x \frac{\partial \rho}{\partial x}%
-\rho \frac{\partial u_x}{\partial x},  \label{Eq.partial-midu-1}
\end{equation}
\begin{equation}
\frac{\partial u_x}{\partial t_1}=-u_x\frac{\partial u_x}{\partial x}-\frac{T%
}{\rho}\frac{\partial \rho}{\partial x}-\frac{\partial T}{\partial x},
\end{equation}
and
\begin{equation}
\frac{\partial T}{\partial t_1}=-u_x\frac{\partial T}{\partial x}-\frac{2}{%
n+1}T\frac{\partial u_x}{\partial x}.  \label{Eq.partial-wendu-1}
\end{equation}

From Eq. (\ref{Eq.first-order}), the first-order distribution function is
given by
\begin{equation}
{f^{(1)}} = - \tau [\frac{{\partial {f^{eq}}}}{{\partial {t_1}}} + {v_x}
\frac{{\partial {f^{eq}}}}{{\partial x}}] + {f^{s(1)}},  \label{Eq.f1}
\end{equation}
where $f^{s(k)} = f^{eq} \left[ (1 - \Pr) c_x q_x^{(k)} \frac{ \frac{c_x^2 +
\eta^2}{RT} - (n + 3) }{(n + 3) p RT} \right]$. Substituting Eq. (\ref%
{Eq.feq}) into Eq. (\ref{Eq.f1}), and replacing the temporal derivatives
with spatial ones, analytical solutions of $f^{(1)}$, expressed as spatial
derivatives of macroscopic quantities, are obtained.

Along this way, by extracting the second- and third-orders terms of Kn
number from Eq. (\ref{Eq.quan}), we obtain
\begin{equation}
\frac{\partial f^{eq}}{\partial t_2 }+ \frac{\partial f^{(1)}}{\partial t_1 }%
+ v_{x} \frac{\partial f^{(1)}} {\partial x }=-\frac{1}{\tau}(
f^{(2)}-f^{s(2)}),
\end{equation}
and
\begin{equation}
\frac{\partial f^{eq}}{\partial t_3 }+ \frac{\partial f^{(1)}}{\partial t_2 }%
+ \frac{\partial f^{(2)}}{\partial t_1 }+ v_{x} \frac{\partial f^{(2)}} {%
\partial x }=-\frac{1}{\tau}( f^{(3)}-f^{s(3)}).  \label{Eq.third-order}
\end{equation}
The second- and third-orders expressions of distribution function are given
as follows:
\begin{equation}
{f^{(2)}} = - \tau [\frac{{\partial {f^{(1)}}}}{{\partial {t_1}}} + \frac{{%
\partial {f^{eq}}}}{{\partial {t_2}}} + {v_x} \frac{{\partial {f^{(1)}}}}{{%
\partial x}}] + {f^{s(2)}},
\end{equation}
and
\begin{equation}
{f^{(3)}} = - \tau [\frac{{\partial {f^{eq}}}}{{\partial {t_3}}} + \frac{{%
\partial {f^{(1)}}}}{{\partial {t_2}}} + {v_x} \frac{{\partial {f^{(2)}}}}{{%
\partial x}}] + {f^{s(3)}}.
\end{equation}

\subsection{TNE measures}

\label{analysis-TNE}

By integrating the distribution functions $f^{(k)}$ over their velocity and $%
\eta$ spaces, 
higher-order TNE quantities can be defined as:
\begin{equation}
\Delta_{m}^{*(k)} = \int {{f^{(k)}}} {c^2_x}d{v_x}d\eta,
\end{equation}
and
\begin{equation}
\Delta_{m,n}^{*(k)} = \int {\left[ {{f^{(k)}}\frac{{{c}_x^{2}+\eta^{2}}}{2} }
\right]} {c_x}d{v_x}d\eta.
\end{equation}

For convenience, the analytical expressions for different orders of TNE
quantities are listed in Table \ref{table1}. Specifically, expressions for
the first-order to third-order terms of $\Delta_{3,1}^{*}$ are given, while
only the first-order and second-order expressions are provided for the other
TNE quantities. Additionally, the results in Table \ref{table1} correspond
to the case where $n=2$.

\begin{figure*}[h]
\center\includegraphics*
[ width=0.8\textwidth]{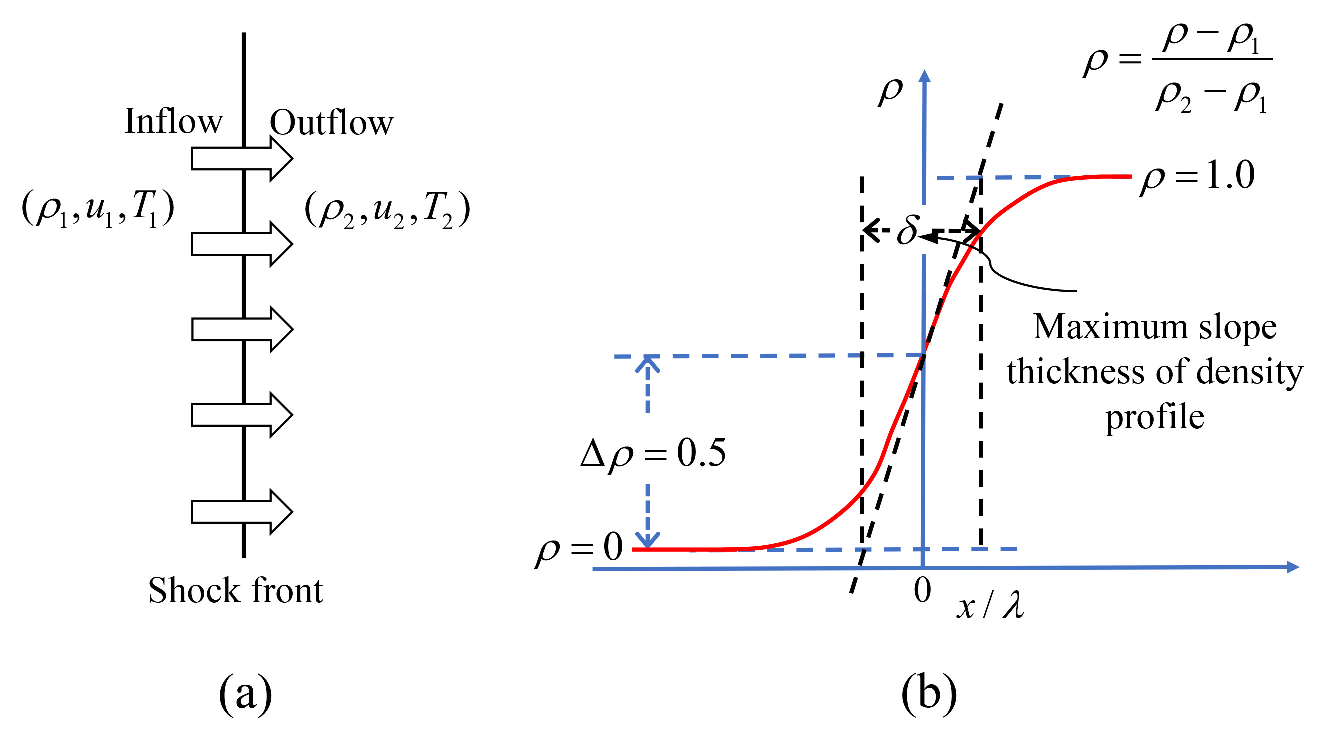}
\caption{ (a) Schematic representation of flows on both sides of the shock
front. (b) Schematic of the normalized density profile within the shock's
internal structure. }
\label{fig1}
\end{figure*}

\section{ DBM simulation and numerical results}

\label{Numerical-results}

\subsection{ Shock wave configuration}

In this section, we analyze the effects of the Mach number on
non-equilibrium characteristics, including both HNE and TNE effects, using
the constructed Shakov-DBM. Figures \ref{fig1}(a) illustrates the
configurations of macroscopic quantities, while Figure \ref{fig1}(b) shows
the density profile of the internal structure of the shock when it has
evolved to a steady state. From Figure \ref{fig1}(b), the following
observations can be made: (i) The position corresponding to a normalized
density of $\widetilde{\rho} = 0.5$ is set as the origin of the horizontal
axis. (ii) The parameter $\delta$ represents the maximum slope thickness of
the density profile, which serves as an approximate measure of the shock's
internal structure thickness \citep{Alsmeyer1976JFM}.

The macroscopic quantities on either side of the shock front satisfy the
Rankine-Hugoniot relations, expressed as,
\begin{equation}
\rho_2 = \rho_1 \cdot a,
\end{equation}
\begin{equation}
T_2 = T_1 \cdot b / a,
\end{equation}
and
\begin{equation}
u_2 = u_1 / a.
\end{equation}
where $a=\gamma_{1}\cdot\mathrm{Ma}^2/(2.0+\gamma_{-1}\cdot\mathrm{Ma}^2)$
and $b=2.0 \cdot \gamma \cdot \mathrm{Ma}^2/\gamma_{1}-\gamma_{-1}/%
\gamma_{1} $, with $\gamma_{-1}=\gamma-1$ and $\gamma_{1}=\gamma+1$. For
argon gas, the specific heat ratio is $\gamma=5/3$.

\subsection{ Molecular interaction models}

\begin{table*}[htbp]
\centering
\begin{tabular}{m{2.5cm}<{\centering}m{2.5cm}<{\centering}m{2.5cm}<{\centering}m{5cm}<{\centering}}
\hline
Ma number & molecular model & $dt$ & $\alpha$, \quad $\omega$, \quad $\chi$
\\ \hline
1.2 , 1.4 & VHS-VSS & \multirow{2}{*}{\vspace{0.2cm}\centering 0.001 } & $%
\alpha=1.4$,$\omega=0.81$,$\chi=0.81$ \\
1.55 - 2.5 & \multirow{5}{*}{\vspace{0.2cm}\centering VHS } &  & $\chi=0.81$
\\
3.8 - 6.0 &  & 0.0005 & $\chi=0.81$ \\
8.0 &  & \multirow{3}{*}{\vspace{0.2cm}\centering 0.0001 } & $\chi=0.75$ \\
9.0 &  &  & $\chi=0.72$ \\
10.0 &  &  & $\chi=0.71$ \\ \hline
\end{tabular}%
\caption{ Parameters used in the simulation of shock waves with various Ma
numbers. }
\label{table2}
\end{table*}

Macroscopic transport characteristics of flows arise from the collective
effects of microscopic molecular collisions. Consequently, different
molecular interaction models can yield distinct transport behaviors. To
facilitate comparison, two molecular models are discussed below.

The first model is the variable hard-sphere (VHS) model, which incorporates
an inverse power law for intermolecular forces. In this model, the collision
frequency (the inverse of the collision relaxation time) between molecules
is given by \citep{Li2007ACTA}:
\begin{equation}
\hat{\nu} = \frac{16}{5} \sqrt{\frac{R}{2\pi}} \frac{T_{\infty}^{\chi-1/2}}{%
\rho_{\infty}} \frac{\hat{\rho}}{T^{\chi-1}} \frac{1}{\lambda_{\infty}},
\label{model1}
\end{equation}
where $\chi $ represents the temperature dependence of viscosity. Physical
quantities marked with ``$\infty $'' and ``$\hat{\quad} $'' represent
reference values and real physical quantities, respectively. $%
\lambda_{\infty} $ denotes the average free path of molecules.

The second model combines the VHS model with the variable soft-sphere (VSS)
model \cite{Li2008IJCFD, Li2015PAS}, and is expressed as:
\begin{equation}
\hat{\nu} = \frac{4\alpha(5-2\omega)(7-2\omega)}{2(\alpha+1)(\alpha+2)}
\sqrt{\frac{R}{2\pi}} \frac{T_{\infty}^{\chi-1/2}}{\rho_{\infty}} \frac{1}{%
\lambda_{\infty}} \frac{\hat{\rho}}{\hat{T}^{\chi-1}},  \label{model2}
\end{equation}
where $\omega $ and $\alpha $ are indices for the VHS and VSS models,
respectively. Their values depend on the gas type and state.

Before performing the simulation, it is essential to nondimensionalize Eqs. %
\eqref{model1} and \eqref{model2}. By substituting the reference velocity $%
c_{\infty} = \sqrt{RT_{\infty}}$ into these equations, the dimensionless
collision frequency becomes:
\begin{equation}
\nu = \frac{16}{5\sqrt{2\pi} \mathrm{Kn} } \rho T^{1-\chi},  \label{model1-1}
\end{equation}
and
\begin{equation}
\nu = \frac{4\alpha(5-2\omega)(7-2\omega)}{5(\alpha+1)(\alpha+2)}\cdot \frac{%
1}{\sqrt{2\pi}\mathrm{Kn}}\rho T^{1-\chi},  \label{model2-1}
\end{equation}
where $\mathrm{Kn}=\lambda_{\infty}/L_{\infty}$. To capture the internal
structure of a shock, the averaged free path is typically taken as the
characteristic $L_{\infty}$, i.e., $L_{\infty}=\lambda_{\infty}$. Under this
condition, the Kn number is $\mathrm{Kn}=1$.

In the subsequent simulations, the second model is used for cases with $%
\mathrm{Ma}=1.2$ and $\mathrm{Ma}=1.4$, while the first is applied to the
other Ma numbers.

\subsection{ Numerical schemes and parameter settings}

This paper continues the direct discretization of particle velocity space as
presented in the Ref. \cite{Zhang2023POF}. The velocity space, ranging from $%
-v_{max}$ to $v_{max}$, is divided non-uniformly, with $v_{max}$
representing the truncation velocity. The discrete method is given by
\begin{equation}
v_{i}=\left(i-\frac{N_{vx}+1}{2}\right)^{\lambda}/\left(\frac{N_{vx}-1}{2}%
\right)^{\lambda} v_{max} + v_{0} ,
\end{equation}
where $i$ represents the index of grid points in the velocity space, and $%
N_{vx}$ denotes the total number of grids. The parameter $\lambda$ is a
positive odd number that refines the velocity space near the initial
velocity $v_0$. 
The difference between two adjacent velocity grids is
expressed as
\begin{equation}
\delta v(i) = \frac{\lambda}{(i-\frac{N_{vx}+1}{2})(v_{i}-v_0)}.
\end{equation}

To numerically solve Eq. (\ref{Eq.gh}), appropriate numerical schemes are
required for both temporal and spatial derivatives. In this study, we use
the first-order forward Euler finite difference scheme for the time
derivative and the fifth-order weighted essentially non-oscillatory (WENO)
scheme for the spatial derivative. Both schemes have been extensively
validated in previous literature.

Shock waves with Mach numbers ranging from 1.2 to 10.0 are simulated. The
spatial domain, spanning from 0 to $200 \cdot \lambda $, is divided into
1000 grid points, resulting in a dimensionless grid size of $\Delta x = 0.2 $%
. The parameters for discretizing the velocity space are as follows: $v_{%
\text{max}} = 50.0 $, $N_{vx} = 300 $, and $\lambda = 5.0 $. For argon gas,
the degrees of freedom is $n = 2 $, leading to $\gamma = 5/3 $. The Prandtl
number for argon is $2/3 $. Other dimensionless parameters depend on the gas
state, as summarized in Table \ref{table2}.

\begin{figure}[htbp]
\center\includegraphics*
[ width=0.54\textwidth]{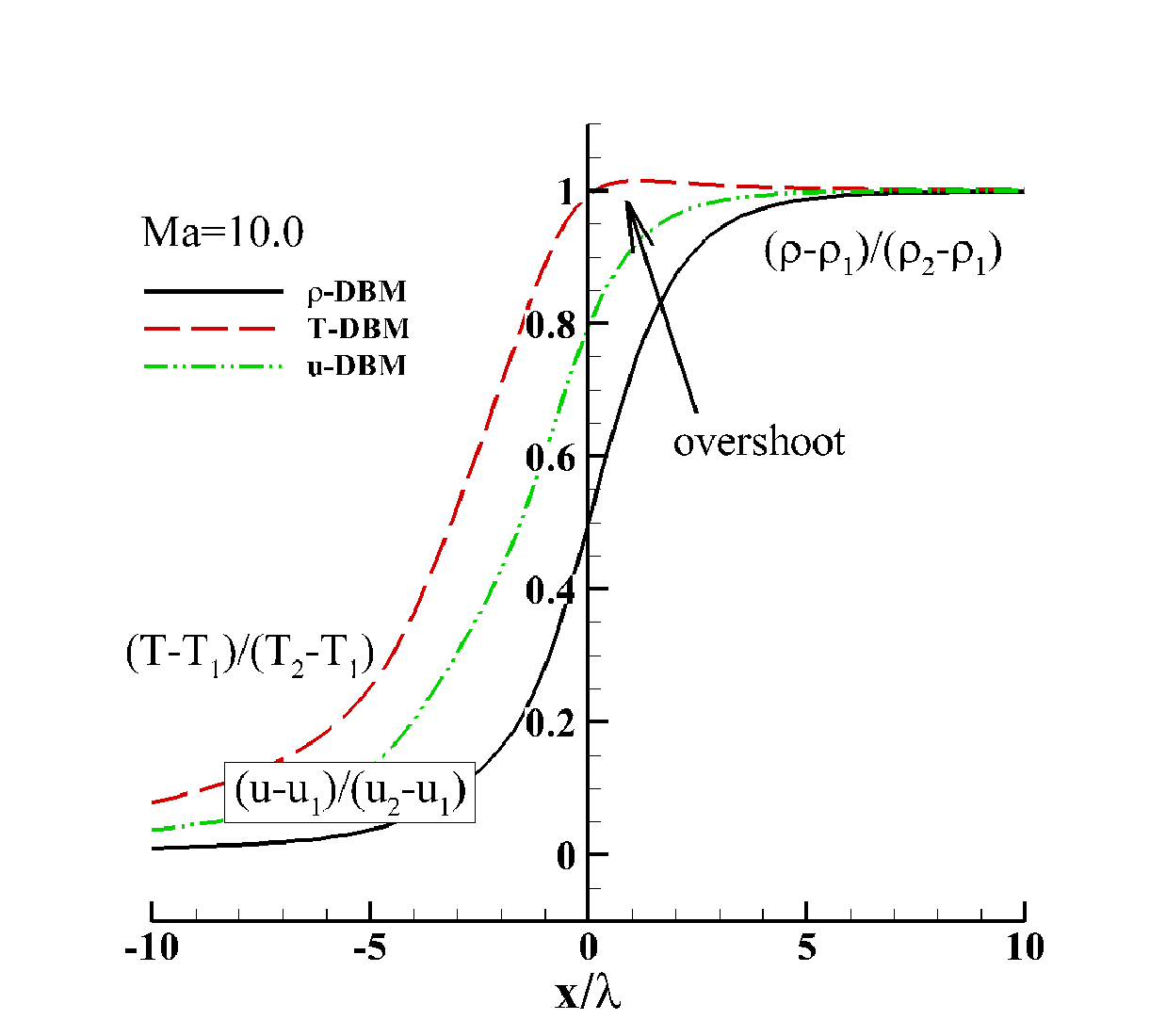}
\caption{ The normalized density, velocity and temperature profiles of a
shock for $\mathrm{Ma} = 10.0$. }
\label{fig2}
\end{figure}

\subsection{ Internal structure of a shock}

For clarity, Fig. \ref{fig2} shows the DBM numerical results for normalized
density, velocity, and temperature profiles of a shock with a Mach number of
10.0. The following observations can be made:

(I) The positions of the three interfaces do not coincide, with a separation
of several $\lambda $.

(II) The temperature interface is located at the front of the shock,
followed by the velocity and density interfaces, respectively.

(III) The shapes of the three interfaces differ in terms of slope,
thickness, and symmetry. Notably, the temperature overshoot is pronounced at
high Mach numbers, while the velocity and density interfaces do not exhibit
such an overshoot.

The impact of these differences in the internal interfaces of the shock on
flow behavior, essentially governed by different TNE quantities, warrants
further investigation.

\subsection{ Comparison with DSMC and experimental results}

\begin{figure*}[tbp]
\centering
\subfigure*{\
\begin{minipage}{8cm}
			\includegraphics*[width=5.5cm]{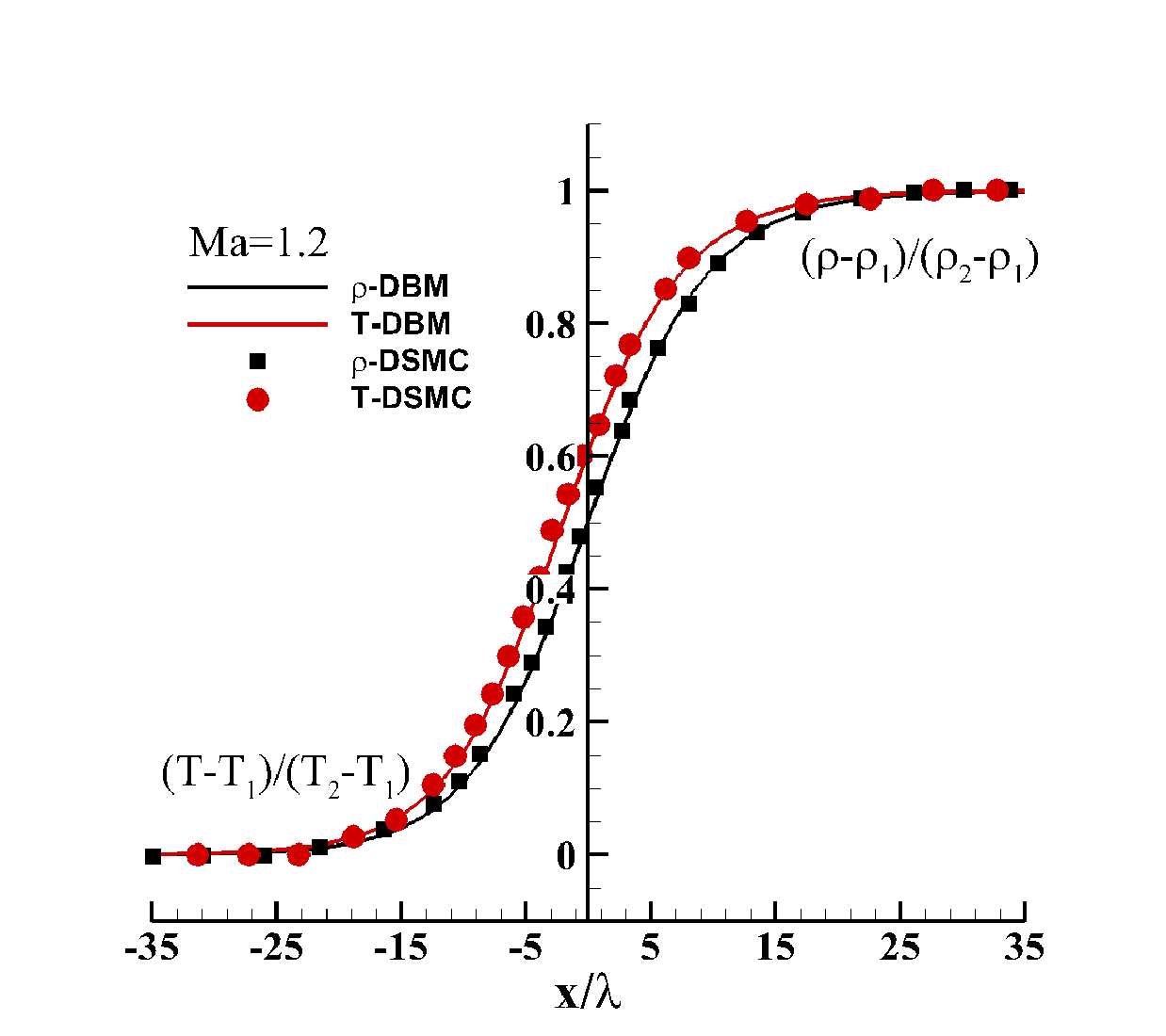}
			\centering \caption*{(a) $\mathrm{Ma} = 1.2$}
		\end{minipage}
\begin{minipage}{8cm}
			\includegraphics*[width=5.5cm]{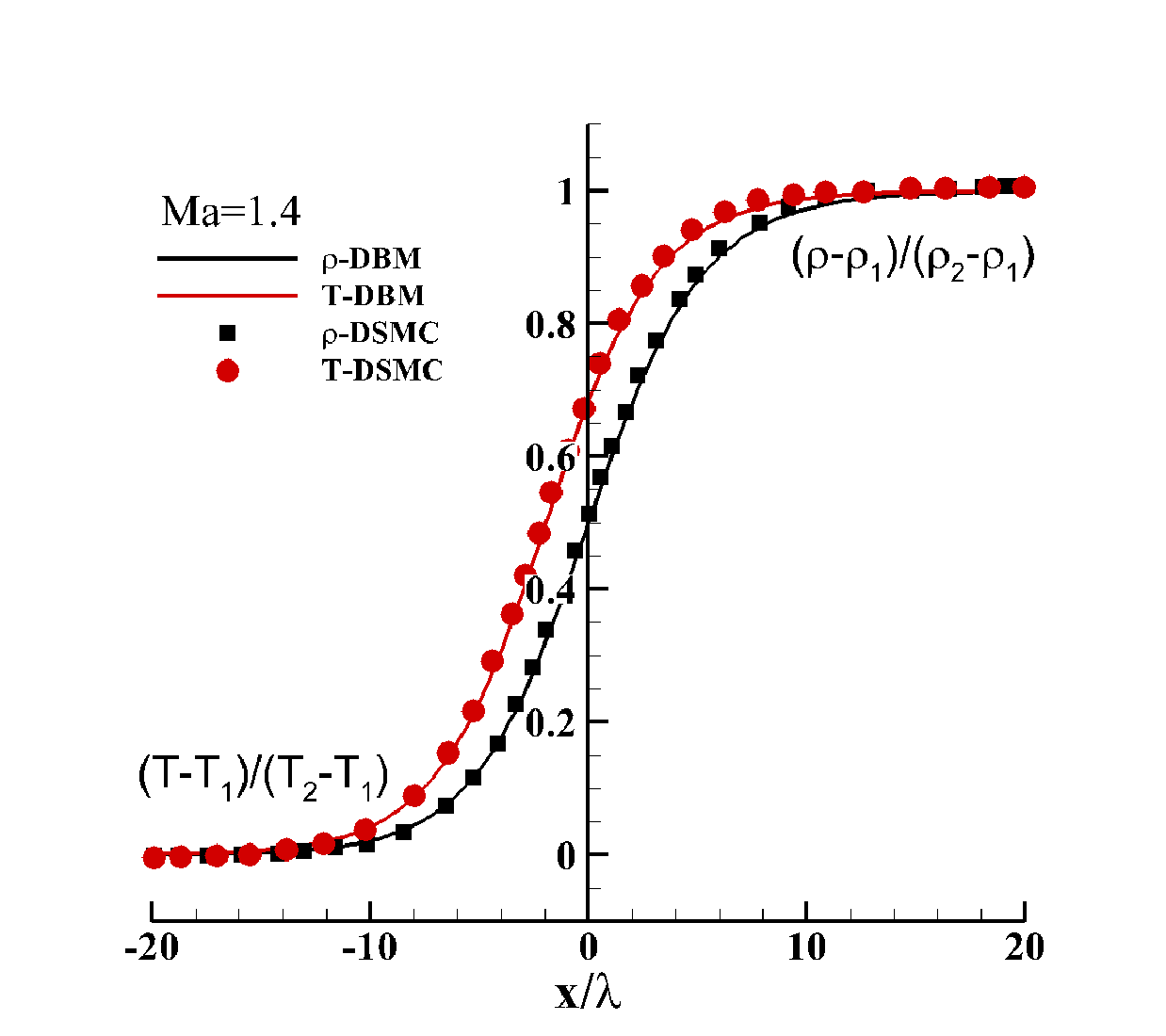}
			\centering \caption*{(b) $\mathrm{Ma} = 1.4$}
		\end{minipage}
}
\par
\subfigure*{\
\begin{minipage}{8cm}
			\includegraphics*[width=5.5cm]{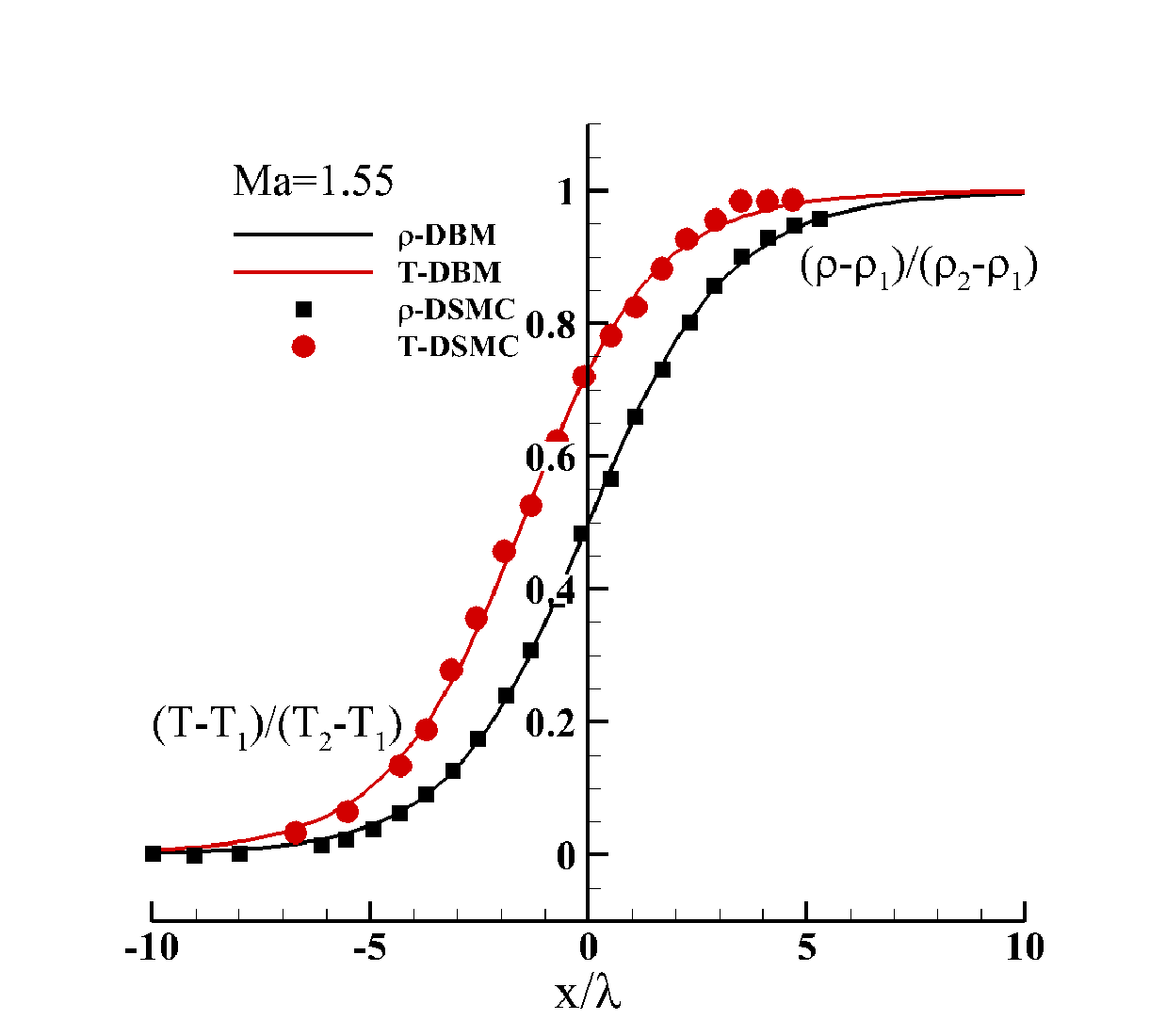}
			\centering \caption*{(c) $\mathrm{Ma} = 1.55$}
		\end{minipage}
\begin{minipage}{8cm}
			\includegraphics*[width=5.5cm]{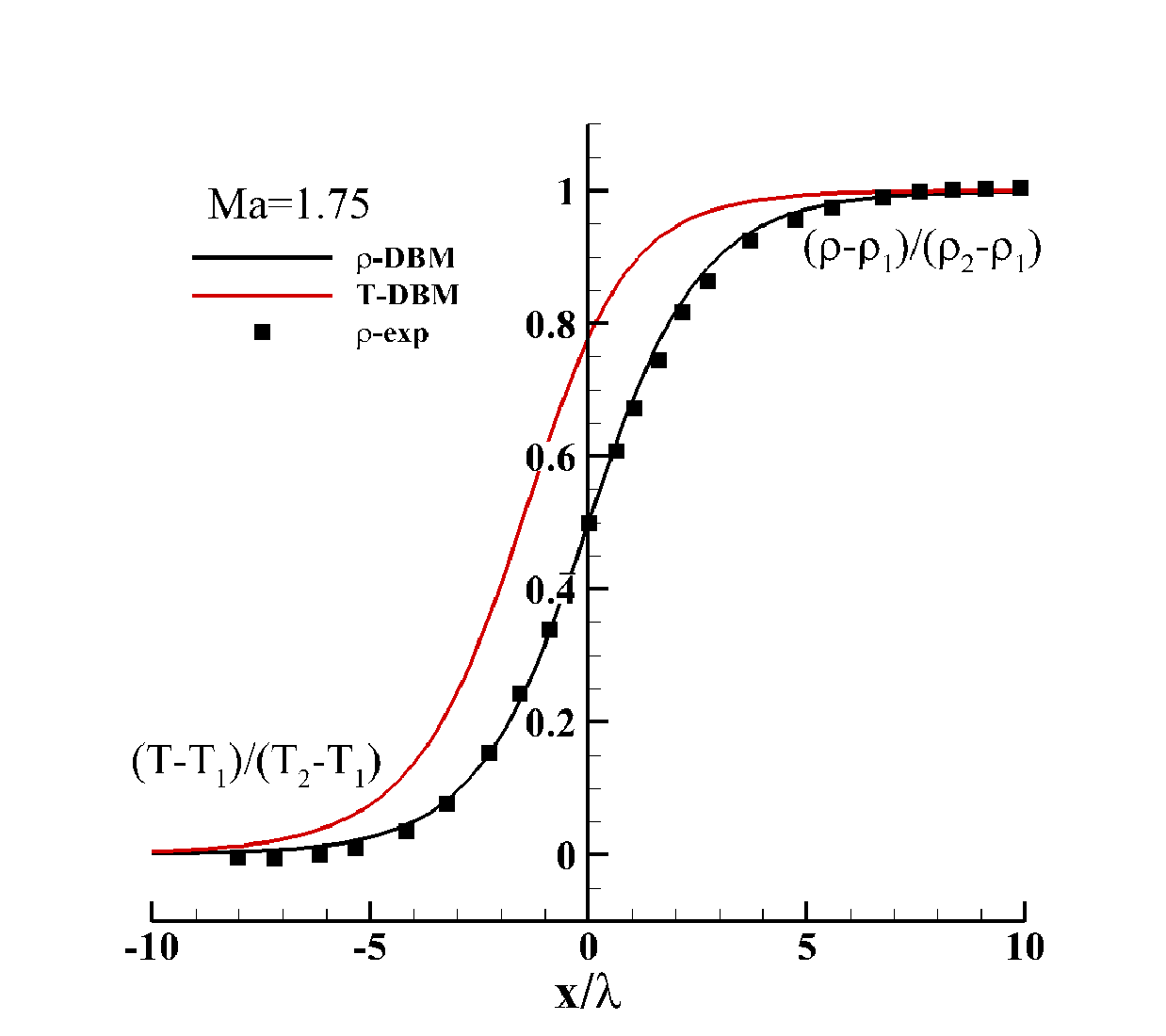}
			\centering \caption*{(d) $\mathrm{Ma} = 1.75$}
		\end{minipage}
}
\par
\subfigure*{\
\begin{minipage}{8cm}
			\includegraphics*[width=5.5cm]{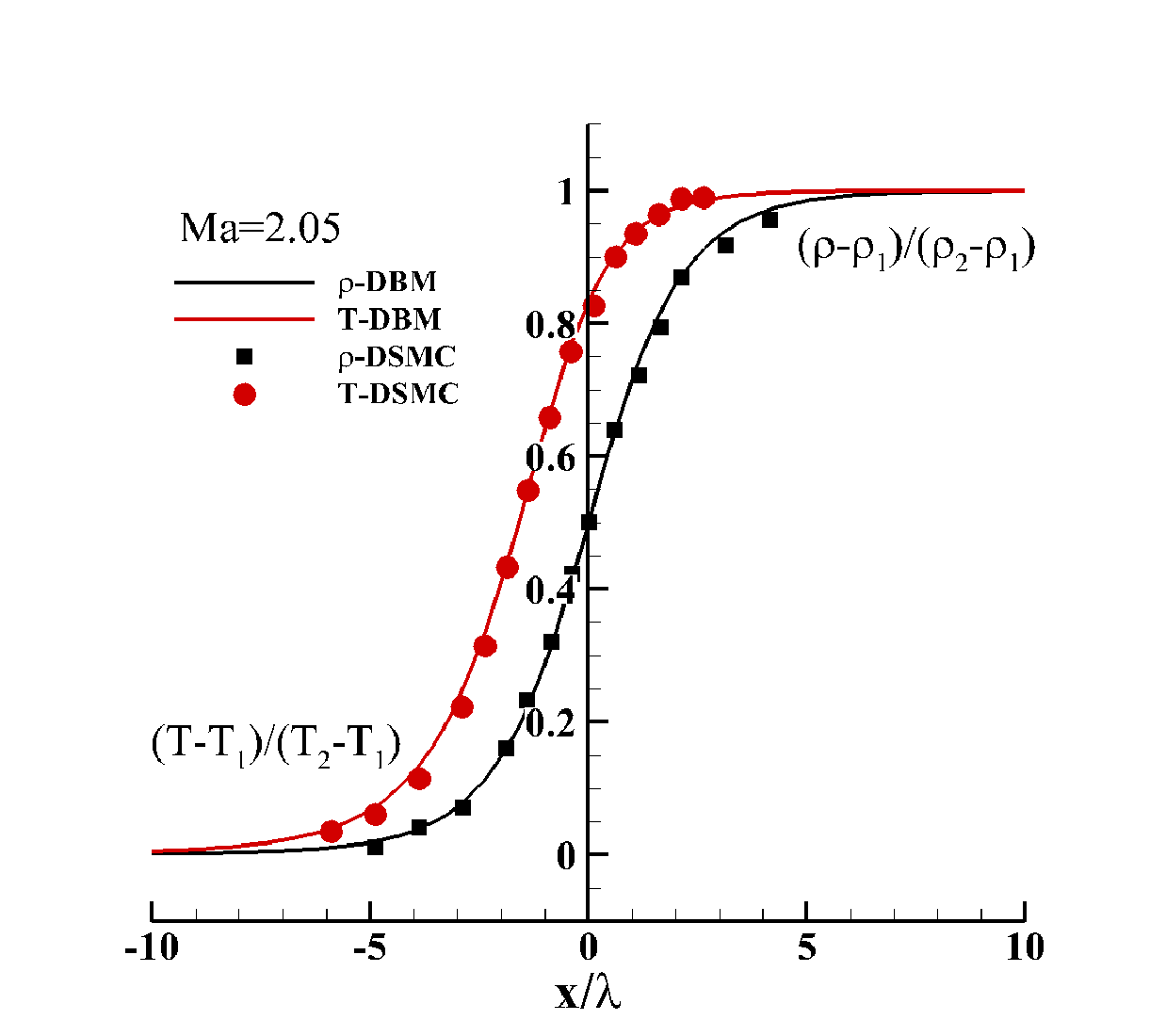}
			\centering \caption*{(e) $\mathrm{Ma} = 2.05$}
		\end{minipage}
\begin{minipage}{8cm}
			\includegraphics*[width=5.5cm]{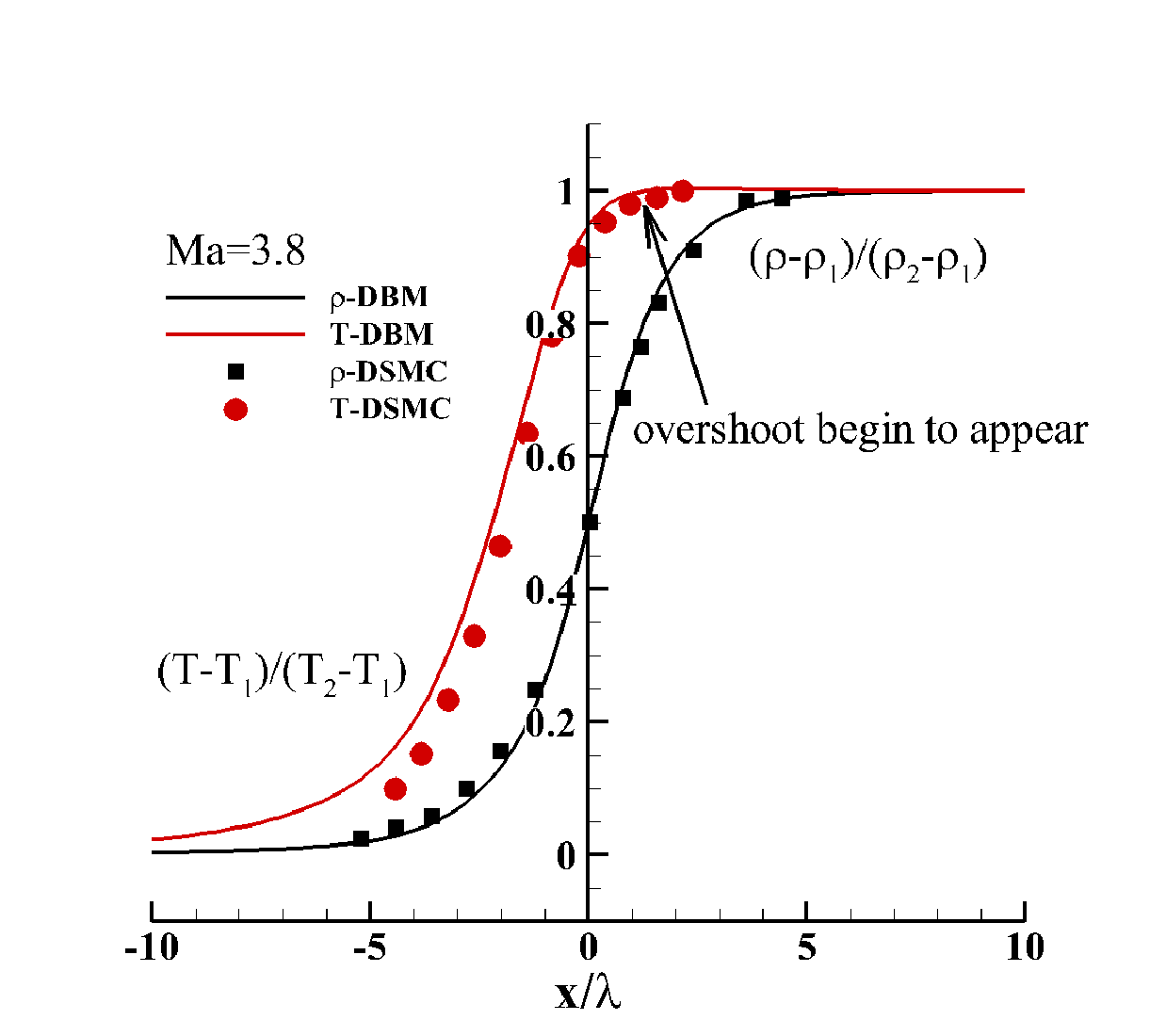}
			\centering \caption*{(f) $\mathrm{Ma} = 3.8$}
		\end{minipage}
}  \subfigure*{\
\begin{minipage}{8cm}
			\includegraphics*[width=5.5cm]{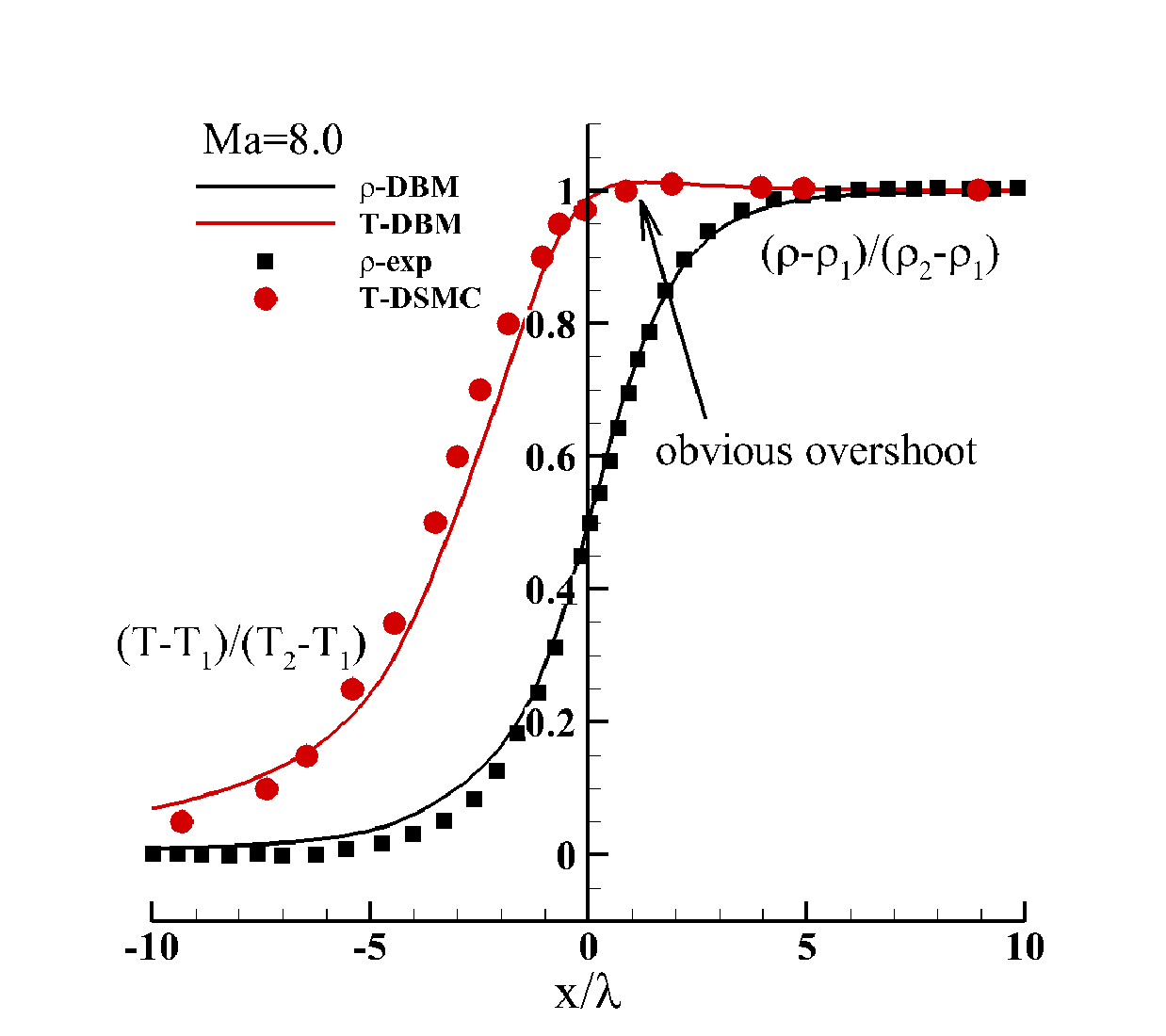}
			\centering \caption*{(g) $\mathrm{Ma} = 8.0$}
		\end{minipage}
\begin{minipage}{8cm}
			\includegraphics*[width=5.5cm]{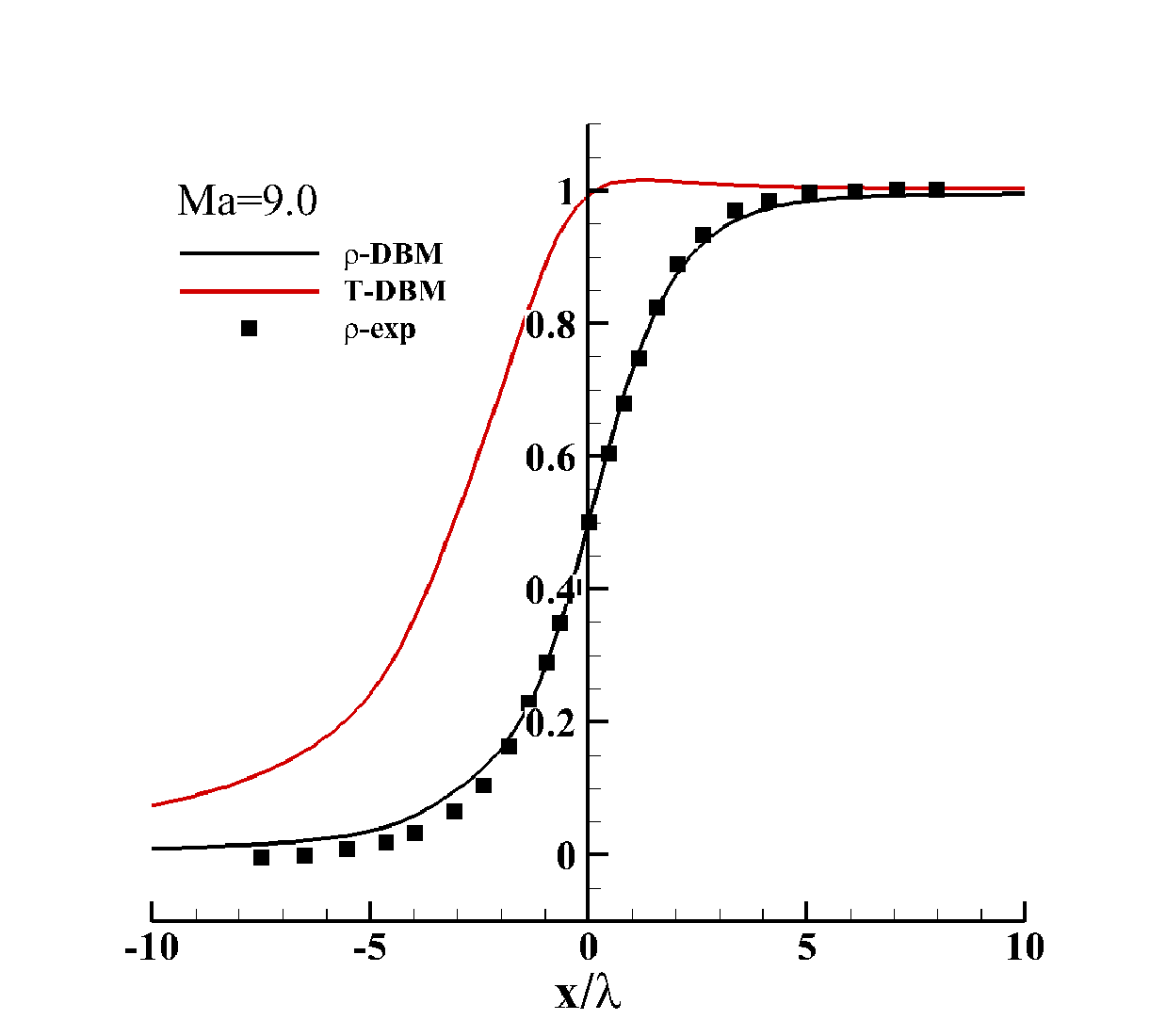}
			\centering \caption*{(h) $\mathrm{Ma} = 9.0$}
		\end{minipage}
}
\caption{ Comparisons of shock structure between DBM simulation and DSMC (or
experimental) results. The DSMC results for $\mathrm{Ma} = 1.2$ and $1.4$
are obtained from Fig. 3 in the Ref. \citep{Li2007ACTA}. The DSMC and
experimental data for $\mathrm{Ma} = 1.55$ to $9.0$ are from the Ref.
\citep{Alsmeyer1976JFM}. Among these, the DSMC results for $\mathrm{Ma} =
1.55$, 2.05, and 3.8 are from Fig. 5. The experimental data sources are: $%
\mathrm{Ma} = 1.75$ from Fig. 3, $\mathrm{Ma} = 8.0$ from Fig. 7, and $%
\mathrm{Ma} = 9.0$ from Fig. 4, respectively. }
\label{fig3}
\end{figure*}

Figure \ref{fig3} compares the density and temperature profiles inside the
shocks from DBM simulations with other results. The profiles for shock waves
with Mach numbers ranging from 1.2 to 9.0 are shown. The DBM simulation
results closely match those from DSMC simulations and experimental data,
demonstrating that the DBM model accurately captures the internal structures
of shock waves, even at high Mach numbers.

Additionally, a temperature overshoot begins to appear as the Mach number
approaches 3.8. As the Mach number increases to 8.0, the temperature
overshoot becomes more pronounced. This overshoot is a typical
nonequilibrium phenomenon caused by the rapid accumulation of heat without
sufficient time for dissipation. Shan \emph{et al.} investigated the
mechanism behind the temperature overshoot, attributing it to higher-order
TNE effects \cite{Shan2025}.

\subsection{ Effects of Mach number on macroscopic quantities}

\subsubsection{ Two-stage Effects on shock shapes and compressibility}

\begin{figure*}[tbp]
\centering
\subfigure*{\
\begin{minipage}{8cm}
			\includegraphics*[width=8cm]{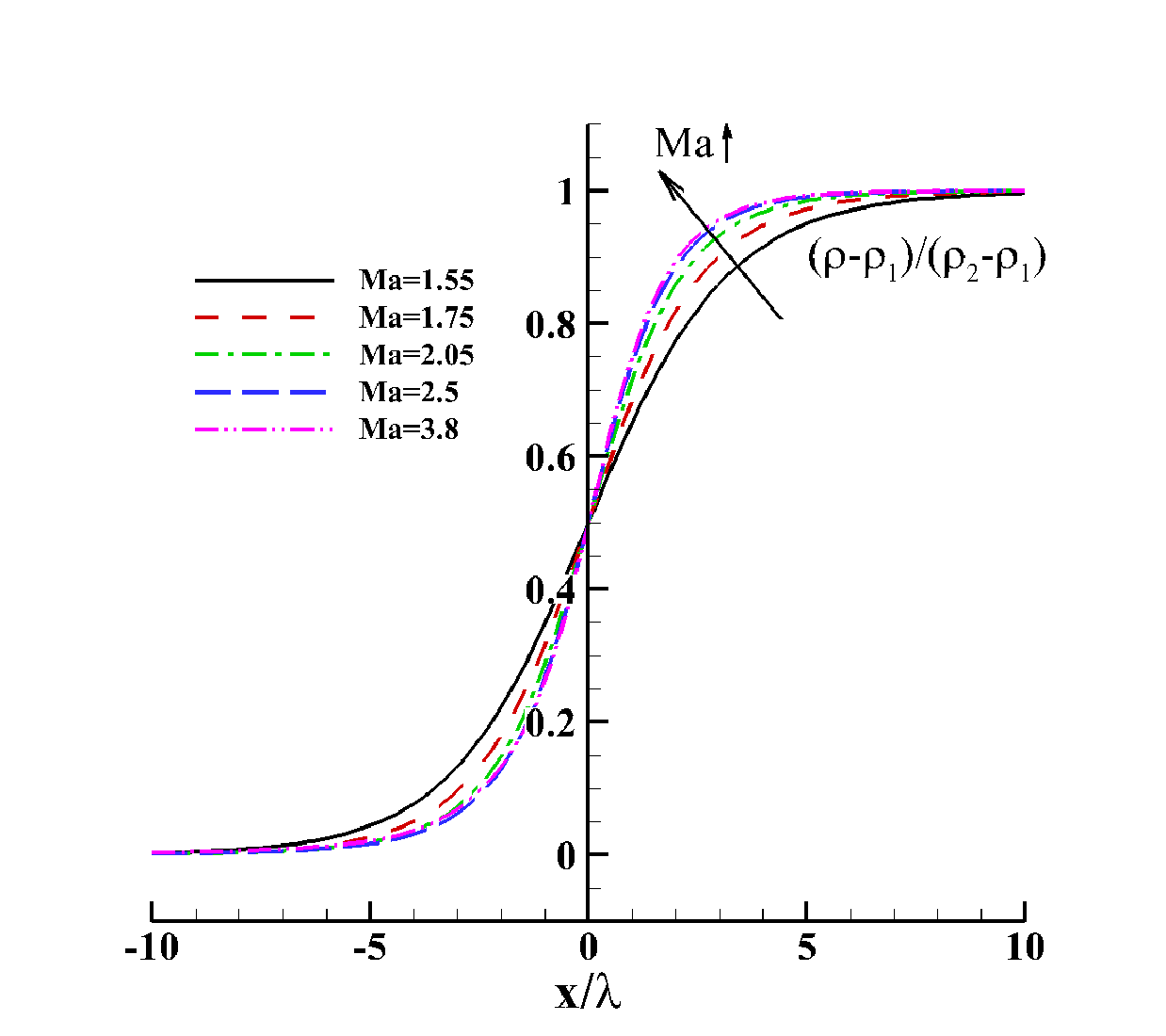}
			\centering \caption*{(a) $\mathrm{Ma}$ = 1.55 $\sim$ 3.8}
		\end{minipage}
\begin{minipage}{8cm}
			\includegraphics*[width=8cm]{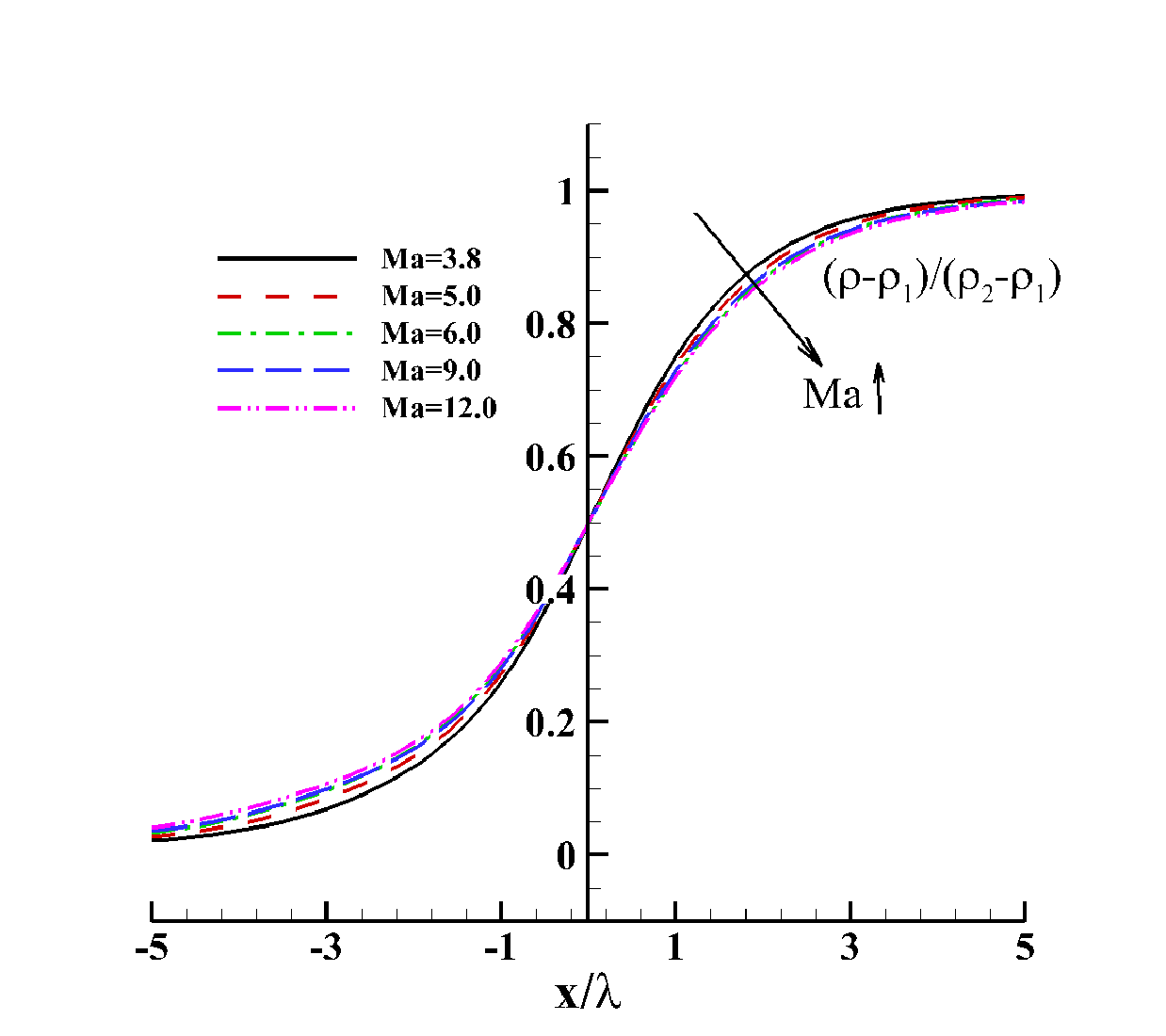}
			\centering \caption*{(b) $\mathrm{Ma}$ = 3.8 $\sim$ 12.0}
		\end{minipage}
}
\par
\subfigure*{\
\begin{minipage}{8cm}
			\includegraphics*[width=8cm]{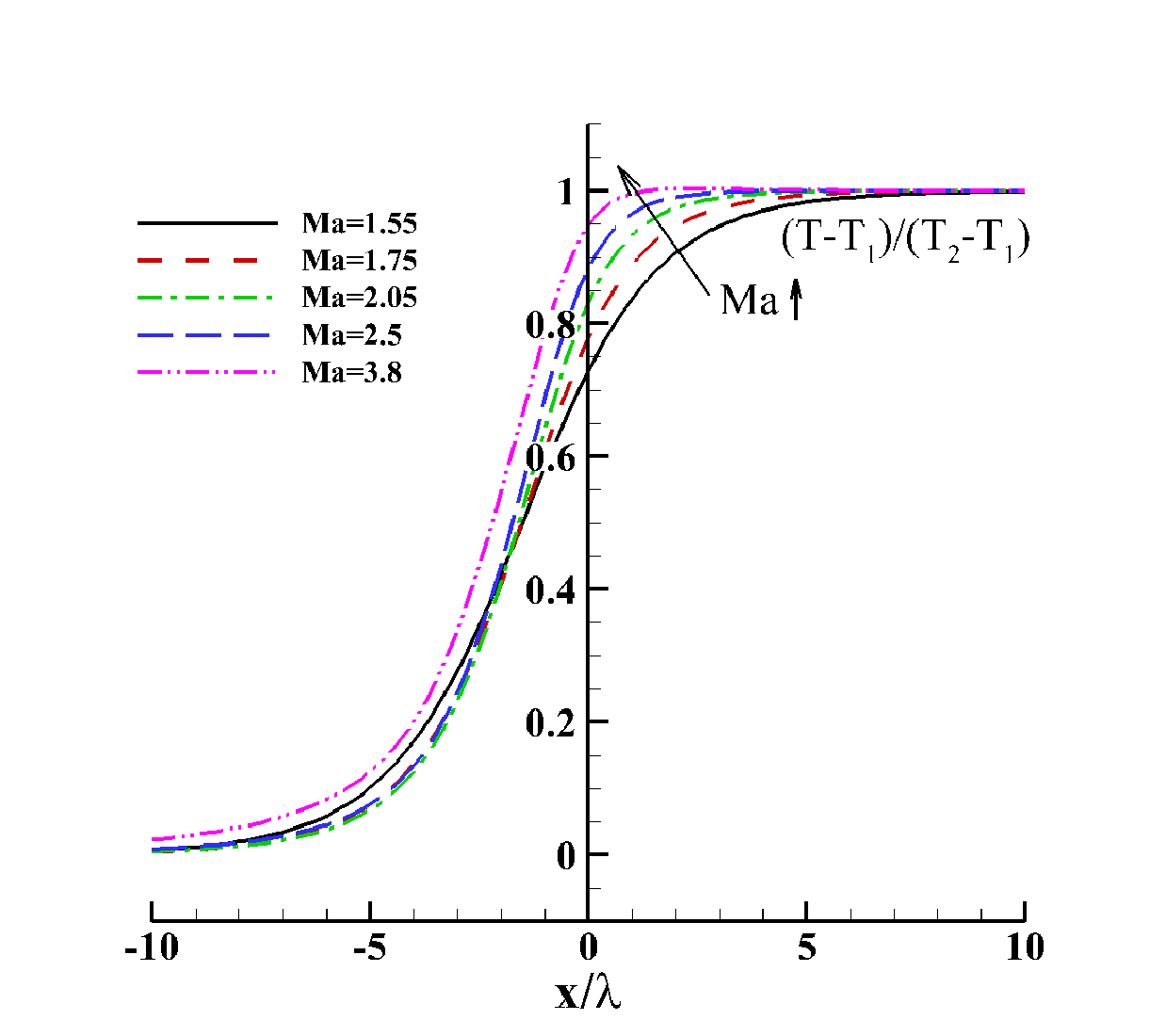}
			\centering \caption*{(c) $\mathrm{Ma}$ = 1.55 $\sim$ 3.8}
		\end{minipage}
\begin{minipage}{8cm}
			\includegraphics*[width=8cm]{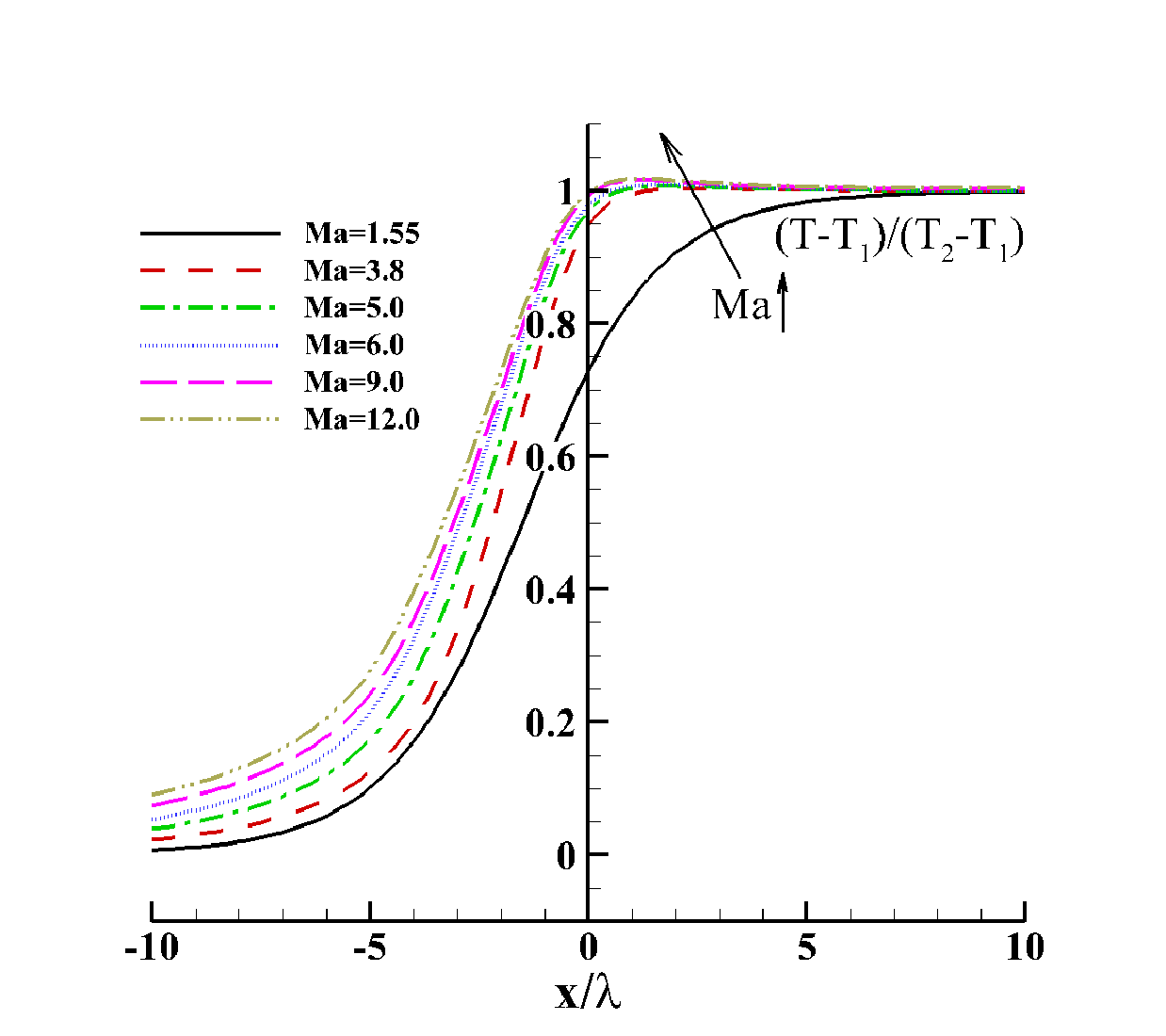}
			\centering \caption*{(d) $\mathrm{Ma}$ = 1.55, 3.8 $\sim$ 12.0}
		\end{minipage}
}
\par
\subfigure*{\
\begin{minipage}{8cm}
			\includegraphics*[width=8cm]{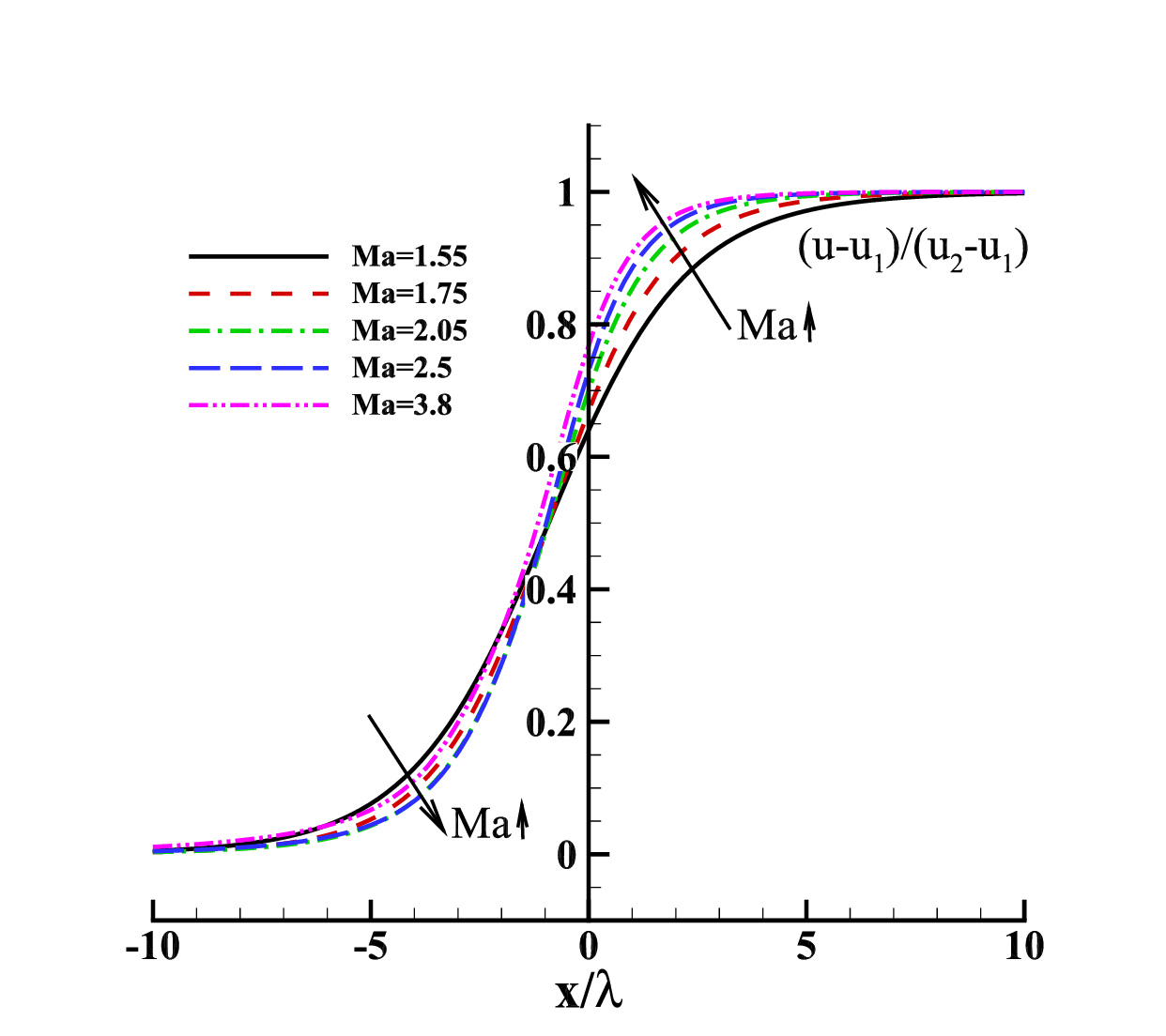}
			\centering \caption*{(e) $\mathrm{Ma}$ = 1.55 $\sim$ 3.8}
		\end{minipage}
\begin{minipage}{8cm}
			\includegraphics*[width=8cm]{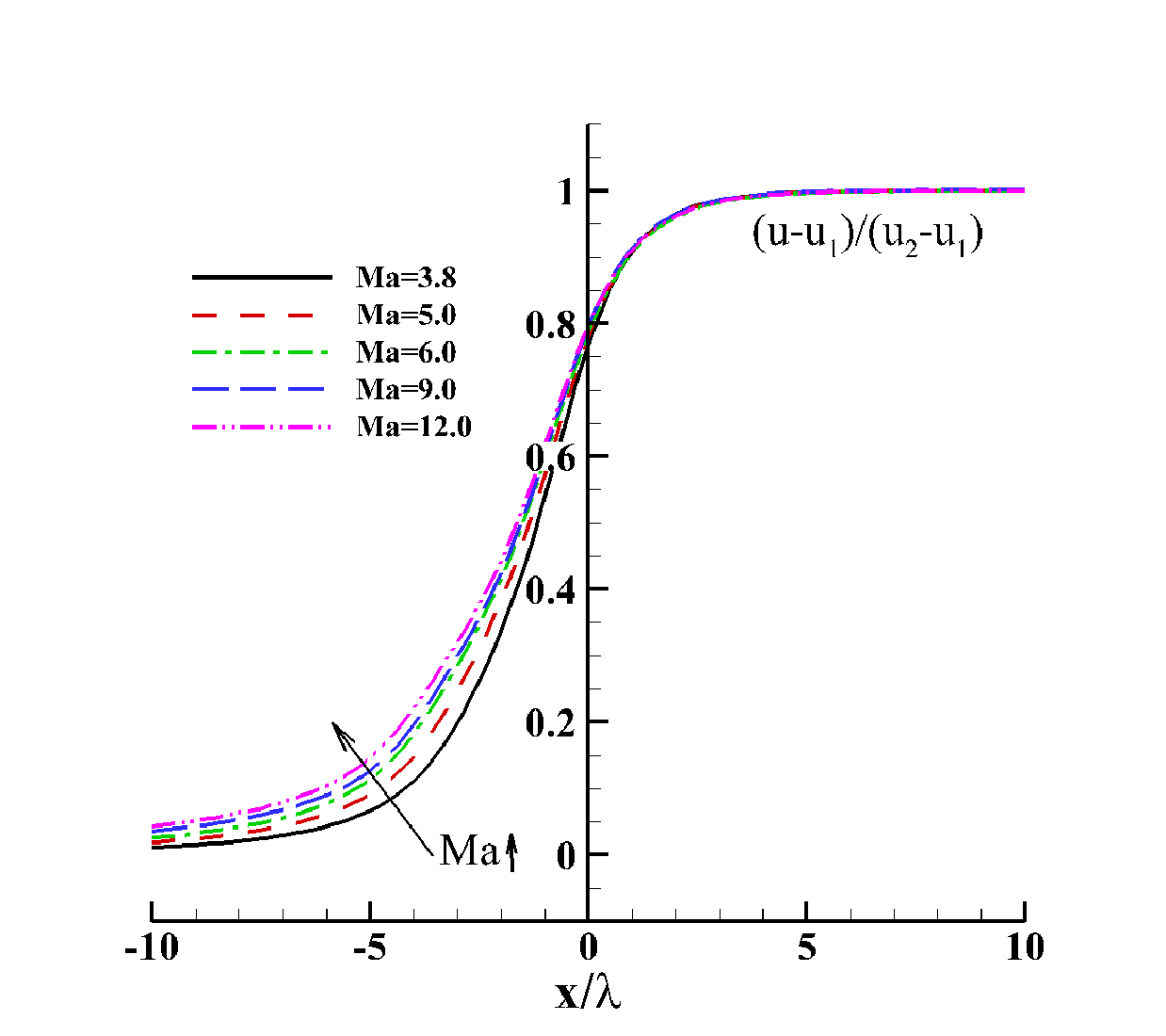}
			\centering \caption*{(f) $\mathrm{Ma}$ = 3.8 $\sim$ 12.0}
		\end{minipage}
}
\caption{ DBM simulation results of density, temperature, and velocity
profiles, respectively. }
\label{fig4}
\end{figure*}

To investigate the effects of Mach number on macroscopic quantities within
the shock structure, Fig. \ref{fig4} presents the DBM simulation results for
the density (first row), temperature (second row), and velocity (third row)
profiles. The left and right columns correspond to cases with lower and
higher Mach numbers, respectively.

For the density profiles, the effects of the Mach number are two-stage.
Specifically, a critical Mach number, $\mathrm{Ma}_{\rho} $ ($\mathrm{Ma}%
_{\rho} \approx 3.8 $), separates two distinct behaviors. When $\mathrm{Ma}<
\mathrm{Ma}_{\rho}$, as shown in Fig. \ref{fig4}(a), the Mach number
steepens the interface. Conversely, for $\mathrm{Ma} > \mathrm{Ma}_{\rho} $,
as shown in Fig. \ref{fig4}(b), the interface becomes gentler. Additionally,
for $\mathrm{Ma} < \mathrm{Ma}_{\rho} $, the density profiles are more
diffuse near the outflow region and more compact near the inflow region.
Beyond the critical value ($\mathrm{Ma} > \mathrm{Ma}_{\rho} $), this trend
reverses. In summary, when $\mathrm{Ma} < \mathrm{Ma}_{\rho} $, the Mach
number primarily sharpens the interface, while for $\mathrm{Ma} >\mathrm{Ma}%
_{\rho} $, it broadens the interface.

This phenomenon is primarily due to the compressibility of the fluid inside
the shock. When the Mach number is below $Ma_{\rho} $, the shock is weaker,
and the fluid's compressibility is stronger, particularly in the region
ahead of the shock, where the fluid undergoes significant compression. This
compression amplifies changes in physical quantities across the shock,
resulting in a steeper interface gradient. In this case, the density
distribution exhibits larger variations near the outflow region and a more
compact arrangement near the inflow region. When the Mach number exceeds $%
\mathrm{Ma}_{\rho} $, the compressibility weakens, the compression effect
saturates, and the changes in physical quantities across the shock become
smoother. The transition region of the shock broadens, and the density
distribution smooths, producing a gentler shock interface.


The effects of Mach number on temperature and velocity interfaces also
exhibit two-stage behavior. Taking the temperature profiles as an example,
an intersection occurs when $\mathrm{Ma} < \mathrm{Ma}_{T} $ [$\mathrm{Ma}%
_{T} \approx 2.5 $, see Fig. \ref{fig4}(c)], but no intersection is observed
for $\mathrm{Ma} > \mathrm{Ma}_{T} $ [see Fig. \ref{fig4}(d)]. Similar to
the density profiles, the temperature profiles near the outflow region are
more diffuse for $\mathrm{Ma} < \mathrm{Ma}_{T} $ and more compact for $%
\mathrm{Ma} > \mathrm{Ma}_{T} $. This phenomenon indicates that, as the Mach
number increases, the region with strong compressibility in the fluid shifts
from the near-outflow region to the near-inflow region.


\begin{figure}[h]
\center\includegraphics*
[ width=0.45\textwidth]{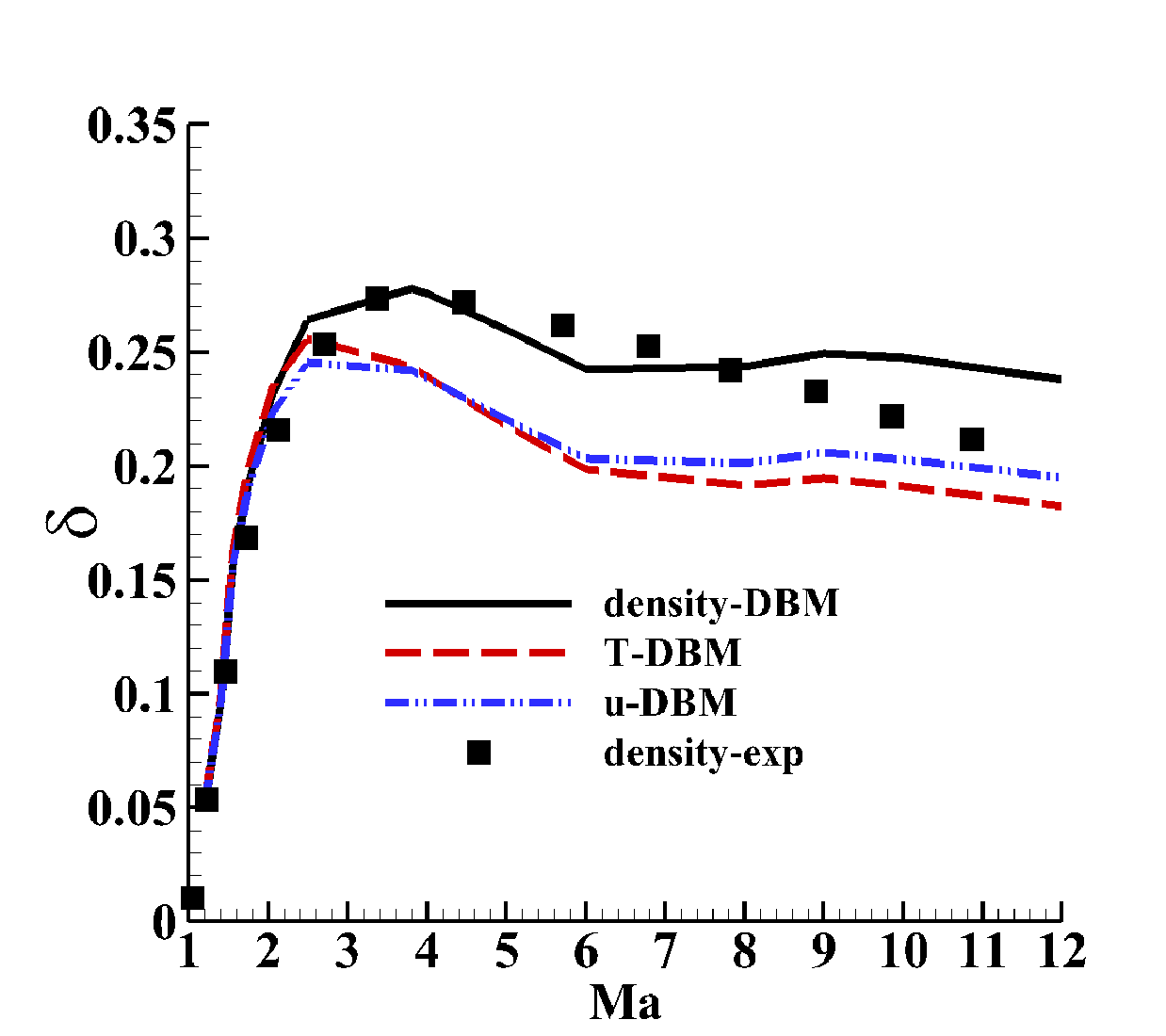}
\caption{ Thicknesses of various types of interfaces. The experimental data
are taken from Fig. 2 in Ref. \citep{Alsmeyer1976JFM}.}
\label{fig5}
\end{figure}

\subsubsection{ Effects of Mach number on shock thickness}

To quantitatively characterize the thickness of the shock structure, the
maximum slope thickness [see Fig. \ref{fig1}(b)] is defined. Figure \ref%
{fig5} illustrates the thicknesses of the density, temperature, and velocity
interfaces obtained from DBM simulations. For comparison, experimental
results for density interface thicknesses are also included. Across the Mach
number range from 1.1 to 9.0, the DBM simulation results align with the
experimental data, staying within the error margins. However, when $\mathrm{%
Ma} > 9 $, the discrepancy between the simulation and experimental results
gradually increases.

The effects of Mach number on interface thicknesses also exhibit two-stage
behavior. As Mach number increases, the thicknesses of all three interfaces
first increase sharply and then decrease gradually, reaching their maximum
at critical Mach numbers. The critical Mach numbers for the three types of
interfaces differ. Specifically, they are $\mathrm{Ma}_{\rho} \approx 3.8 $,
$\mathrm{Ma}_{T} \approx 2.5 $, and $\mathrm{Ma}_{u} \approx 2.5 $,
respectively, consistent with results in Fig. \ref{fig4}.

Key observations include: (i) For $\mathrm{Ma} < 2.0 $, the thicknesses of
all three interfaces are nearly identical. (ii) For $\mathrm{Ma} > 2.0 $,
the density interface thickness becomes greater than those of the velocity
and temperature interfaces. As $\mathrm{Ma} $ increases, the gap between the
density interface and the other two interfaces widens. This is because
density changes are more directly influenced by compression effects, leading
to a faster growth rate for the density interface thickness compared to the
velocity and temperature interfaces. (iii) For $\mathrm{Ma} < 3.8 $, the
temperature interface thickness exceeds that of the velocity interface, but
this trend reverses for $\mathrm{Ma} > 3.8 $.

Each interface exhibits distinct structural features and driving mechanisms.
The density interface reflects the compressibility of the fluid, making it
the most affected by shock compression effects. In contrast, temperature
changes depend more on thermal conduction than compressibility. As the Mach
number increases, the time for fluid to pass through the shock decreases,
limiting the time available for heat diffusion. This weakens the thermal
conduction effect, resulting in a relatively smaller temperature interface
thickness. The velocity interface is primarily influenced by viscous
effects. As the Mach number increases, the velocity gradient within the
shock steepens. However, the shorter transit time across the shock limits
the relaxation time required for viscous effects to fully smooth the
velocity gradient. For compressible flows, changes in density, temperature,
and velocity interfaces are interconnected. These effects combine to create
maximum interface thickness at critical Mach numbers, corresponding to the
strong nonequilibrium state of the shock.


\subsection{ Effects of Ma number on distribution function}

In kinetic methods, nonequilibrium effects are reflected not only
macroscopically in the spatio-temporal gradients of macroscopic quantities,
such as constitutive relations, but also mesoscopically in the distribution
function. Analyzing the characteristics of the distribution function is
essential for model selection and provides insights into both HNE and TNE
phenomena. As mentioned in Section \ref{higer-order-TNE}, directly
discretizing the velocity space enables DBM models to accurately capture
higher-order TNE effects. This approach also facilitates the direct
acquisition of the true distribution functions. However, a significant
challenge of this method is identifying which order of TNE dominates in
practical simulations.

To address this issue, CE analysis is often employed to systematically
increase the degree of non-equilibrium. The CE analysis provides analytical
expressions for distribution functions at different TNE orders, enabling a
more comprehensive understanding of the dominant nonequilibrium effects. For
example, as shown in Section \ref{CE-analysis}, the first three orders of
the distribution function $g $ are:
\begin{equation}
{g^{(1)}} = - \tau [\frac{{\partial {g^{eq}}}}{{\partial {t_1}}} + {v_x}
\cdot \frac{{\partial {g^{eq}}}}{{\partial x}}] + {g^{s(1)}},  \label{Eq.g1}
\end{equation}
\begin{equation}
{g^{(2)}} = - \tau [\frac{{\partial {g^{(1)}}}}{{\partial {t_1}}} + \frac{{%
\partial {g^{eq}}}}{{\partial {t_2}}} + {v_x} \cdot \frac{{\partial {g^{(1)}}%
}}{{\partial x}}] + {g^{s(2)}} ,
\end{equation}
and
\begin{equation}
{g^{(3)}} = - \tau [\frac{{\partial {g^{eq}}}}{{\partial {t_3}}} + \frac{{%
\partial {g^{(1)}}}}{{\partial {t_2}}} + {v_x} \cdot \frac{{\partial {g^{(2)}%
}}}{{\partial x}}] + {g^{s(3)}} ,
\end{equation}
where
\begin{equation}
g^{s(k)}={g^{eq}} + {g^{eq}}\left\{ {(1 - \Pr )\cdot{c_x }{q_x^{(k)} } \cdot
\frac{{[\frac{{{c_x^2} }}{{RT}} - 3 ]}}{{[( n + 3)pRT]}}} \right\} .
\end{equation}

It is worth noting that the term $\frac{{\partial g^{(2)}}}{{\partial t_1}}
= 0 $, as changes in $g^{(2)} $ cannot be observed on the $t_1 $ time scale.
Otherwise, the zeroth, first, and second contracted moments of $g^{(3)} $
would be nonzero, violating the conservation laws \cite{Gan2025JFM}.


\begin{figure*}[tbp]
\centering
\subfigure*{\
\begin{minipage}{8cm}
			\includegraphics*[width=8cm]{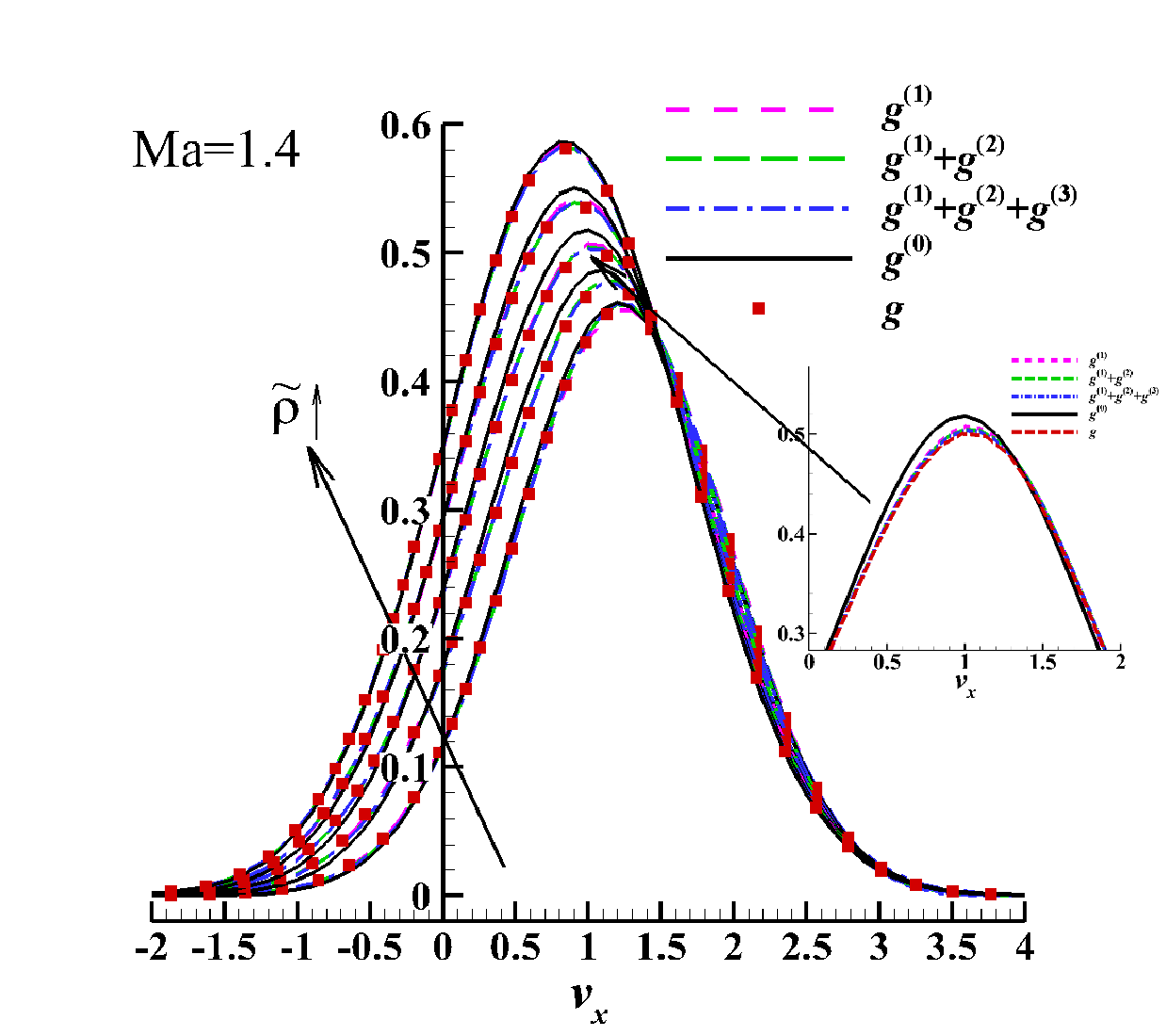}
			\centering \caption*{(a) $\mathrm{Ma} = 1.4$}
		\end{minipage}
\begin{minipage}{8cm}
			\includegraphics*[width=8cm]{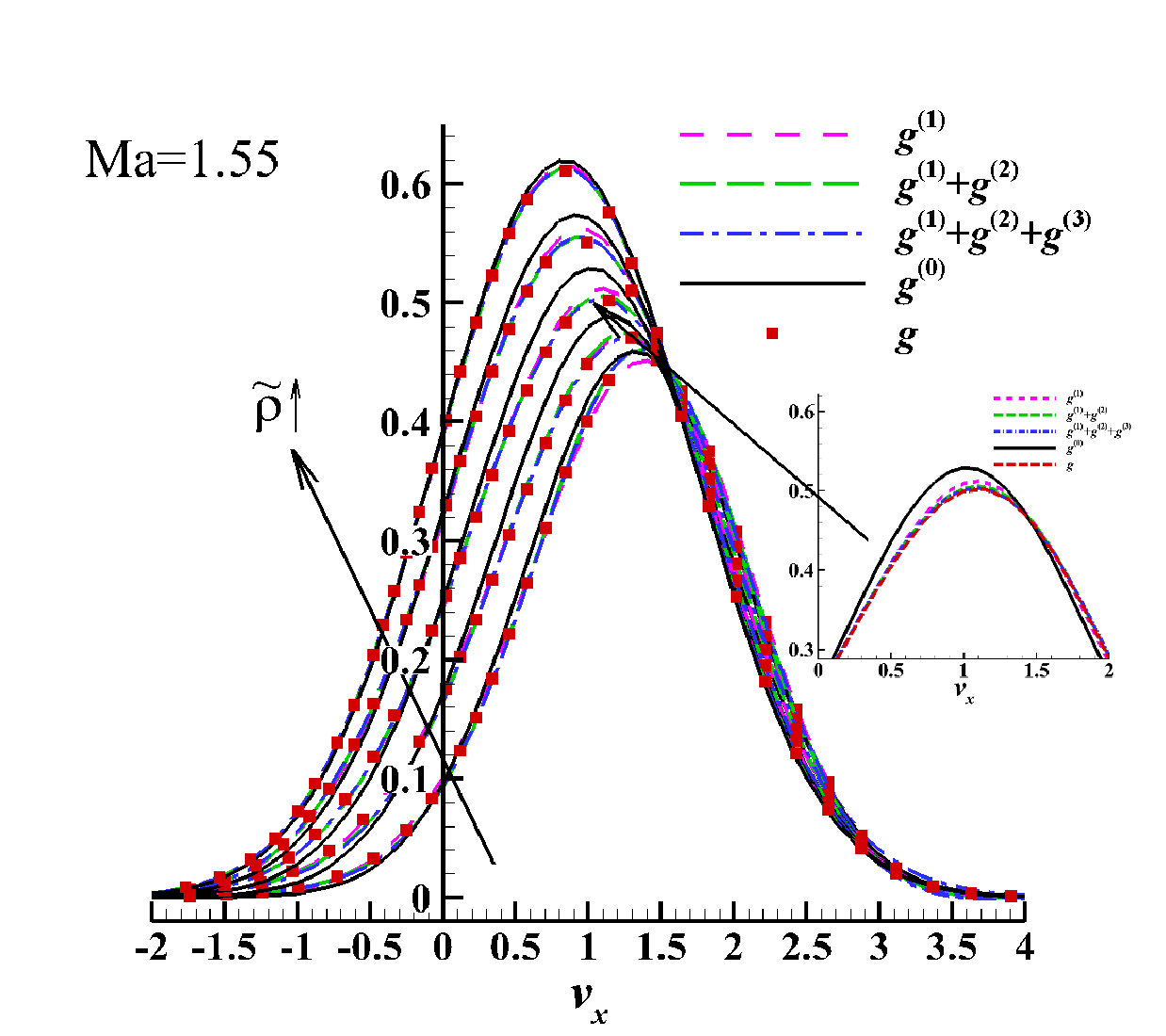}
			\centering \caption*{(b) $\mathrm{Ma} = 1.55$}
		\end{minipage}
}
\par
\subfigure*{\
\begin{minipage}{8cm}
			\includegraphics*[width=8cm]{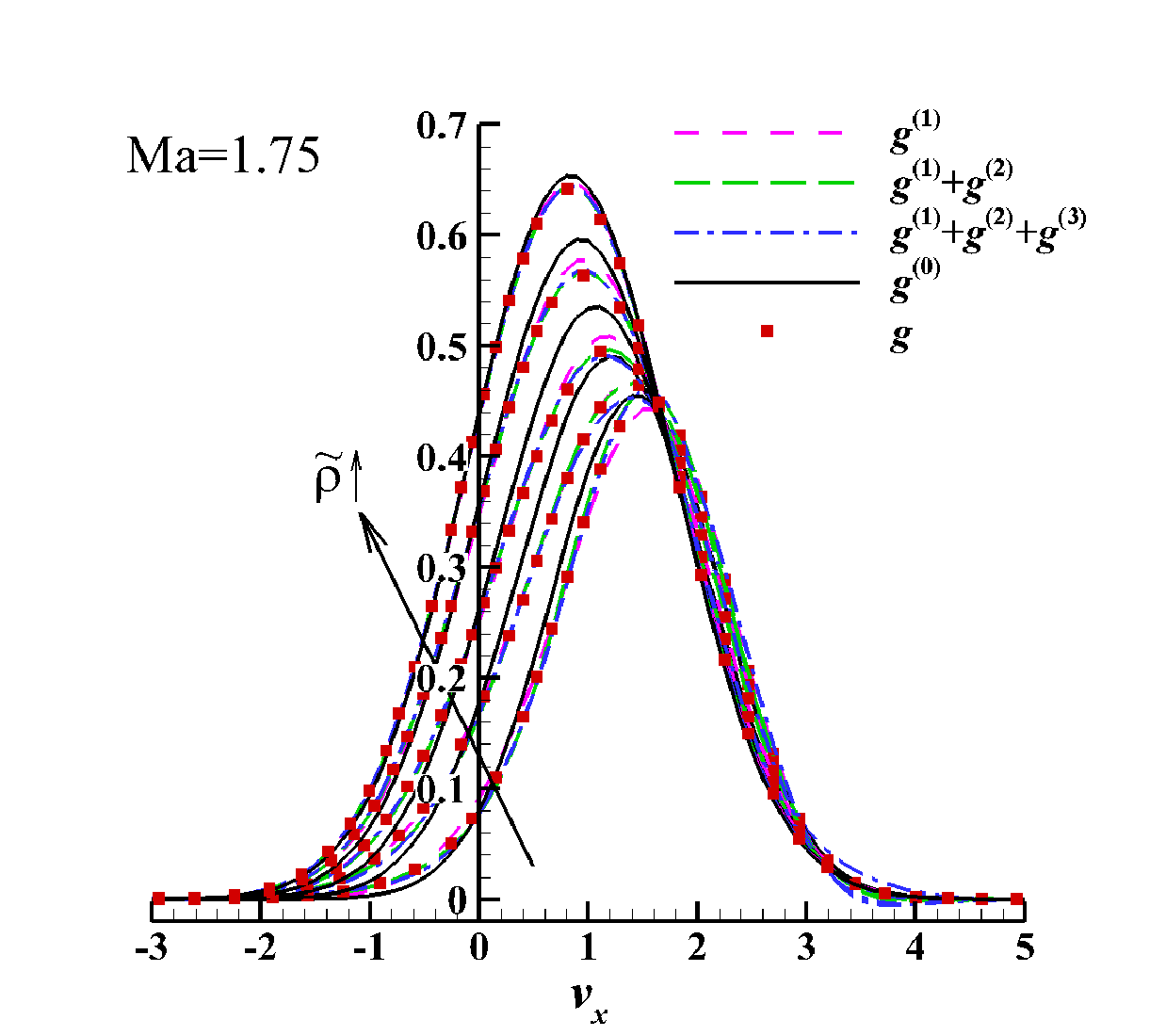}
			\centering \caption*{(c) $\mathrm{Ma} = 1.75$}
		\end{minipage}
\begin{minipage}{8cm}
			\includegraphics*[width=8cm]{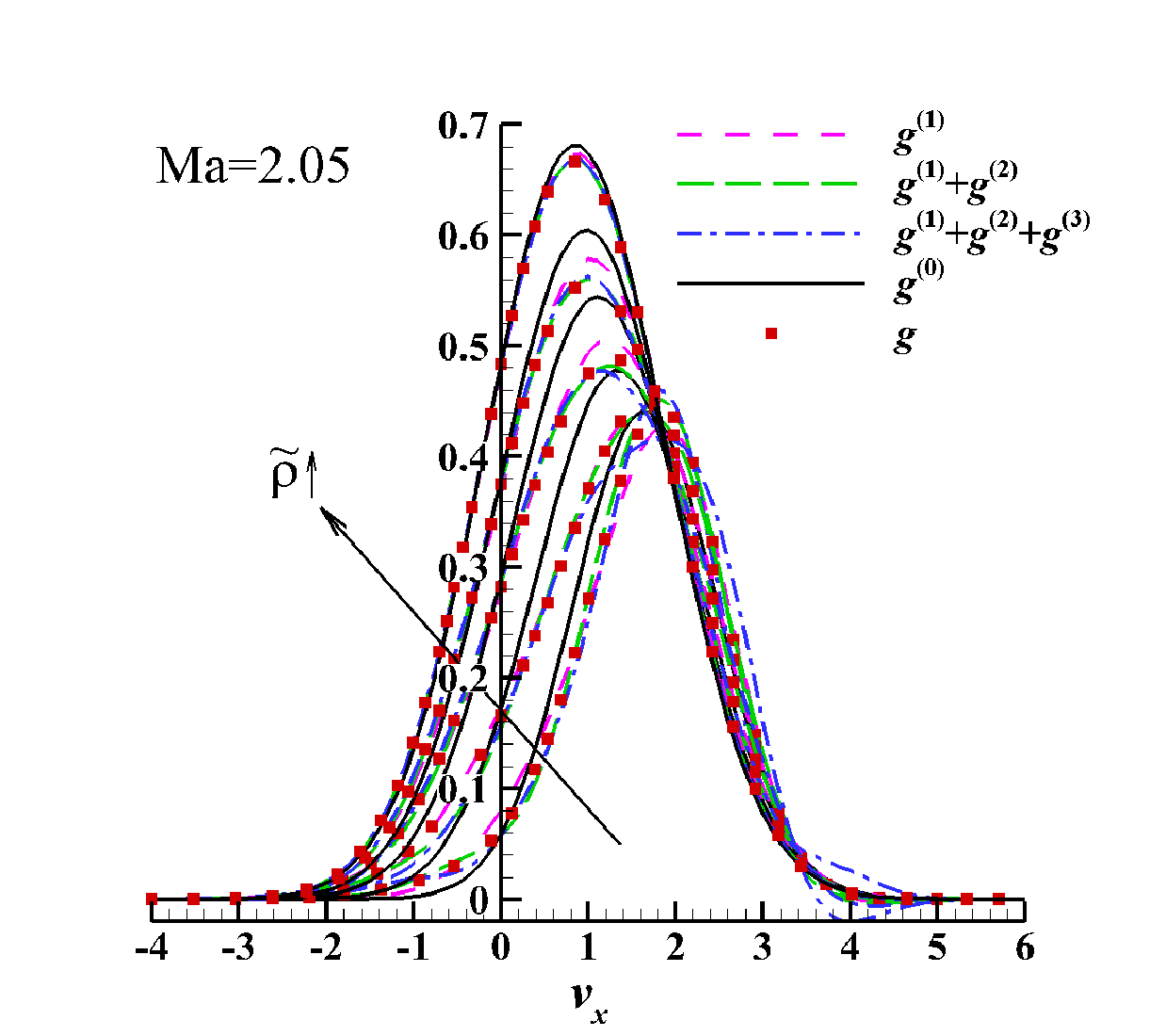}
			\centering \caption*{(d) $\mathrm{Ma} = 2.05$}
		\end{minipage}
}
\par
\subfigure*{\
\begin{minipage}{8cm}
			\includegraphics*[width=8cm]{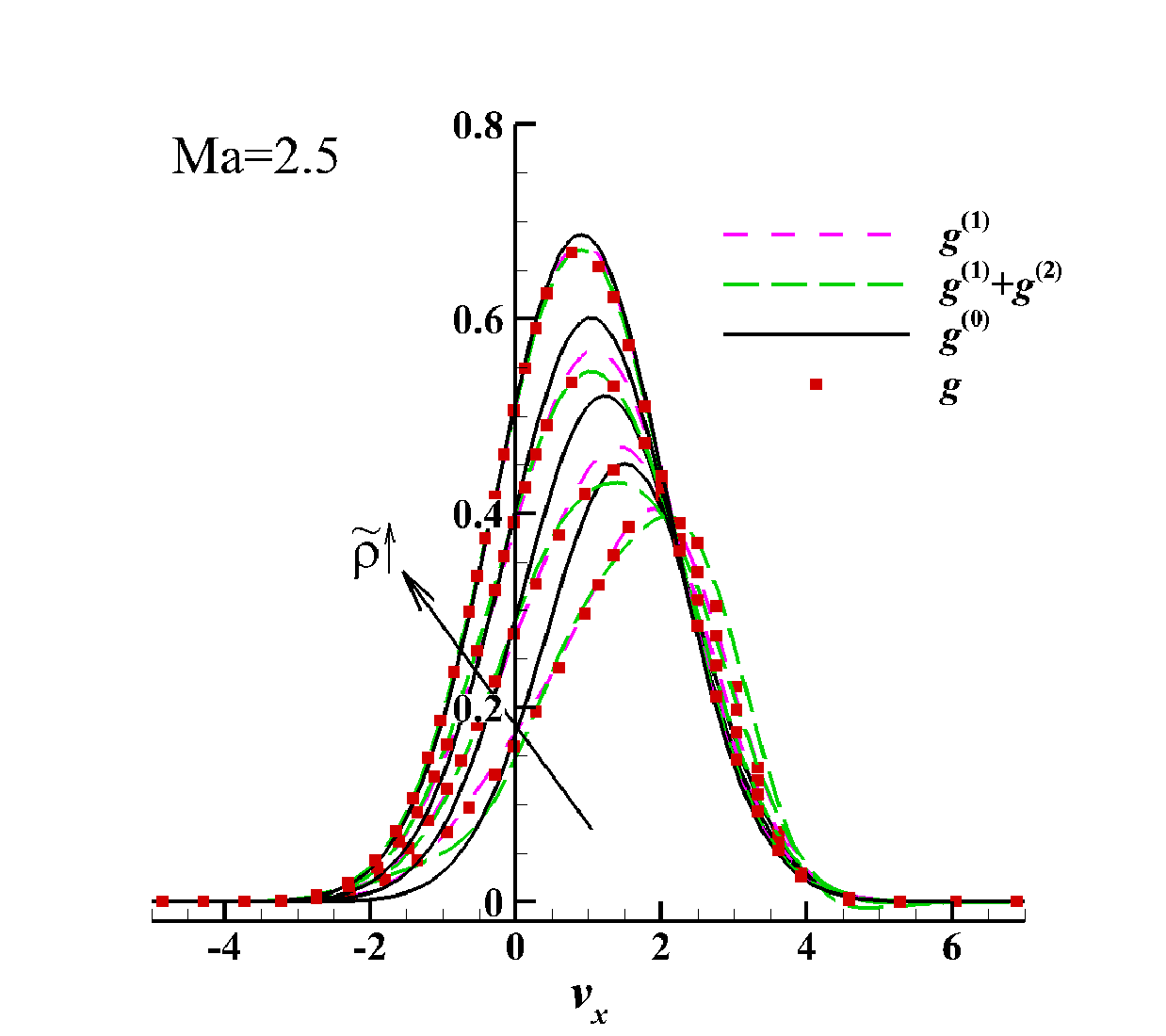}
			\centering \caption*{(e) $\mathrm{Ma} = 2.5$}
		\end{minipage}
\begin{minipage}{8cm}
			\includegraphics*[width=8cm]{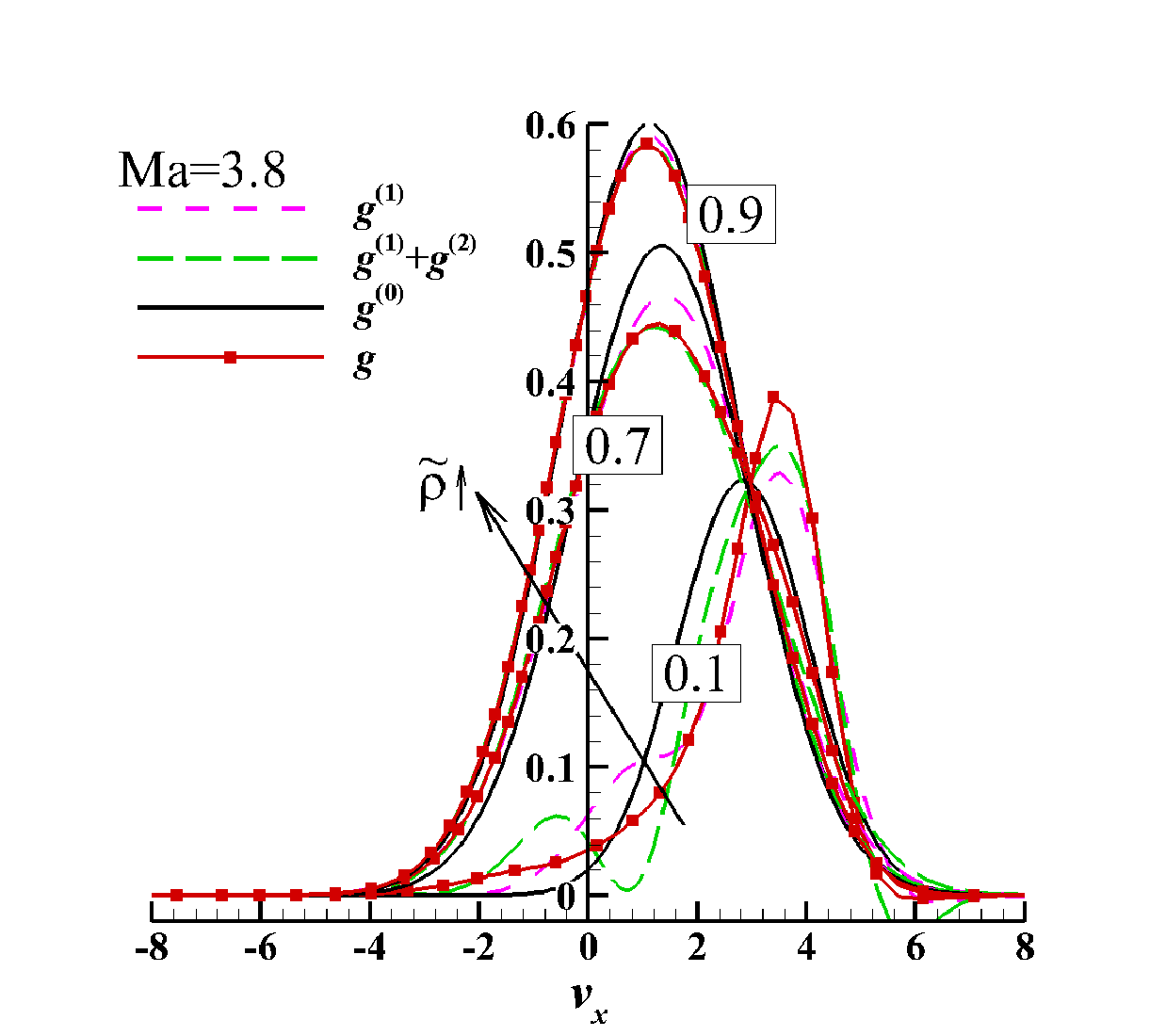}
			\centering \caption*{(f) $\mathrm{Ma} = 3.8$}
		\end{minipage}
}
\caption{ Comparisons of distribution functions between DBM simulation and
analytical solutions inside the shock structure. }
\label{fig6}
\end{figure*}

To analyze non-equilibrium states within the shock, Fig. \ref{fig6} compares
the distribution functions $g$ obtained from DBM simulations with analytical
results of various accuracy orders. For Mach numbers ranging from 1.4 to
2.5, distribution functions are evaluated at positions across the internal
structure, spanning from the inflow to outflow regions. Specifically, five
positions with normalized densities $\widetilde \rho $ = 0.1, 0.3, 0.5, 0.7
and 0.9, are considered, respectively. Larger (smaller) normalized density
values $\widetilde \rho $ indicate positions closer to the outflow (inflow)
region. Black lines represent the equilibrium distribution function $g^{(0)}$%
. Red points are BM simulation results with $g$ preserving sufficient
non-equilibrium orders. Pink (${g^{(1)}}$), green (${g^{(1)}}+{g^{(2)}}$)
and blue (${g^{(1)}}+{g^{(2)}}+{g^{(3)}}$) lines denote analytical solutions
considering up to the first-order, second-order, and third-order TNE,
respectively.

\subsubsection{ Appearances of distribution function}

Several key observations can be made:

(I) For $\mathrm{Ma} = 1.4 $, as shown in Fig. \ref{fig6}(a), the
distribution function is non-zero within $v_x = (-2, 4) $, and zero outside
this range. As the Mach number increases, the non-zero region of the
distribution function expands, compared to cases with lower Mach numbers.

(II) As $\widetilde{\rho} $ increases (i.e., closer to the outflow), the
values of the distribution function increase due to the corresponding
macroscopic quantities.

(III) The peak of the distribution function shifts towards lower particle
velocities $v_x $ as $\widetilde{\rho} $ increases.

(IV) For $\mathrm{Ma} = 1.55 $, as shown in Fig. \ref{fig6}(b), the peak
values of the distribution function are higher than those observed for $%
\mathrm{Ma} = 1.4 $. As the Mach number further increases, the peak values
of the distribution function gradually rise. Interestingly, for $\mathrm{Ma}
= 3.8 $ [see Fig. \ref{fig6}(a)], the peak values of the distribution
function are significantly lower than those observed at smaller Mach
numbers. This suggests that the influence of Mach number on the distribution
function exhibits a two-stage effect, similar to its impact on macroscopic
quantity interfaces. For more details, refer to Sec. \ref{TWE}.

\subsubsection{ Non-equilibrium degree from perspective of distribution
function}

In the case of $\mathrm{Ma} = 1.4 $ [see Fig. \ref{fig6}(a)], the following
observations can be made:

(I) At $\widetilde{\rho} = 0.1 $ and $0.9 $, the system is near equilibrium,
while intermediate positions ($\widetilde{\rho} = 0.3, 0.5, 0.7 $) deviate
from equilibrium.

(II) The strongest TNE occurs at $\widetilde{\rho} \approx 0.5 $ and
gradually weakens towards both ends. The location with the greatest degree
of TNE is determined collectively by density, temperature, and velocity
gradients, rather than solely by the density gradients.

(III) At the intermediate positions $\widetilde{\rho} = 0.3, 0.5, $ and $0.7
$, the distribution functions, including first-order, second-order, and
third-order TNE effects, align closely with the DBM results. This suggests
that the fluid's deviation from equilibrium is predominantly first-order.

(IV) The subfigure showing an enlarged view of the distribution functions
around the peak at $\widetilde{\rho} = 0.5 $ reveals that incorporating
higher-order TNE effects brings the analytical results closer to the DBM
simulation. This demonstrates the enhanced physical accuracy of higher-order
analytical solutions.

\begin{figure*}[h]
\centering
\subfigure*{\
\begin{minipage}{8cm}
			\includegraphics*[width=8cm]{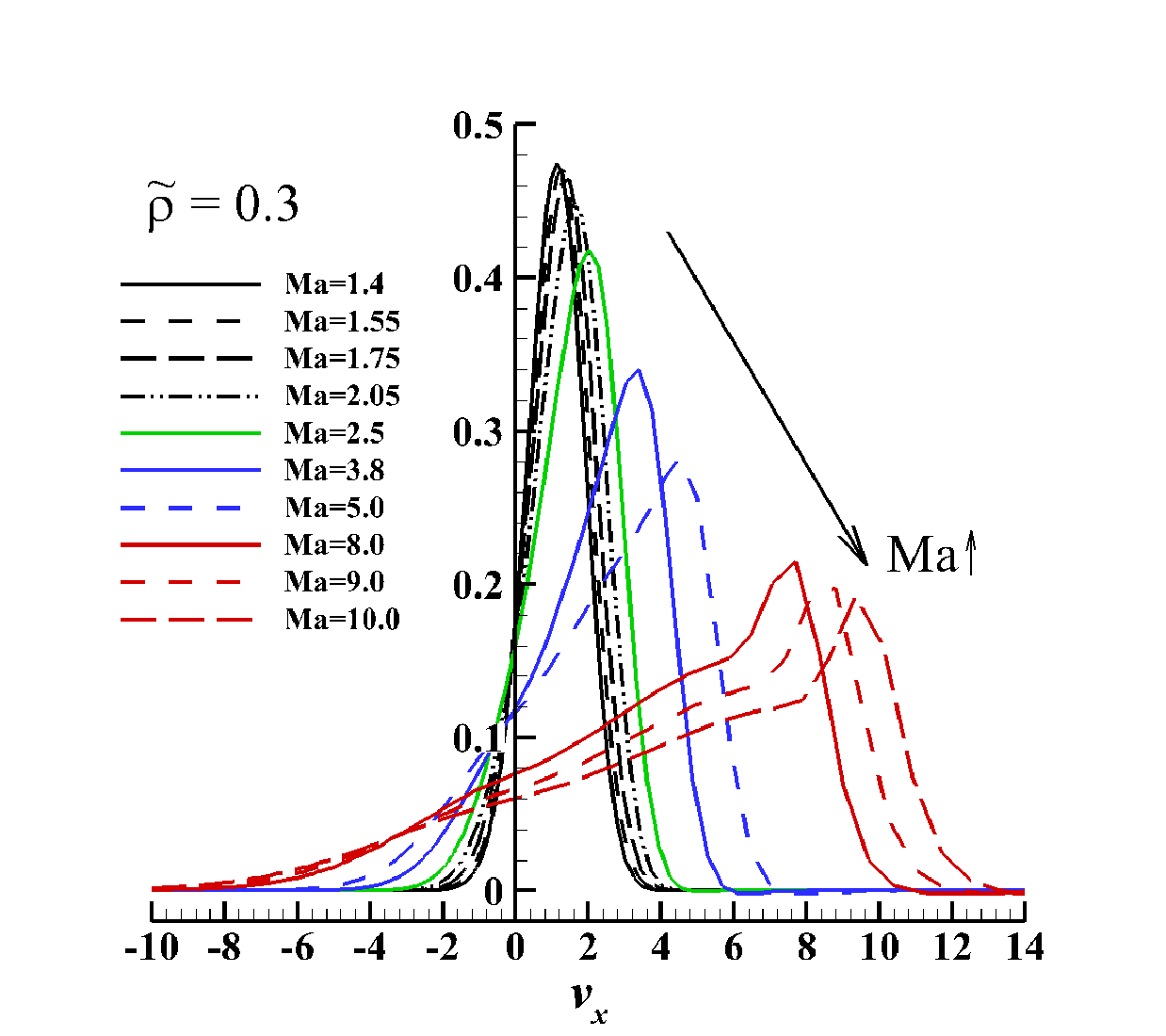}
			\centering \caption*{(a) $\widetilde \rho $ = 0.3}
		\end{minipage}
\begin{minipage}{8cm}
			\includegraphics*[width=8cm]{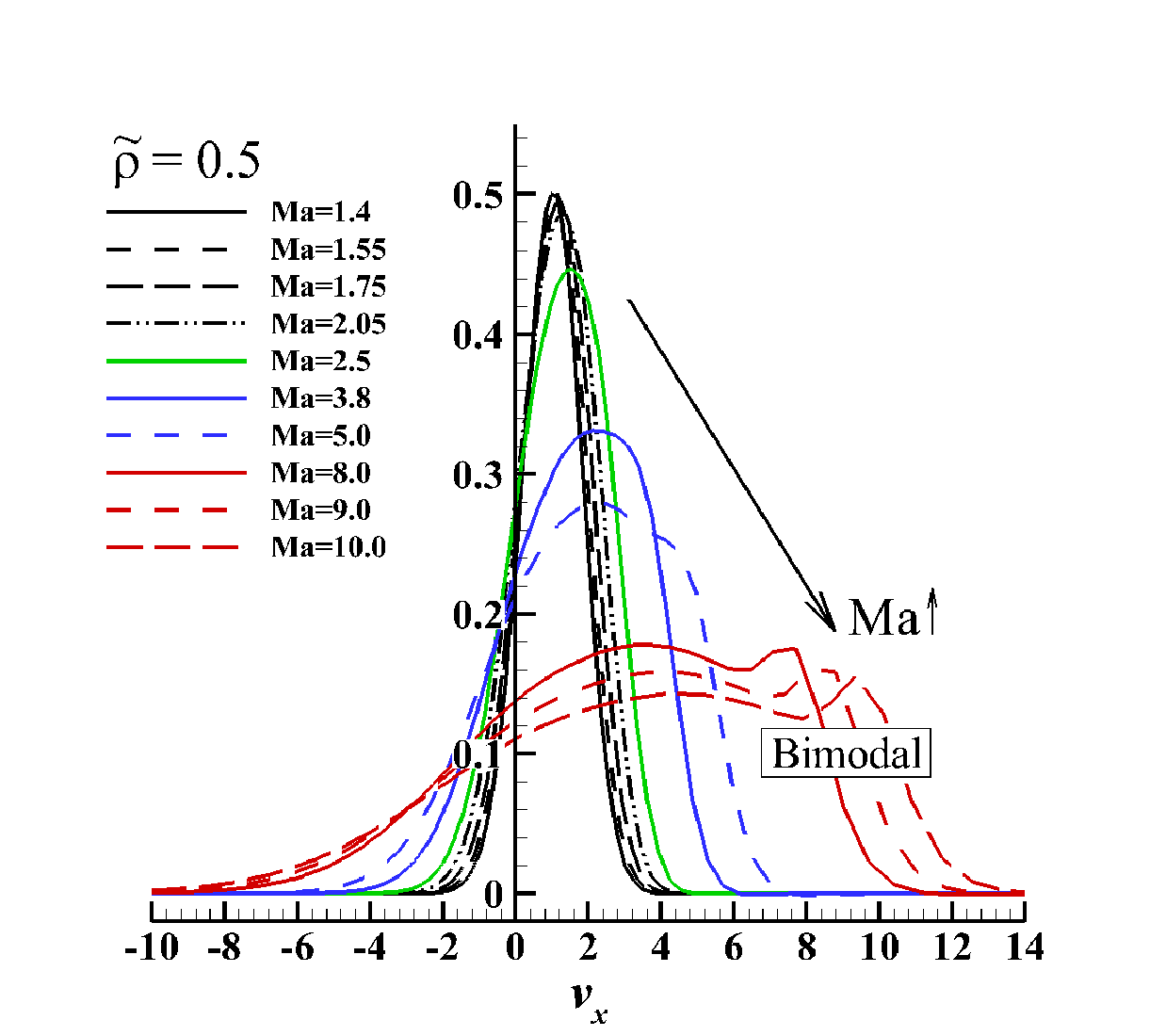}
			\centering \caption*{(b) $\widetilde \rho $ = 0.5}
		\end{minipage}
}
\par
\subfigure*{\
\begin{minipage}{8cm}
			\includegraphics*[width=8cm]{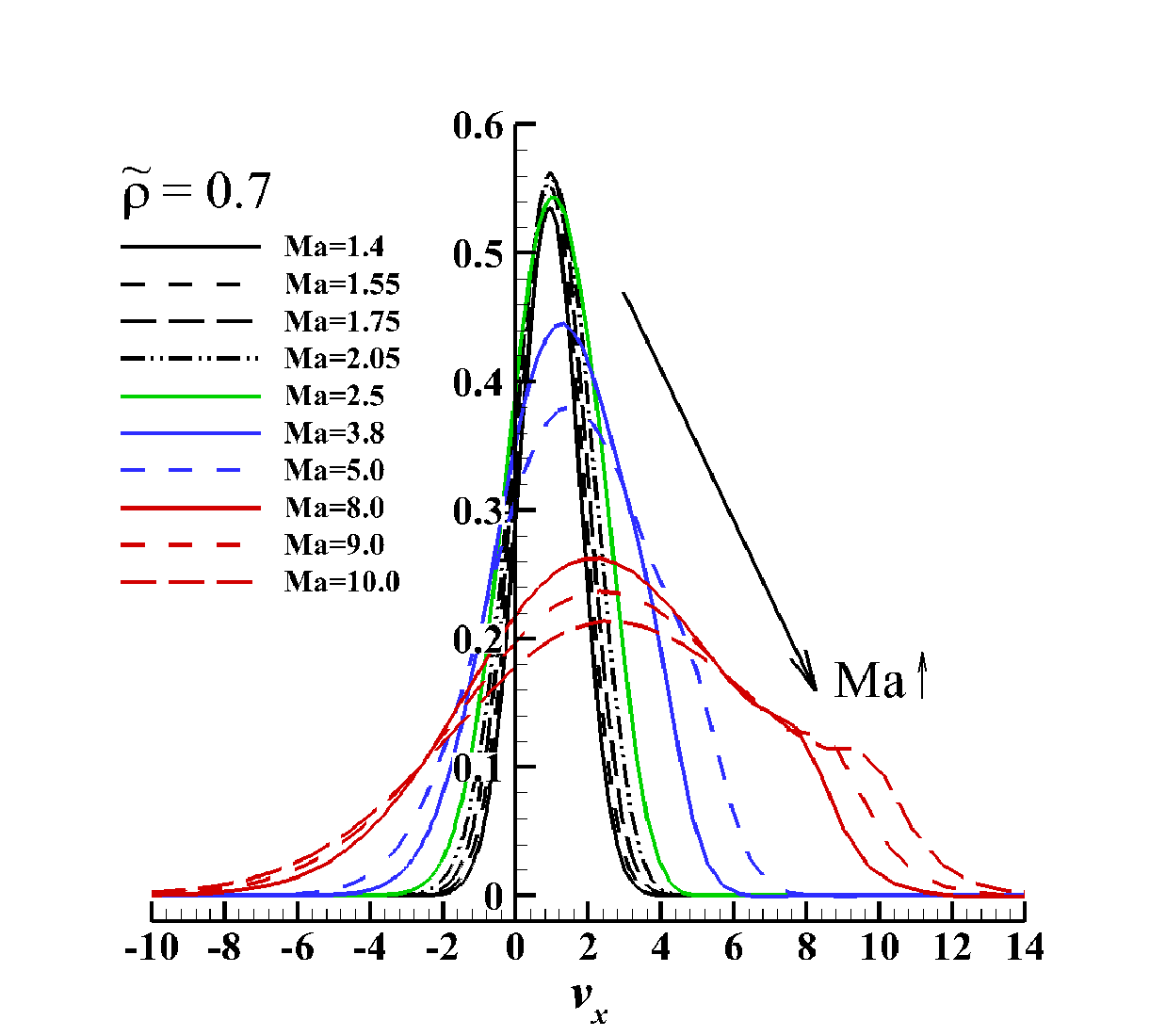}
			\centering \caption*{(c) $\widetilde \rho $ = 0.7, Ma = 1.4 $\sim$ 10.0}
		\end{minipage}
\begin{minipage}{8cm}
			\includegraphics*[width=8cm]{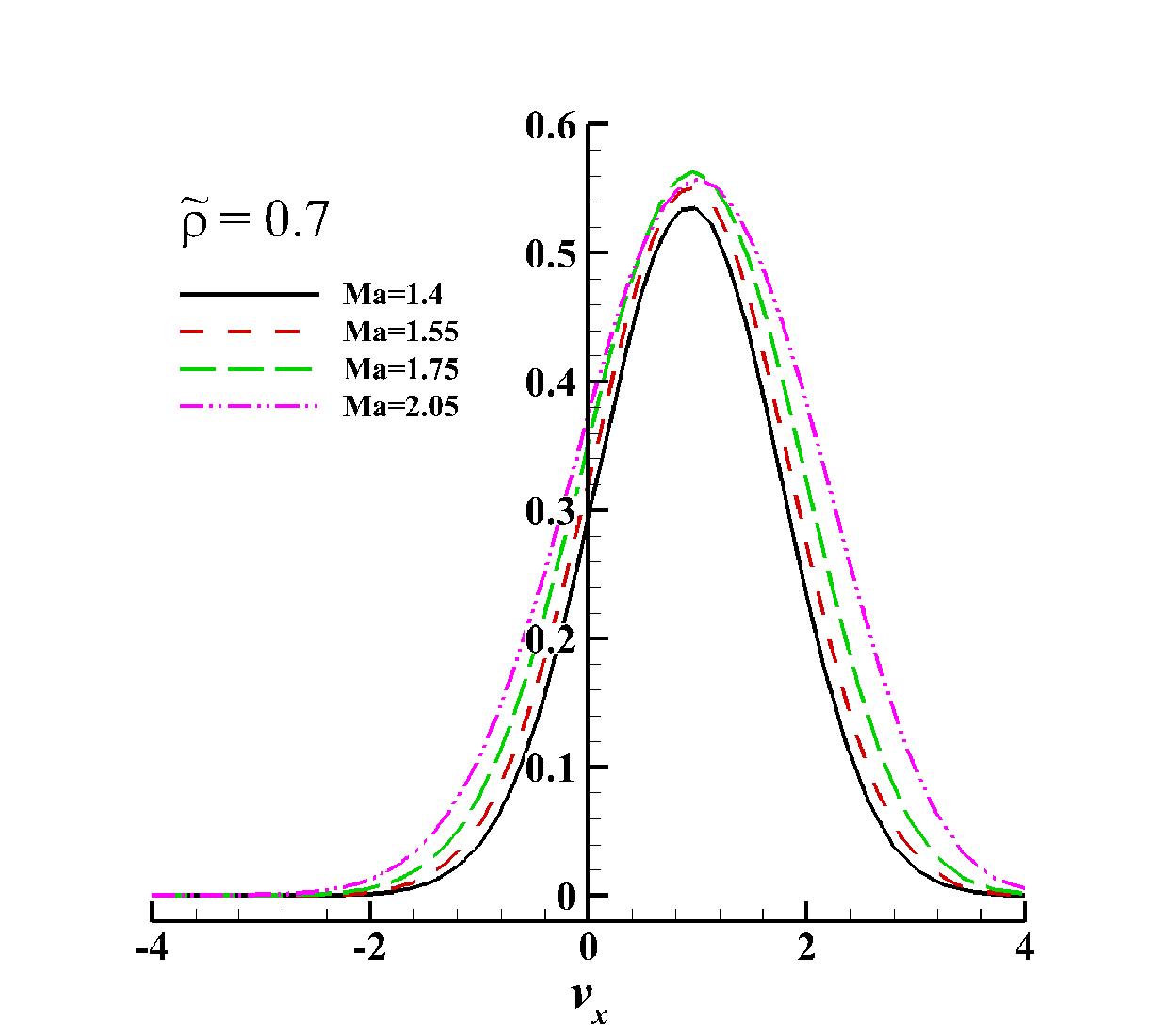}
			\centering \caption*{(d) $\widetilde \rho $ = 0.7, Ma = 1.4 $\sim$ 2.05}
		\end{minipage}
}
\caption{DBM simulations of the distribution functions $g$ at positions with
$\widetilde{\protect\rho}$ = 0.3, 0.5, and 0.7, for Ma numbers ranging from
1.4 to 10.0 }
\label{fig7}
\end{figure*}

For $\mathrm{Ma} = 1.55 $ [see Fig. \ref{fig6}(b)], the following
observations are made:

(I) The positions at $\widetilde{\rho} = 0.1 $ and $0.9 $ are near
equilibrium, while the positions at $\widetilde{\rho} = 0.3, 0.5, $ and $0.7
$ deviate from equilibrium. However, the deviations between $g^{(0)} $ and
other results are more pronounced compared to the case with $\mathrm{Ma} =
1.4 $.

(II) At $\widetilde{\rho} = 0.1 $, the deviation from equilibrium is greater
for $\mathrm{Ma} = 1.55 $ than for $\mathrm{Ma} = 1.4 $. Despite these
differences, the fluid still exhibits primarily a first-order deviation from
equilibrium, consistent with the observations for $\mathrm{Ma} = 1.4 $.

For $\mathrm{Ma} = 1.75 $, as shown in Fig. \ref{fig6}(c), the fluid at all
five positions clearly exhibits a non-equilibrium state. At positions $%
\widetilde{\rho} = 0.1, 0.3, 0.7, $ and $0.9 $, only first-order deviations
from equilibrium are observed. However, at $\widetilde{\rho} = 0.5 $, the
first-order analytical distribution function $g^{(1)} $ deviates noticeably
from the higher-order results, while the second-order and third-order
solutions align closely. This indicates that at $\widetilde{\rho} = 0.5 $,
the fluid exhibits a second-order deviation from equilibrium.

For $\mathrm{Ma} = 2.05 $ [see Fig. \ref{fig6}(d)], the TNE intensity
increases compared to the previous cases. At positions $\widetilde{\rho} =
0.5 $ and $0.7 $, the fluid exhibits second-order deviations from
equilibrium. However, as spatial gradients of macroscopic quantities
increase, numerical errors in the analytical solutions become more
significant. As a result, third-order TNE solutions do not necessarily match
the simulation results better than second-order solutions, despite their
higher physical accuracy.

The second-order TNE analytical solutions align well with DBM simulation
results for $\mathrm{Ma} = 2.5 $ and $3.8 $ at most positions, except at $%
\widetilde{\rho} = 0.1 $ for $\mathrm{Ma} = 3.8 $. However, third-order
solutions (not shown here) lose accuracy due to numerical errors.
Consequently, as the Mach number increases further, even the second-order
analytical solution may become less effective.

A detailed analysis of the distribution function's characteristics is
crucial for adjusting simulation parameters and fully capturing the TNE
characteristics of the fluid. Additionally, analyzing the deviation between
the distribution function and the equilibrium distribution function helps
determine the appropriate order for fluid models.

\subsubsection{ Two-stage effects on distribution functions}

\label{TWE}

To further investigate the influence of Mach number on distribution
functions, Fig. \ref{fig7} presents the DBM simulation results for $g $ at
positions $\widetilde{\rho} = 0.3, 0.5, $ and $0.7 $ for Mach numbers
ranging from 1.4 to 10.0. The following observations can be made:

(I) The influence of Mach number on the DBM results for distribution
functions exhibits a two-stage effect. Overall, as the Mach number
increases, the peak value of the distribution function decreases. However,
for $\mathrm{Ma} < \text{Ma}_{f} $ ($\text{Ma}_{f} \approx 2.05 $), as shown
in Fig. \ref{fig7}(d), the peak value increases with the Mach number. This
phenomenon is primarily due to the combined effects of various macroscopic
quantities and their gradients.

(II) For $\mathrm{Ma} < 2.05 $, the distribution function retains a higher
degree of symmetry. At $\mathrm{Ma} = 2.5 $, this symmetry is significantly
reduced. When the Mach number reaches 3.8, the distribution function clearly
deviates from a normal distribution. For $\mathrm{Ma} = 8.0 $, a pronounced
bimodal character emerges. In fact, the symmetry of the distribution
function also serves as a coarse-grained measure of the TNE degree.

(III) As the Mach number increases, the peak of the distribution function
gradually shifts toward higher particle velocities $v_x $. This shift is
related to the increase in macroscopic velocity. Based on this property, the
corresponding velocity space can be refined to ensure more accurate
simulation results.

\subsection{ Effects of Ma number on TNE measures}

\begin{figure*}[tbp]
\centering
\subfigure*{\
\begin{minipage}{8cm}
			\includegraphics*[width=8cm]{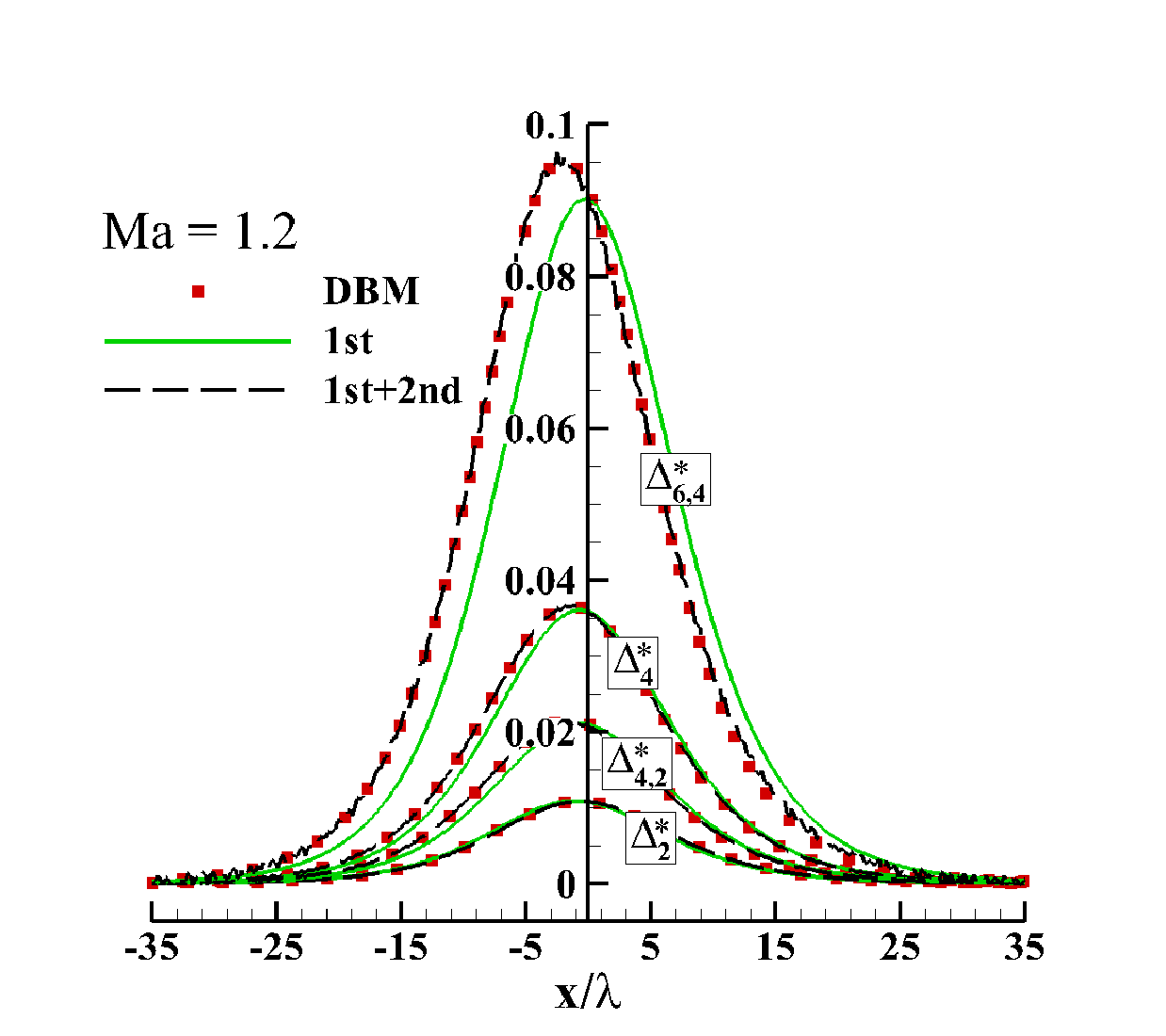}
			\centering \caption*{(a) $\mathrm{Ma} = 1.2$, even-order: $\Delta_{2}^{*}$, $\Delta_{4}^{*}$, $\Delta_{4,2}^{*}$,  and $\Delta_{6,4}^{*}$. }
		\end{minipage}
\begin{minipage}{8cm}
			\includegraphics*[width=8cm]{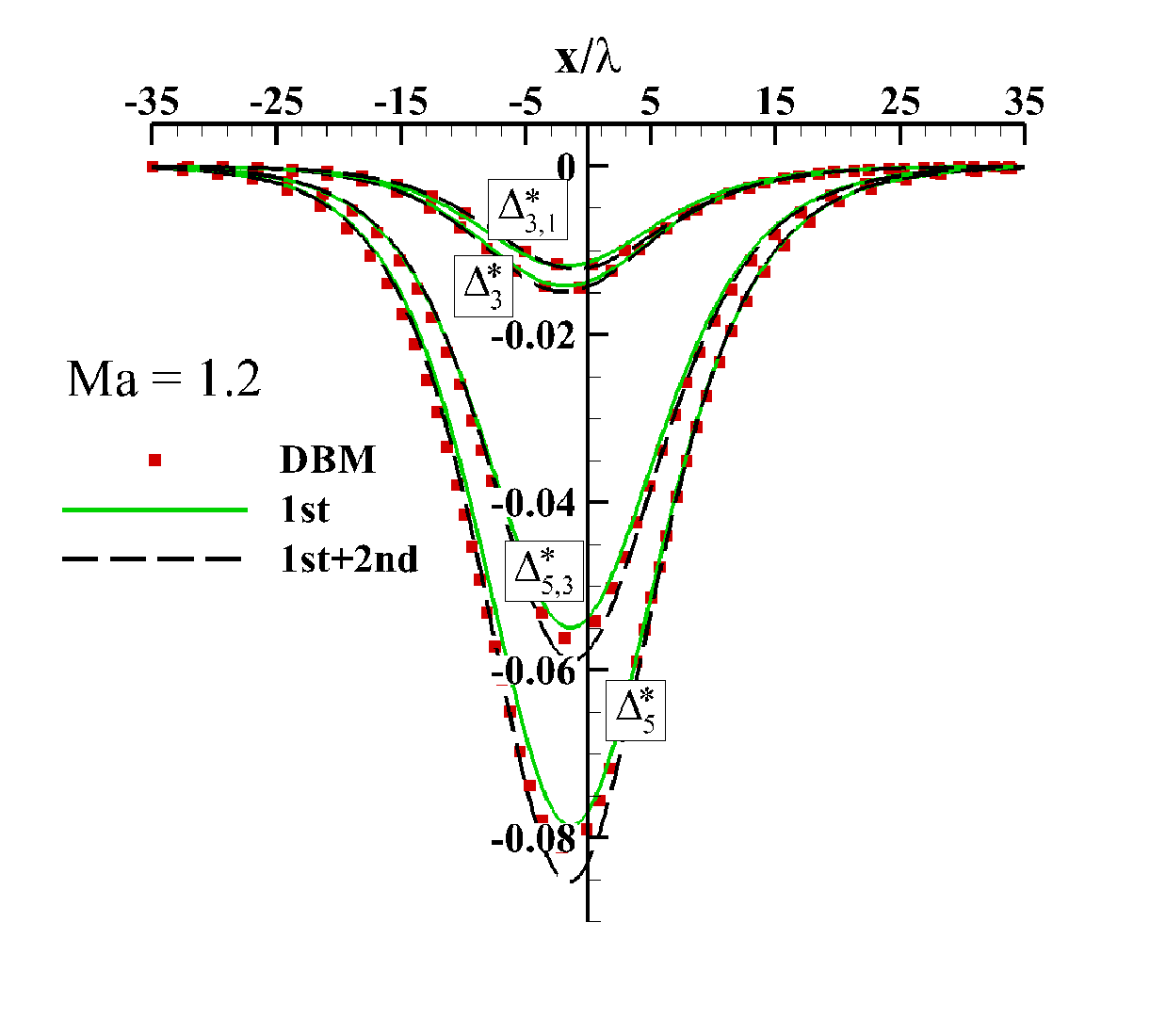}
			\centering \caption*{(b) $\mathrm{Ma} = 1.2$, odd-order: $\Delta_{3}^{*}$, $\Delta_{5}^{*}$, $\Delta_{3,1}^{*}$, and $\Delta_{5,3}^{*}$ }
		\end{minipage}
}
\par
\subfigure*{\
\begin{minipage}{8cm}
			\includegraphics*[width=8cm]{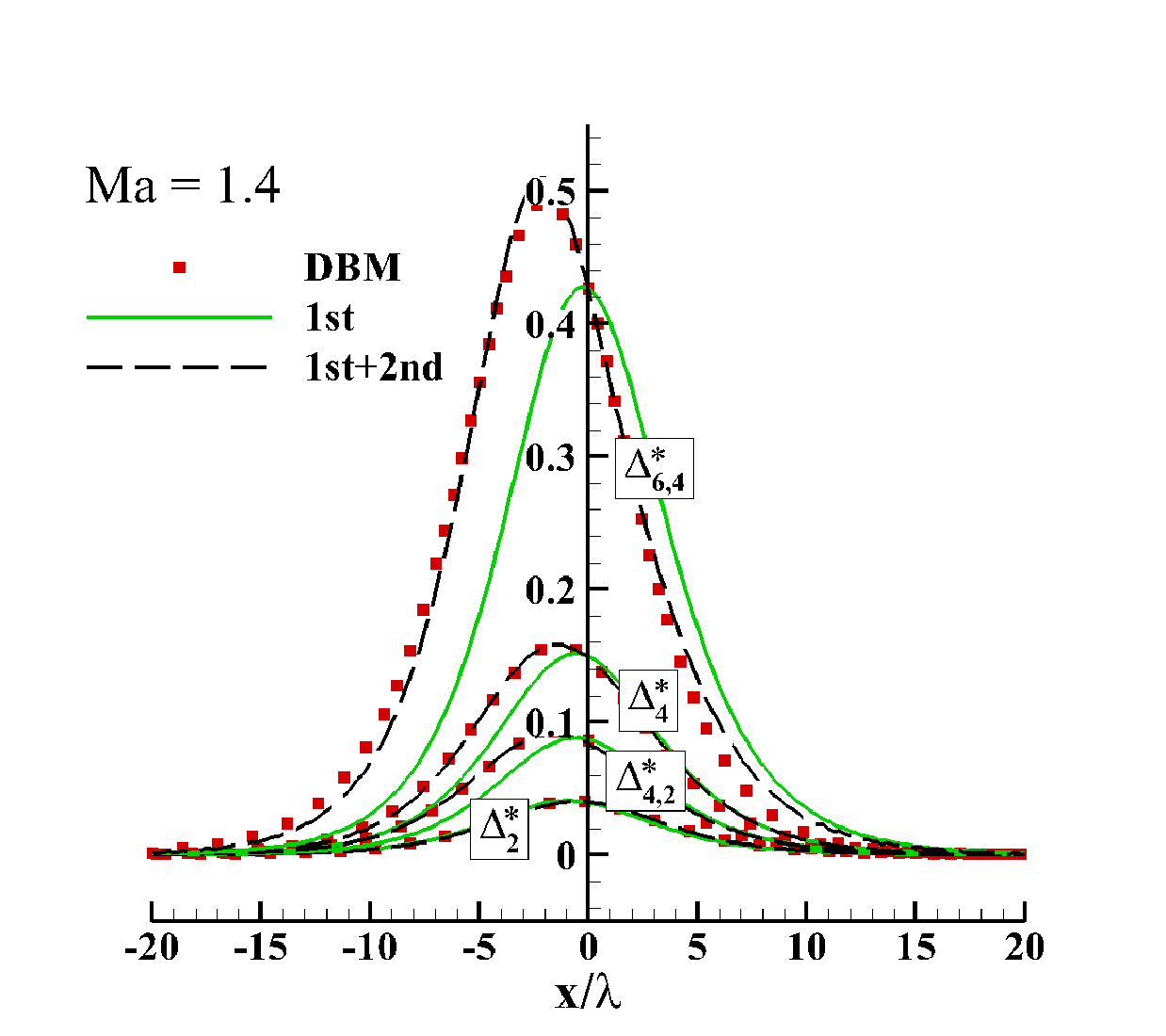}
			\centering \caption*{(c) $\mathrm{Ma} = 1.4$, even-order: $\Delta_{2}^{*}$, $\Delta_{4}^{*}$, $\Delta_{4,2}^{*}$,  and $\Delta_{6,4}^{*}$}
		\end{minipage}
\begin{minipage}{8cm}
			\includegraphics*[width=8cm]{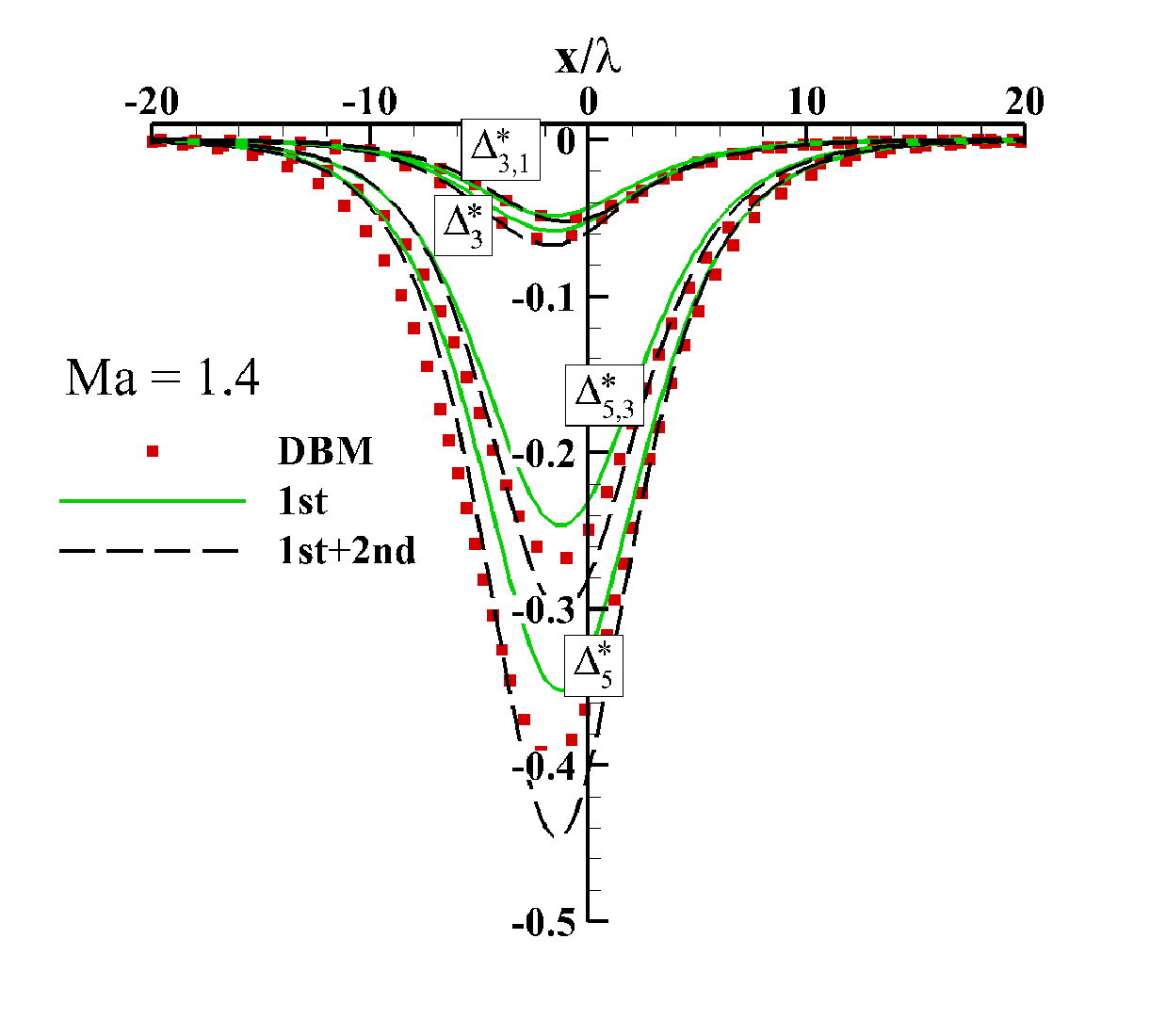}
			\centering \caption*{(d) $\mathrm{Ma} = 1.4$, odd-order: $\Delta_{3}^{*}$, $\Delta_{5}^{*}$, $\Delta_{3,1}^{*}$, and $\Delta_{5,3}^{*}$}
		\end{minipage}
}
\par
\subfigure*{\
\begin{minipage}{8cm}
			\includegraphics*[width=8cm]{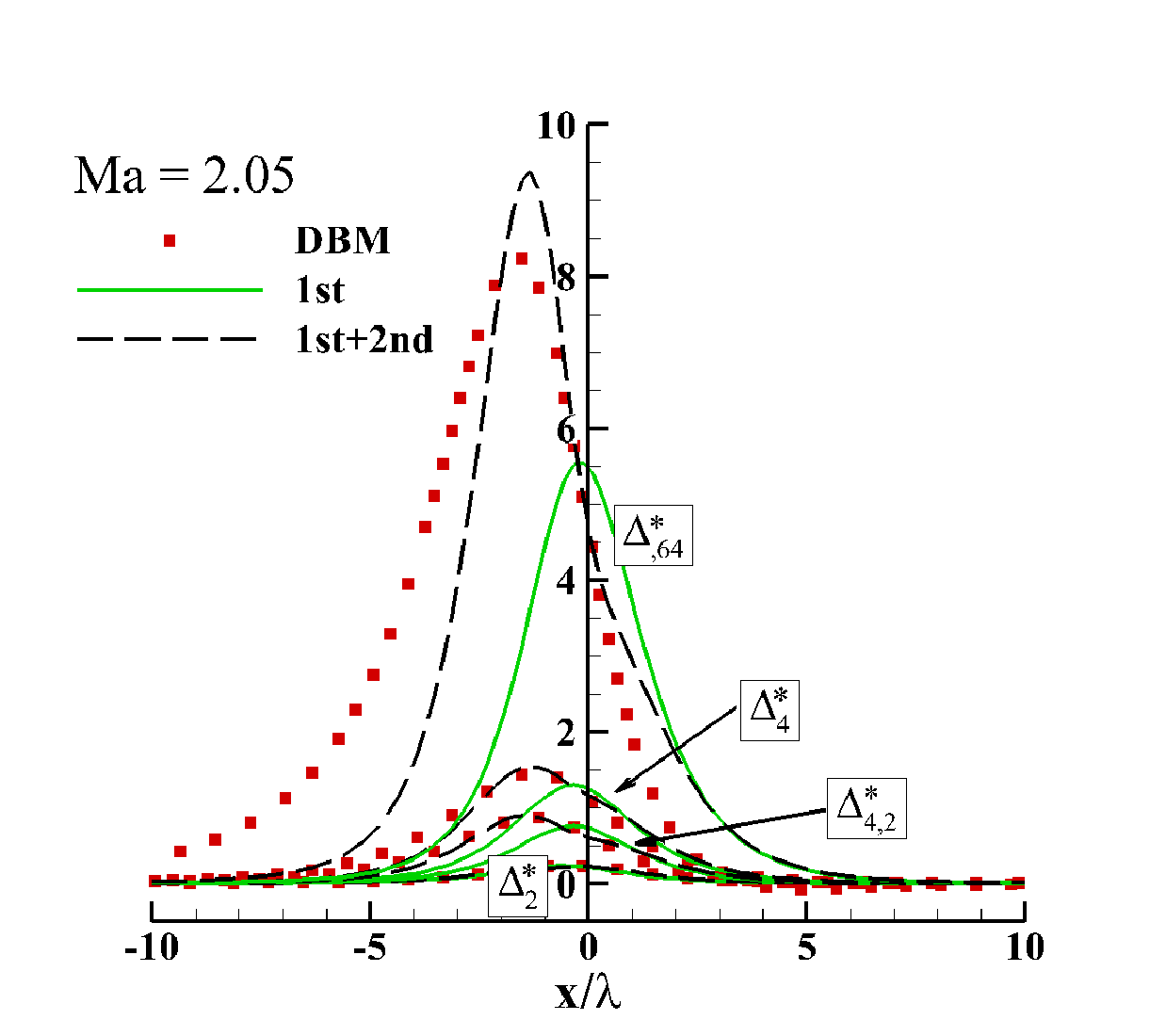}
			\centering \caption*{(e) $\mathrm{Ma} = 2.05$, even-order: $\Delta_{2}^{*}$, $\Delta_{4}^{*}$, $\Delta_{4,2}^{*}$,  and $\Delta_{6,4}^{*}$}
		\end{minipage}
\begin{minipage}{8cm}
			\includegraphics*[width=8cm]{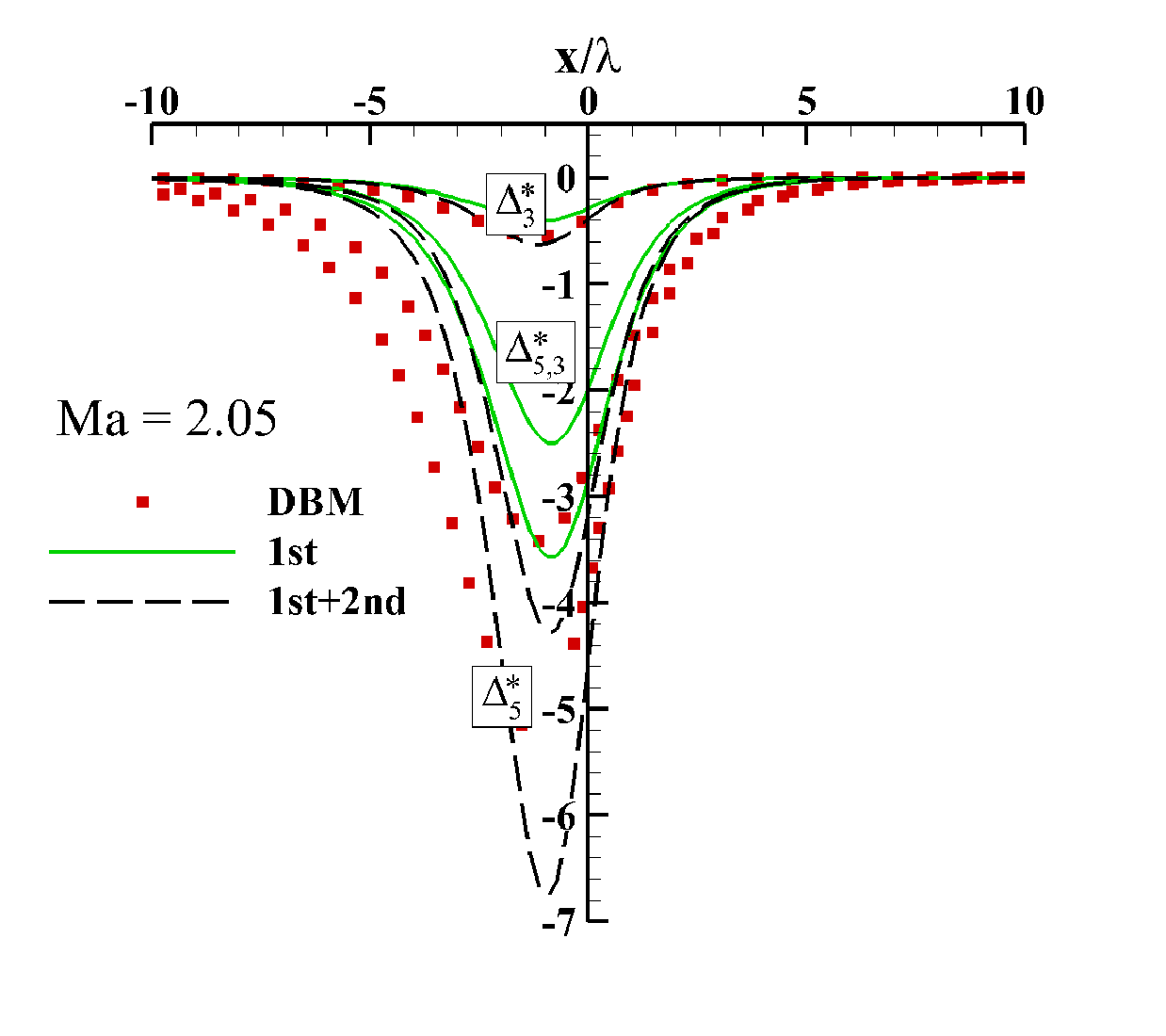}
			\centering \caption*{(f) $\mathrm{Ma} = 2.05$, odd-order:  $\Delta_{3}^{*}$, $\Delta_{5}^{*}$, $\Delta_{3,1}^{*}$, and $\Delta_{5,3}^{*}$}
		\end{minipage}
}
\caption{ Comparisons of various TNE quantities between DBM simulation and
analytical solutions for Ma numbers 1.2, 1.4, and 2.05, respectively. Here,
``1st'' refers to $\bm{\Delta}_{m}^{*(1)}$ ($\bm{\Delta}_{m,n}^{*(1)}$),
``1st + 2nd'' refers to $\bm{\Delta}_{m}^{*(1)} + \bm{\Delta}_{m}^{*(2)}$ ($
\bm{\Delta}_{m,n}^{*(1)} + \bm{\Delta}_{m,n}^{*(2)}$).}
\label{fig8}
\end{figure*}

In the framework of DBM, non-conserved kinetic moments of $(f - f^{eq})$
provide an effective approach for characterizing the states, modes, and
amplitudes of fluids deviating from equilibrium. Figure \ref{fig8} displays
the TNE states within shocks for Mach numbers $\mathrm{Ma} = 1.2, 1.4, $ and
$2.05 $, focusing on various TNE measures, including $\Delta_{2}^{*}$, $%
\Delta_{3}^{*}$, $\Delta_{4}^{*}$, $\Delta_{5}^{*}$, $\Delta_{3,1}^{*}$, $%
\Delta_{4,2}^{*}$, $\Delta_{5,3}^{*}$, and $\Delta_{6,4}^{*}$. As
illustrated, the left column represents even-order TNE quantities, while the
right column shows odd-order quantities. Comparisons between DBM simulations
and analytical solutions with first- and second-order accuracy (see Table %
\ref{table1}) are presented.

These TNE quantities offer insights into non-equilibrium characteristics
from various perspectives:

(I) As the Mach number increases, the profiles of TNE quantities shift, with
higher Mach numbers showing more pronounced deviations from equilibrium. At $%
\mathrm{Ma} = 1.2$, the TNE quantities exhibit relatively small deviations.
However, as the Mach number increases, the internal non-equilibrium states
of the shock become more evident, particularly at the shock core.

(II) Even-order TNE quantities are positive, while odd-order TNE quantities
are negative. The positive values of $\Delta_{2}^{*}$ [see Fig. \ref{fig8}%
(a)] indicate a positive deviation from equilibrium in momentum transfer,
suggesting that momentum transport within the shock generally moves toward
the compression region. Conversely, $\Delta_{3,1}^{*}$ shows that heat is
transferred in the negative direction, while the heat conduction flux $%
\Delta_{4,2}^{*}$ is positive.

(III) For the same situation, as the order $m$ of the TNE quantities
increases, their magnitude significantly increases. For instance, $%
\Delta_{4}^{*}$ indicates a much higher TNE degree compared to that inferred
from $\Delta_{2}^{*}$.

By analyzing these quantities, especially through higher-order TNE effects,
we can obtain a more detailed understanding of the shock structure and the
dominant non-equilibrium mechanisms. The shift in distribution functions and
their interaction across different Mach numbers underscores the importance
of incorporating higher-order TNE effects for accurate shock modeling.


\begin{figure*}[h]
\centering
\subfigure*{\
\begin{minipage}{9cm}
			\includegraphics*[width=9cm]{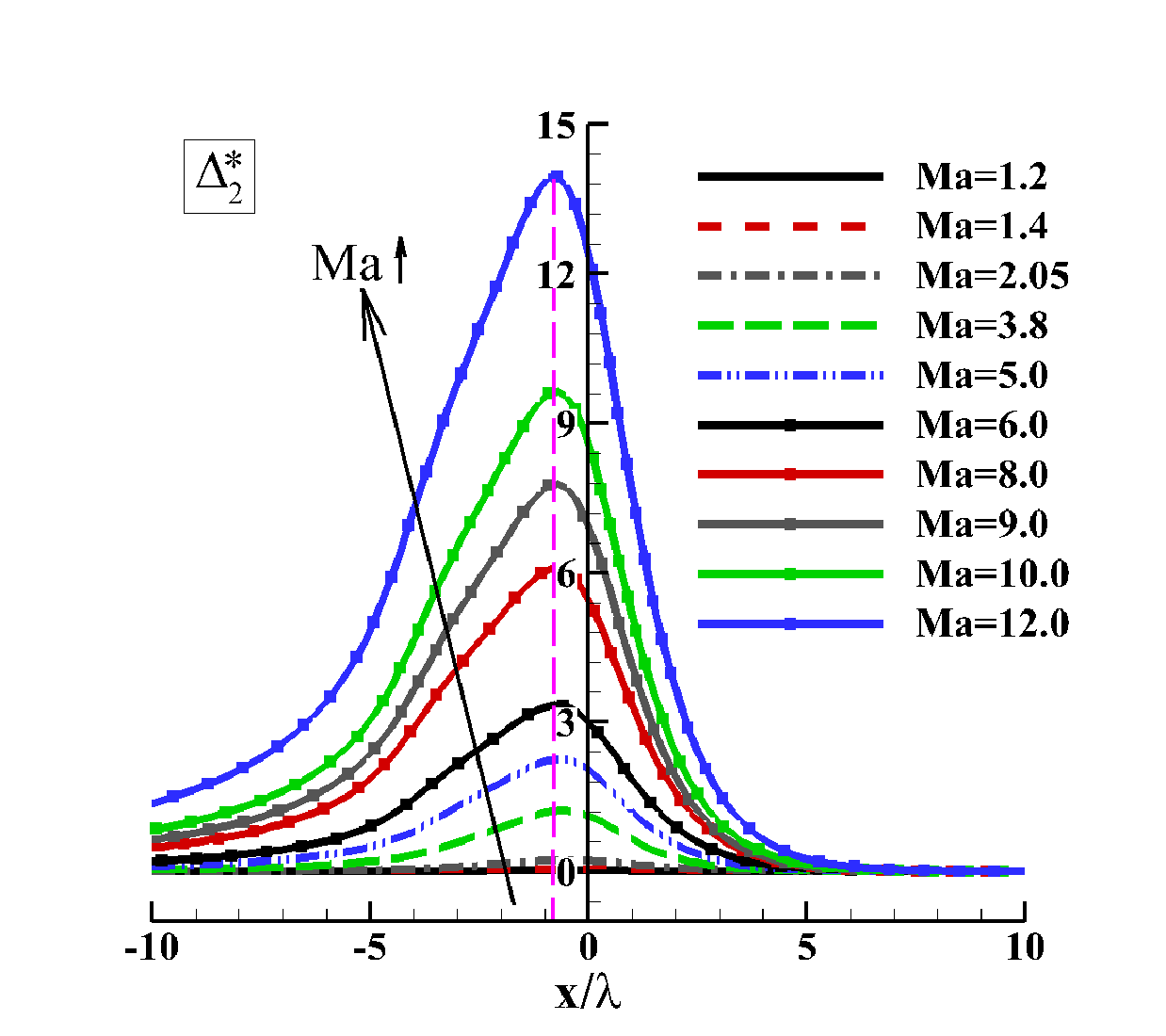}
			\centering \caption*{(a) $\Delta_2^*$}
		\end{minipage}
\begin{minipage}{9cm}
			\includegraphics*[width=9cm]{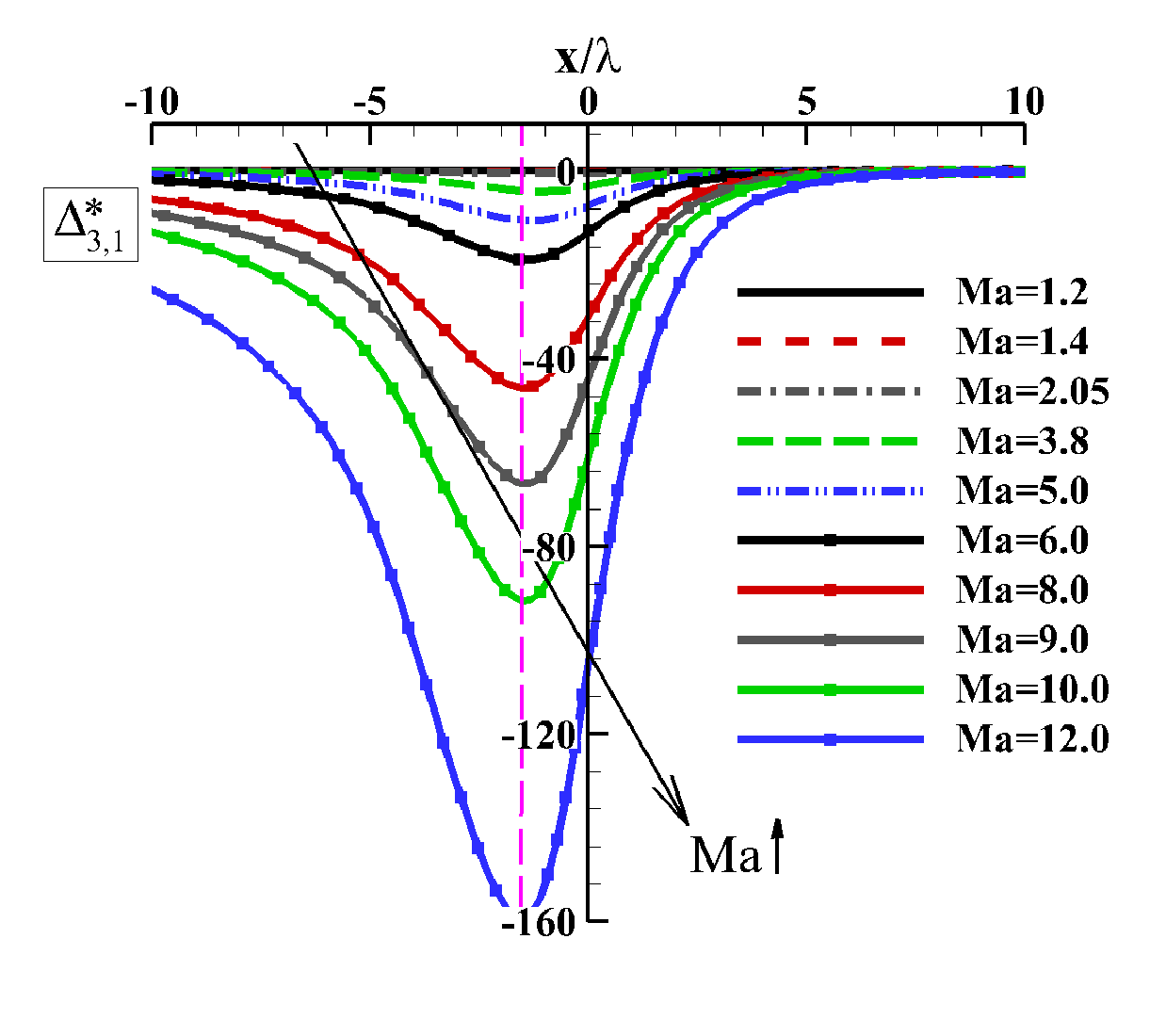}
			\centering \caption*{(b) $\Delta_{3,1}^*$}
		\end{minipage}
}
\caption{ DBM simulations of $\Delta_2^*$ and $\Delta_{3,1}^*$ for Ma
numbers ranging from 1.2 to 12.0.}
\label{fig10}
\end{figure*}

\subsubsection{ Non-equilibrium degree from perspective of TNE quantities}

The non-equilibrium order considered in physical modeling depends on two
factors: the order of relevant TNE quantities and the required precision.
The higher the order of the TNE quantities or the required precision, the
higher the TNE order the model must consider. For example, as shown in Fig. %
\ref{fig8}(a), if only the $\Delta_{2}^{*}$ quantity is of interest, a
first-order non-equilibrium model suffices. However, if higher-order TNE
quantities are involved, a second-order model is necessary. Therefore, it is
important to compare the TNE quantities of different orders between DBM
simulations and analytical solutions. Such comparisons are essential for
selecting the appropriate model order for fluid simulations.

For Ma = 1.2 [see Figs. \ref{fig8}(a) and (b)], the following observations
are made:

(I) The DBM simulation results closely align with the analytical solutions
for both even- and odd-order TNE quantities.

(II) Compared to the odd-order TNE quantities, the even-order ones show
better agreement with the analytical solutions. This improved agreement is
attributed to the stronger isotropy of even-order TNE quantities.

(III) As the order of TNE quantities increases, the differences between the
first-order analytical and DBM simulation results also grow. Second-order
(1st+2nd) analytical results show better alignment with DBM simulations
compared to first-order (1st) solutions, reflecting their higher physical
accuracy. For example, the second-order analytical solution for $%
\Delta_{6,4}^{*}$, as listed in Table \ref{table1}, provides improved
precision by incorporating second-order effects, including the reciprocal of
density, velocity, and temperature profiles, rather than relying solely on
first-order velocity effects.

(IV) As the TNE order increases, the peak location of TNE quantities shifts
closer to the inflow region. This spatial shift highlights the differences
in non-equilibrium descriptions captured by TNE quantities of varying orders.

For Ma = 1.4 [see Figs. \ref{fig8}(c) and (d)], the values of TNE quantities
increase compared to Ma = 1.2, indicating a greater deviation from
equilibrium. It is observed that:

(I) As the TNE degree intensifies, the first-order analytical solution shows
significant deviations from the DBM simulation results. However, the
second-order analytical solution maintains satisfactory agreement with the
DBM simulations.

(II) As the order of TNE quantities increases, the discrepancies between the
second-order analytical results and DBM simulations become more pronounced.

When the Mach number increases to 2.05, the TNE degree of the fluid further
intensifies, resulting in even more significant differences between the
second-order analytical results and DBM simulations. For lower-order TNE
quantities, the second-order analytical results matain their physical
accuracies, while they lose accuracies for higher-order TNE quantities.

For convenience, some analysis results from various perspectives are listed
in Table \ref{table3}. It is clear that when analyzing TNE from different
angles, the results may vary. This emphasizes the importance of describing
non-equilibrium from multiple perspectives.

\begin{table*}[htbp]
\centering
\begin{tabular}{m{3cm}<{\centering}m{7cm}<{\centering}m{5cm}<{\centering}}
\hline
Mach number & Perspective & Required TNE order \\ \hline
\multirow{6}{*}{\vspace{0.0cm}\centering $\mathrm{Ma} = 1.2$ } & $g-g^{(0)}$
& First order \\
& $\Delta_2^{*}$ & First order \\
& $\Delta_{3,1}^{*}$ & Second order \\
& $\Delta_{3}^{*}$ & Second order \\
& Higher-order even-order TNE quantities & Second order \\
& Higher-order odd-order TNE quantities & Beyond second order \\ \hline
\multirow{6}{*}{\vspace{0.0cm}\centering $\mathrm{Ma} = 1.4$ } & $g-g^{(0)}$
& First order \\
& $\Delta_2^{*}$ & First order \\
& $\Delta_{3,1}^{*}$ & Second order \\
& $\Delta_{3}^{*}$ & Second order \\
& Higher-order even-order TNE quantities & Second order \\
& Higher-order odd-order TNE quantities & Beyond second order \\ \hline
$\mathrm{Ma} = 1.55$ & $g-g^{(0)}$ & First order \\ \hline
$\mathrm{Ma} = 1.75$ & $g-g^{(0)}$ & Second order \\ \hline
\multirow{6}{*}{\vspace{0.0cm}\centering $\mathrm{Ma} = 2.05$ } & $g-g^{(0)}$
& Second order \\
& $\Delta_2^{*}$ & First order \\
& $\Delta_{3}^{*}$ & Second order \\
& $\Delta_{4,2}^{*}$ and $\Delta_{4}^{*}$ & Second order \\
& $\Delta_{6,4}^{*}$ & Beyond second order \\
& Higher-order odd-order TNE quantities & Beyond second order \\ \hline
$\mathrm{Ma} = 2.5$ & $g-g^{(0)}$ & Second order \\ \hline
$\mathrm{Ma} = 3.8$ & $g-g^{(0)}$ & Beyond third order \\ \hline
\end{tabular}%
\caption{TNE order required for different Mach numbers under various
nonequilibrium perspectives in physical modeling.}
\label{table3}
\end{table*}

\subsubsection{ Effects of Mach numbers on TNE measures}

To further examine the effects of Mach number on NOMF and NOEF inside
shocks, Figs. \ref{fig10}(a) and (b) present the profiles of $\Delta_2^*$
and $\Delta_{3,1}^*$, respectively, for Mach numbers ranging from 1.2 to
12.0. Both $\Delta_2^*$ and $\Delta_{3,1}^*$ increase as the Mach number
rises, indicating an intensification of non-equilibrium effects. From
another perspective, it can also explain as that momentum transport and heat
conduction over the $x$-direction intensify as the Mach number increases.

In Fig. \ref{fig10}(a), as the Mach number increases, the non-equilibrium
region expands, with the peak of $\Delta_2^*$ shifting progressively closer
to the inflow region. It is seen that effects of Mach numbers are not
two-stage. The increase in Mach number results in a broader shock front and
a thicker transition region, as reflected in the widening of the
non-equilibrium region. In Fig. \ref{fig10}(b), the peak of $\Delta_{3,1}^*$
consistently shifts closer to the outflow region compared to $\Delta_2^*$,
with the gap between the two profiles widening at higher Mach numbers. This
shift highlights the differing dynamics of momentum and heat transport in
the shock.

Furthermore, as the Mach number increases, the peaks of both $\Delta_2^*$
and $\Delta_{3,1}^*$ shift progressively towards the inflow region. This
trend indicates that, as the Mach number increases, the non-equilibrium
effects are more pronounced in the shock's front region, emphasizing the
increased influence of high-speed flow on the shock's internal structure.

\section{\label{sec:conclusion} Conclusion}

Shock waves, as representative non-equilibrium flow phenomena, exhibit pronounced hydrodynamic nonequilibrium (HNE) and thermodynamic nonequilibrium (TNE) effects due to their inherently small-scale structures and rapid kinetic modes. Despite their ubiquity in natural and engineering systems, the underlying mechanisms driving HNE and TNE effects within shock structures remain insufficiently understood.

This study examines HNE and TNE characteristics in argon shock structures using a higher-order discrete Boltzmann method (DBM), a kinetic modeling framework for simulating discrete/ non-equilibrium effects and analyzing complex physical phenomena.
A key feature of the DBM in this study is its direct discretization of velocity space, which preserves high-order HNE and TNE effects. To extract non-equilibrium manifestations, HNE and TNE quantities of different orders are defined using the non-conserved kinetic moments of $(f - f^{eq})$. To hierarchically investigate non-equilibrium mechanisms, higher-order analytical solutions for both distribution functions and TNE quantities are derived using Chapman-Enskog multiscale analysis, though DBM simulations do not depend on these theoretical results.

First, the accuracy of the multiscale DBM is validated by comparing macroscopic characteristics, such as interface profiles and thickness, from DBM simulations with experimental data and DSMC results. Next, the agreement between DBM simulations and analytical solutions is assessed at the mesoscopic level by comparing kinetic quantities such as distribution functions and TNE measures. Results indicate that analytical solutions incorporating higher-order TNE effects align more closely with DBM simulations than those restricted to lower-order effects.

The impact of Mach number on HNE characteristics is analyzed by examining the shape and thickness of density, temperature, and velocity interfaces. Key findings include: (i) The influence of Mach number on macroscopic quantities follows a two-stage trend, affecting both interface smoothness and thickness. (ii) As the Mach number increases, the region of strong compressibility shifts from the outflow vicinity to the inflow region. The effect of Mach number on TNE characteristics is further investigated through different TNE measures. The deviation patterns and amplitudes of these measures at different shock positions are analyzed from multiple perspectives. Results show that a higher Mach number significantly intensifies TNE and enlarges the non-equilibrium region.

As a hallmark of multiscale system complexity, non-equilibrium effects manifest differently across analytical perspectives. Understanding HNE and TNE characteristics in shock waves is crucial for multiscale model selection and provides kinetic insights into cross-scale coupling mechanisms governing complex macroscopic and mesoscopic phenomena. Future research will focus on developing high-order DBM models in two and three dimensions to explore richer and more intricate mesoscopic HNE and TNE characteristics, mechanisms, and governing principles. These studies will incorporate increased degrees of freedom, diverse non-equilibrium driving forces, and enhanced spatiotemporal multiscale coupling.

\section*{ Declaration of competing interest}

The authors declare that they have no known competing financial interests or
personal relationships that could have appeared to influence the work
reported in this paper.

\section*{ Acknowledgments}

The authors would like to thank Prof. Yudong Zhang and Dr. Jiahui Song for their insightful discussions. We also acknowledge support from the National Natural Science Foundation of China (Grant Nos. 11875001 and 12172061), the Foundation of the National Key Laboratory of Computational Physics (Grant No. SYSQN2024-10), the High-level Talents Research Start-up Grant from Guangxi University (Grant No. ZX01080031224009), the Hebei Outstanding Youth Science Foundation (Grant No. A2023409003), the Central Guidance on Local Science and Technology Development Fund of Hebei Province (Grant No. 226Z7601G), the
Opening Project of State Key Laboratory of Explosion Science and Safety Protection (Beijing Institute of Technology) (Grant No. KFJJ25-
02M), and the Foundation of the National Key Laboratory of Shock Wave and Detonation Physics (Grant No. JCKYS2023212003).


\bibliographystyle{elsarticle-num}
\bibliography{shock-wave}

\end{document}